\documentclass[preprint,12pt,number]{elsarticle}

\usepackage{graphicx}
\usepackage{amsmath, amssymb, amsthm}
\usepackage{algorithmic}
\usepackage{array}
\usepackage{amssymb}
\usepackage{multirow}
\usepackage{float}
\usepackage{longtable}
\usepackage{hyperref}
\usepackage{cleveref}
\usepackage{graphicx}
\usepackage{caption}
\usepackage{subcaption}
\usepackage{booktabs}
\theoremstyle{plain}
\newtheorem{lemma}{Lemma}

\journal{Transportation Research Part C}

\begin{document}
\begin{frontmatter}

\title{From Voice to Safety: Language AI Powered Pilot-ATC Communication Understanding for Airport Surface Movement Collision Risk Assessment}

\affiliation[inst1]{organization={Department of Aerospace Engineering and Engineering Mechanics},
addressline={The University of Texas at Austin}, 
city={Austin},
postcode={78712}, 
state={TX},
country={USA}}
            
\author[inst1]{Yutian Pang\corref{mycorrespondingauthor}}
\cortext[mycorrespondingauthor]{Corresponding author.}
\ead{yutian.pang@austin.utexas.edu}
\author[inst1]{Andrew Kendall}
\author[inst1]{Alex Porcayo}
\author[inst1]{Mariah Barsotti}
\author[inst1]{Anahita Jain}
\author[inst1]{John-Paul Clarke}

\begin{highlights}
\item A review of language AI usage in air traffic control and civil aviation risk modeling is provided. 
\item We introduce a novel framework to integrate natural language processing with surface collision risk modeling to enhance aviation safety.
\item A hybrid rule-based NER model using domain-specific rules is developed, improving the recognition of key entities in pilot-ATC communications.
\item Surface movements are modeled with log-normal distributions and employ node-link graph structures to estimate spatiotemporal collision probabilities.
\item Real-time risk assessment at overlapping nodes with different warning thresholds and lead time analysis.
\item We validate the log-normal link travel speed assumptions by conducting data analysis and statistical tests on ASDE-X ground movement data. 
\item Three case studies, the Haneda runway collision, the KATL taxiway collision, and the Tenerife airport disaster, demonstrate the effectiveness in detecting high-risk nodes.
\end{highlights}

\begin{abstract}

Surface movement collision risk is critical for airport safety. These models play a vital role in identifying and mitigating potential hazards during airport ground operations by providing warnings of near-miss incidents, thereby reducing the risk of accidents that could jeopardize human lives and financial assets. However, existing models, developed decades ago, have not fully integrated recent advancements in machine intelligence, where incorporating additional functionalities presents promising opportunities for improved risk assessment. This work provides a feasible solution to the existing airport surface safety monitoring capabilities (i.e., Airport Surface Surveillance Capability (ASSC)), namely language AI-based voice communication understanding for collision risk assessment. The proposed framework consists of two major parts, (a) rule-enhanced Named Entity Recognition (NER); (b) surface collision risk modeling. NER module generates information tables by processing voice communication transcripts, which serve as references for producing potential taxi plans and calculating the surface movement collision risk. We first collect and annotate our dataset based on open-sourced video recordings and safety investigation reports. Additionally, we refer to FAA Order JO 7110.65W and FAA Order JO 7340.2N to get the list of heuristic rules and phase contractions of communication between the pilot and the Air Traffic Controller (ATCo). Then, we propose the novel ATC Rule-Enhanced NER method, which integrates the heuristic rules into the model training and inference stages, resulting in a hybrid rule-based NER model. We show the effectiveness of this hybrid approach by comparing different setups with different token-level embedding models. For the risk modeling, we adopt the node-link airport layout graph from NASA FACET and model the aircraft taxi speed at each link as a log-normal distribution and derive the total taxi time distribution. Then, we propose a spatiotemporal formulation of the risk probability of two aircraft moving across potential collision nodes during ground movement. Furthermore, we propose the real-time implementation of such a method to obtain the lead time, with a comparison with a Petri-Net based method. We show the effectiveness of our approach through case studies, (a) the Haneda airport runway collision accident happened in January 2024; (b) the KATL taxiway collision happened in September 2024; (c) the Tenerife airport disaster in March 1977. We show that, by understanding the pilot-ATC communication transcripts and analyzing surface movement patterns, the proposed model estimates the surface movement collision probability within machine processing time, thus enabling proactive measures to possible collisions at a certain node, which improves airport safety. A study on validating the log-normal assumption of aircraft taxi speed distributions is also given. We provide the link to code and data repository \href{https://github.com/YutianPangASU/ATC-Text-Communication}{HERE}. 

\end{abstract}


\end{frontmatter}

\section{Introduction \label{sec: introduction}}

\begin{figure}
\centering
    \begin{subfigure}[t]{0.85\textwidth}
        \includegraphics[width=\textwidth]
        {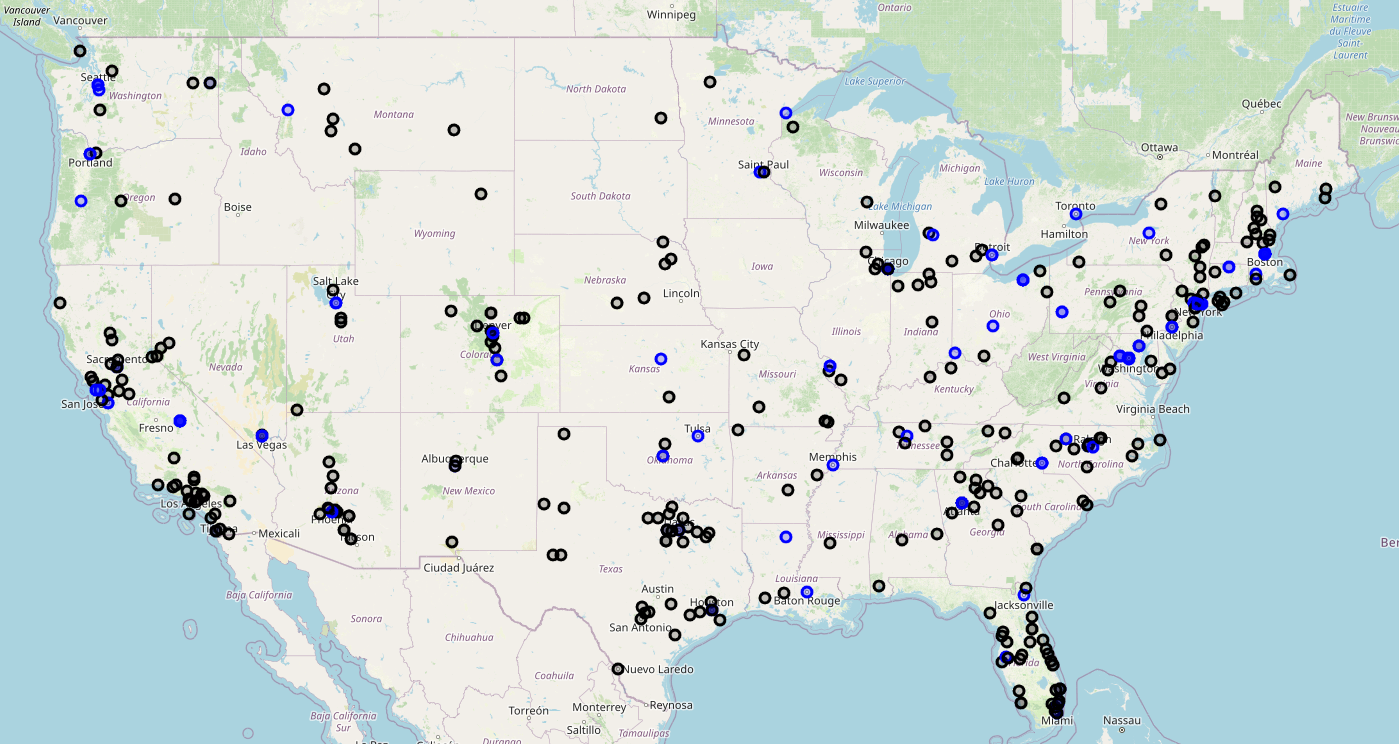}
        \caption{}
        \label{fig: accident-map}
    \end{subfigure}
    \begin{subfigure}[t]{0.95\textwidth}
        \centering
        \includegraphics[width=\textwidth]
        {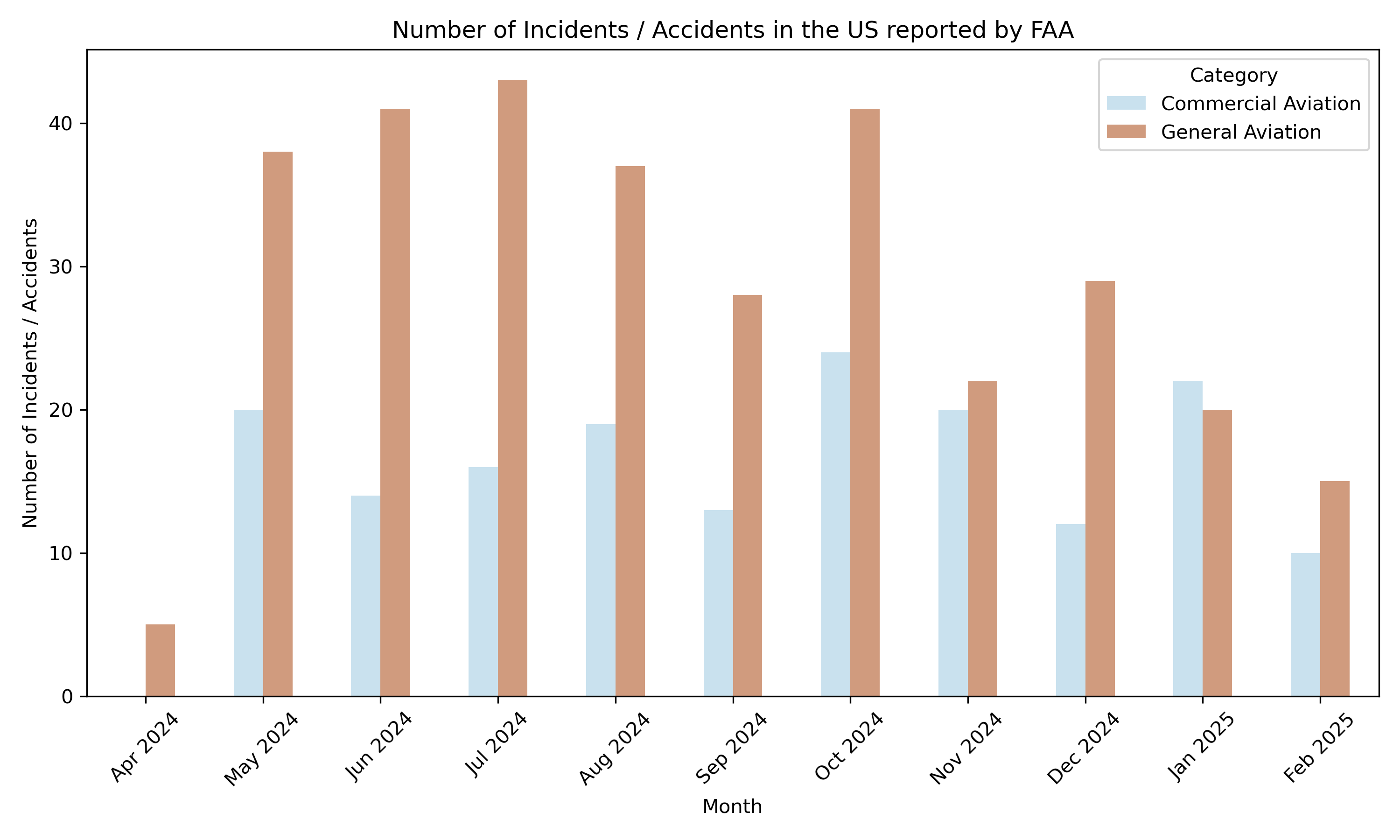}
        \caption{}
        \label{fig: accident-hist}
    \end{subfigure}
\caption{Accidents and Incidents in the continental U.S. from April 2024 to February 2025, documented by the FAA \citep{FAA_Accidents_Incidents}. In \Cref{fig: accident-map}, blue dots indicate commercial aviation events, and black dots indicate general aviation events. }
\label{fig: faa-accident}
\end{figure}

The United States is currently facing an alarming escalation in aviation accidents. In the first six weeks of 2025 alone, there have been over 30 commercial aviation incidents/accidents, including four major plane crashes that have tragically claimed 85 lives (\Cref{fig: faa-accident}). This surge in accidents is unprecedented, especially considering that prior to 2025, the most recent fatal crash involving a U.S. airliner dated back in 2009 \citep{bush2011crash}. This disturbing trend not only deters passengers but also poses significant economic threats to the industry, which is already grappling with financial instability. Immediate and robust research and development efforts are imperative to enhance safety protocols, rectify systemic deficiencies, and restore public trust in air travel. The safety of airport surface operations remains a critical challenge, particularly with the increasing complexity and traffic volume of modern airports \citep{attaccalite2012risk}. The safety of airport surface operations encompasses both runway and taxiway environments, where the movement of aircraft and ground vehicles occurs in close proximity \citep{icao2010faa, icao2012strategic}. The complexity of surface operations stems from the diverse and dynamic nature of airport environments. For instance, the Flight Safety Foundation (FSF) reported that 30\% of commercial aviation accidents between 1995 and 2008 were runway-related, resulting in 973 fatalities \citep{wilke2015modelling}. 

Runway incursions, defined as the unauthorized presence of an aircraft, vehicle, or person on a runway \citep{icao2007runway}, represent one of the most significant threats to safe airport operations, where a substantial portion is attributed to human error in communication between pilots and controllers. Runway incursions can be attributed to pilot deviations, operational incidents, and vehicle-related anomalies, collectively challenging the effectiveness of conventional safety protocols \citep{rankin2012faa}. Federal Aviation Administration (FAA) recorded 7,864 runway incursions from 2012 and July 2017, with 63\% (4,947 incidents) attributed to pilot deviations, 20\% (1,529 incidents) to air traffic control (ATC) errors, and 17\% (1,388 incidents) to vehicle or pedestrian deviations \citep{werfelman2017tracking}. In 2024, FAA reported a total of 1,757 runway incursions, indicating a persistent safety concern \citep{faa2024runwayincursions}. Additionally, FAA identified an annual occurrence rate of 0.200 Category A and B incursions per million operations in 2013, signaling the persistent risk posed by ground-based collisions \citep{faa2014par}. One recent runway incursion case is the Haneda airport runway collision happened on January 2nd 2024, where an Airbus A350 collided with a Japan Coast Guard airplane during landing. The collision resulted in both aircraft catching fire. Although all 379 occupants of the Japan Airlines flight were safely evacuated, five of the six crew members on the Coast Guard aircraft perished. The accident was due to human error, citing miscommunication and misunderstanding between the Coast Guard pilot and air traffic control as a primary factor \citep{bisset2024379}.

While runway incursions have historically garnered significant attention due to their potential for catastrophic outcomes, the number of taxiway collisions is also increasing and gaining public attention. Taxiway collisions occur when aircraft or vehicles inadvertently occupy the same space on the ground, leading to significant safety hazards, operational disruptions, and financial losses. Studies have shown that areas with higher taxiway occupancy rates are more prone to collisions, especially during peak operational periods. For instance, research indicates that increased workload during busy times can elevate collision risks, making it imperative to focus on these high-density areas, or potential collision spots, for effective risk management \citep{sun2024collision}. To address taxiway collisions, the Surface Safety Risk Index (SSRI) is developed by the FAA, which assigns risk weights to various outcomes, such as aircraft damage or injuries, based on their proximity to potential fatalities. Another recent taxiway collision case happened on September 10th 2024, at the Hartsfield-Jackson Atlanta International Airport (KATL) involved an Airbus A350's wing striking the tail of a CRJ-900 regional jet during taxiing, resulting in significant damage to both aircraft \citep{NTSB_DCA24FA299_2024}. 

These statistics and accident cases underscore the persistent challenges associated with both runway incursions and taxiway collisions, highlighting the critical need for enhancing surface monitoring and risk assessment capabilities. Effort has been made towards this direction. A comprehensive, technology-driven efforts have been made towards enhancing runway safety, such as integrating advanced surveillance, alerting systems, and infrastructural improvements. Runway Status Lights (RWSL) serve as a critical safety layer by utilizing real-time surveillance data to alert aircrews and vehicle operators to hazardous runway conditions, while surface surveillance systems such as ASDE‑X or ASSC provide air traffic controllers with highly accurate, integrated displays of surface movements. Enhancements like Taxiway Arrival Prediction (TAP) further refine these systems by alerting controllers when aircraft deviate from their assigned paths, thereby reducing misalignment risks \citep{FAA_AC_150_5300_13B_2024}. However, neither ASDE-X nor ASSC considers the input from the ATC radio communication signals, making it hard to check for pilot-ATC communication induced risks. The rapid advancement of artificial intelligence (AI) paves the way to understand and extract key information from the voice communications through machine intelligence. Natural Language Processing (NLP), as a sub-field of AI, focuses on understanding and analyzing the contextual meaning of human spoken language, and is thus considered in this research work. Specifically, the NLP model working on understanding the voice communication transcripts is also known as Automatic Speech Recognition (ASR). On the tactical level, extracting keywords from the textual format transcripts is one of the major interests. Information retrieval takes keywords and useful contexts from ASR-converted text using either token-based or contextual-based language model. This process is known as Named Entity Recognition (NER) where entities are the extracted keywords. A detailed review of the various usage of NER is given in \Cref{subsec: language-ai-review}.

Despite the demonstrated effectiveness of NLP to other domains, air traffic control is heavily regulated by communication rules and standards to ensure clarity, consistency, and safety in all ATC interactions. For instance, the air traffic control manual FAA JO 7110.65 defines the standardized phraseology and terminology that controllers are required to use during pilot-ATC communications \cite{FAA_ATC}. \cite{FAA_7340.2N_2024} provides the word and phrase contractions used by all of the air traffic management related parties. This leads to the question: \textit{how can these heavily regulated ATC rules be incorporated into a data-driven NLP framework?} On the other side, although English is the standard language used for air traffic control, multilingualism in aviation communications warrants careful consideration because the inherently international nature of air travel means that pilots and controllers often operate with diverse linguistic backgrounds, miscommunications stemming from accents, code-switching, and non-standard phraseology can lead to critical errors that compromise safety \citep{lin2021atcspeechnet, krejvcikovamiscommunication}. This poses the challenge: \textit{how can the model be ensured to be multilingual for potential usage?}. Moreover, recent advancements of large language models (LLMs) in NLP enables the capability of performing various tasks through prompt engineering. These models can also be adapted to different domains using either fine-tuning or domain-specific Retrieval-Augmented Generation (RAG) \citep{wang2024aviationgpt}. However, these LLM-based solutions require a significant amount of computer power during both the training and inference stages \citep{wang2024aviationgpt, andrade2023safeaerobert, tikayat2023aerobert}, which are extremely critical for onboard applications in near real-time. In \citep{wang2024aviationgpt}, the llama2-based fine-tuning process requires hundreds of gigabyte (GB) of CPU and GPU memory, and at least 25 GB of GPU memory during inference. Impressively, \citep{andrade2023safeaerobert} mentioned the time required for pre-training of SafeAeroBERT takes over two months. This leads to the third focus: \textit{how can we ensure the model is of suitable size for efficient real-time information extraction on-prem?} Lastly, the existing collision risk models fail to consider the spatial layout of the airport in a spatiotemporal fashion \citep{sun2024collision}, which presents the challenge: \textit{how can the ground risk be formulated in a spatiotemporal sense, based on the estimated travel time distribution to the potential collision spots, which can utilize the output from the novel rule-enhanced machine learning model?}

To address these aforementioned challenges, we propose an approach to expand the existing Airport Surface Surveillance Capability (ASSC) with Language AI-based voice communication understanding for surface movement collision risk assessment. The proposed framework comprises of two main components, (a) Automatic Speech Recognition (ASR); (b) surface collision risk modeling. The ASR module processes voice communication transcripts to generate information tables, which serve as references for developing potential taxi plans and estimating the risk of surface movement collisions. We collected and annotated our own Named Entity Recognition (NER) dataset using open-source video recordings and safety investigation reports. We reference FAA Order JO 7110.65W and FAA Order JO 7340.2N to obtain heuristic rules and phase contractions that are commonly used in routine communications between pilots and air traffic controllers. Building on this, we propose the novel ATC Rule-Enhanced NER method that integrates these heuristic rules during both model training and inference, resulting in a hybrid rule-based ASR model. We validate the effectiveness of our ASR approach by comparing various setups employing different token-level embedding models. For the risk modeling component, we utilize the node-link airport layout graph from NASA FACET \citep{bilimoria2001facet}. We model the aircraft taxi speed on each link as a log-normal distribution and derive the overall taxi time distribution. The collision risk is then determined by convolving the total travel time distributions at nodes where potential collisions between aircraft might occur. 

Our hybrid learning model alone provides the necessary information for an aircraft surface movement compliance check, which happens when a pilot misinterprets the given clearance. Moreover, by adding a risk module, our work enables the estimation of airport surface movement collision risk assessment capability, which happens either if the ATCo misjudges the speed of the aircraft, or if the ATCo forgets a previous issued clearance. The contributions are highlighted here as,

\begin{itemize}
    \item We created our own ATC communication transcript dataset and annotated in NER training format, as well as our approach of encoding ATC rules as regular expressions. 
    \item We provide the hybrid-learning approach of incorporating the ATC-rules into NER training. Through evaluating and comparing different setups and token-level embedding models, we show the complexity and sensitivity studies between setups. 
    \item We propose the link travel speed based spatiotemporal airport ground collision risk formulation, and provide the real-time implementation of risk warning system. 
    \item We show effectiveness of the proposed framework through the reconstruction of three real-world accident scenarios. 
\end{itemize}

The rest of the paper is organized as follows: 
\Cref{sec: review} gives a literature review of related literature, where a review of the usage of language AI in aviation \Cref{subsec: language-ai-review}, along with a review of surface collision risk model is provided in \Cref{subsec: risk-review}. \Cref{sec: methodology} discusses the proposed methodology, which is composed of two sub-modules, the ATC rule-enhanced learning module \Cref{subsec: ner-learning} and the risk formulation in \Cref{subsec: riskmodel}. We demonstrate our approach with three case studies in \Cref{sec: case-study}. \Cref{sec: conclusion} concludes the paper.

\section{Related Work \label{sec: review}}
\subsection{Language AI in Aviation \label{subsec: language-ai-review}}

Natural language processing (NLP) applies computational techniques to learn, understand, and generate human language, evolving from early rule‐based and symbolic approaches to modern statistical and deep learning methods \citep{manning2014stanford, hirschberg2015advances}. Early systems relied on handcrafted grammars and rules, but the availability of large annotated corpora and advances in machine learning spurred breakthroughs in tasks such as machine translation \citep{koehn2003statistical, brown1993mathematics}, where phrase‐based models have gradually been supplanted by neural network architectures \citep{sutskever2014sequence, bahdanau2014neural}, as well as in speech recognition, dialogue systems, and sentiment analysis \citep{young2013pomdp, hinton2012deep}. These advancements have enabled real-world applications like conversational agents and real-time translation services, yet challenges remain in modeling complex semantic nuances and extending robust NLP capabilities to low-resource languages \citep{angeli2014naturalli}. Two major research directions in NLP for Aviation safety are, (a) aviation safety analysis on incident/accident reports; (b) air traffic control (ATC) communications transcripts analysis \citep{yang2023natural}. These research directions are essential for enhancing safety, operational efficiency, and decision-making processes. By leveraging NLP, we can achieve more accurate and timely analysis of safety data, improve the clarity and effectiveness of communication between pilots and air traffic controllers, and better manage the growing complexity of air traffic systems \citep{badrinath2022automatic, helmke2016reducing}. 

The key research direction of NLP usage in aviation safety reports include root cause analysis and critical factors identification related to aviation incidents/accidents, giving risk assessment insights to inform risk management strategies and proactive operations. Topic modeling and pattern analysis is also a way to uncover and understand the latent behavior and common pattern of multiple incidents/accidents, such that operators can identify emerging risks with similar patterns from historical reports. U.S. National Aeronautics and Space Administration (NASA) Aviation Safety Report System (ASRS) database and the U.S. National Transportation Safety Board (NTSB) are two commonly used databases for aviation safety/accident reports. Machine learning classification models such as SVM, CNN, RNN, LSTM are adopted in the literature. For instance, \citep{abedin2010cause} proposes a method for incident root course analysis using the weakly supervised semantic lexicon learning and support vector machine (SVM) for root cause identification from ASRS. On the technical level, researchers propose various methods to identify risk-related causes from flight status \citep{zhang2019ensemble, shi2017data, andrzejczak2012application}, human factors \citep{perboli2021natural, ahadh2021text, tanguy2016natural}, spatial-temporal relations \citep{robinson2019temporal}, and anomalies \citep{jiao2022classification, dong2021identifying}. However, contextual ambiguity and multilingual scenarios usually lead to misinterpretations or errors in these studies.  

For communication transcripts, Automatic Speech Recognition (ASR), information retrieval, error detection, and speaker classification are major research directions \citep{yang2023natural}. ASR processes audio signals to identify words and phrases spoken by a human speaker, transcribing them into a textual format that can be analyzed and used by various applications \citep{badrinath2022automatic, amin2023low}. Information retrieval takes keywords and useful contexts from ASR converted text using either token-based or contextual-based language model. This process is known as Named Entity Recognition (NER) where entities are the extracted keywords. In air traffic communication transcripts understanding, the ability to accurately extract ATC domain‐specific entities (such as aircraft identifiers, altitudes, call signs, and route information) is critical for both operational safety and real-time decision making, where flight callsigns, and the intended destination are of critical interests \citep{chen2015closed, abedin2010cause, dongyue2024multi}. NER-based information extraction communication transcripts are the key step of building a ATC deviation warning system, where the computer can determine of the aircraft is showing the expected behavior based on the extracted keywords and real-time location. Recent work also emphasizes that AI-driven advisory systems in ATC must also address explainability and trust calibration, which reveals the need for explanations varies with operational goals and explanation design should begin with controllers’ reasoning needs rather than researcher assumptions \citep{fennedy2025atcos}.

ATC communication transcript understanding tasks currently face a data scarcity challenge, with available datasets comprising a mix of open-source and paid subscriptions. Command-related databases, such as MALORCA \citep{srinivasamurthy2017semi, kleinert2018semi}, HIWIRE \citep{segura2007hiwire}, ATCOSIM \citep{hofbauer2008atcosim}, UWB ATCC \citep{vsmidl2019air}, and AIRBUS \citep{delpech2018real}, collectively offer approximately 176 hours of speech data, capturing the highly constrained, standardized phraseology of ATC communications (i.e., limited vocabularies, specific callsigns, and technical jargon) under challenging acoustic conditions. These datasets are typically derived from lengthy recordings where only brief command segments (approximately 10–15 minutes per hour of raw data) are usable after extensive manual transcription efforts. The ATCO2 project \citep{zuluaga2020automatic, zuluaga2022atco2, zuluaga2023lessons} further enriches this landscape by developing a unique platform to automatically collect, organize, and pre-process ATC speech data from diverse sources, including publicly accessible radio frequency channels such as \href{https://www.liveatc.net/}{LiveATC.net} and indirect feeds from Air Navigation Service Providers. In contrast, widely available out-of-domain corpora like Librispeech \citep{panayotov2015librispeech} and Commonvoice \citep{ardila2019common} are also adopted for transfer learning to mitigate these in-domain data limitations.

NER methodologies in aviation communication transcript understanding have evolved from rule-based systems to deep learning-based models. Early work in aviation NER relied on expert-crafted rules using regular expressions and domain-specific lexicons to capture the complex, standardized language of ATC communications. These systems offer high interpretability and precision under controlled conditions; however, their rigidity limits adaptability to variations in speaker accents, noise levels, and spontaneous deviations. Statistical techniques such as Conditional Random Fields (CRFs) and Support Vector Machines (SVMs) are used to overcome the inflexibility of rule-based methods. These models effectively capture contextual dependencies but demand significant feature engineering (i.e., parts-of-speech, orthographic and positional features). Recent advances have seen a marked shift towards deep learning techniques for NER in aviation transcripts. These models automatically learn contextual representations from raw input text, reducing the need for extensive manual feature engineering \citep{chandra2024aviation}. \Citep{badrinath2022automatic} employed Mozilla’s Deep Speech implementation in combination with the Spacy library’s NER module to process ATC communication transcripts. Their system leverages deep learning to capture both the sequential context of spoken instructions and the unique syntactic patterns of ATC language. Moreover, transformer-based models are emerging as strong candidates due to their capacity for modeling long-range dependencies and handling variable-length sequences, which is particularly beneficial given the noisy and often rapidly spoken nature of ATC communications. Pre-trained language models (such as BERT or its derivatives) can be fine-tuned on domain-specific datasets, allowing these models to adapt to the peculiarities of aviation language switching phenomena \citep{chandra2023aviation}. The recent advancement of large language models (LLMs) has enabled the possibility of unified solutions of the above two directions. LLMs can achieve multiple tasks from document understanding to communication transcripts extraction thanks to prompt engineering, and are able to be generated to different domains using either fine-tuning or Retrieval-Augmented Generation (RAG) \citep{wang2024aviationgpt}. However, these LLM-based solutions require a significant amount of computer power during both the training and inference stage \citep{wang2024aviationgpt, andrade2023safeaerobert, tikayat2023aerobert}, which are drawbacks for onboard applications in near real-time. In \citep{wang2024aviationgpt}, the Llama2-based tuning process requires hundreds of gigabytes (GBs) of CPU and GPU memory to fine-tune, and requires at least 25 GB of GPU memory for machine learning inference. Impressively, \citep{andrade2023safeaerobert} mentioned the time required for pre-training of SafeAeroBERT takes over two months. The computational demands associated with the use of generative AI render these models impractical for real-time aviation decision support in the current stage.

As a summary, these approaches offer distinct advantages, (a) rule-based methods provide high precision in controlled settings; (b) statistical machine learning methods offer robust performance with careful feature engineering; (c) deep learning techniques deliver state-of-the-art results in complex, real-world conditions; (d) LLMs are powerful and adaptable but require substantial computational effort. The research trends indicate a clear movement toward end-to-end, hybrid systems that not only address the inherent challenges of noisy, rapid, and accented ATC communications but also integrate domain-specific knowledge to enhance safety and operational efficiency. These systems often incorporate expert-defined rules to pre-filter the input or to post-process the output of statistical or deep learning models. This approach not only leverages the precision of expert rules but also benefits from the adaptability and generalization capabilities of modern neural architectures.

In this work, we address the aforementioned issues by first building our own data processing pipeline to process open source communication audios and scanned documents using the state-of-the-art speech-to-text engine. Then, we propose a hybrid ATC domain specific rule-enhanced NER to extract key information (i.e., callsign and destination intent) from the pilot-ATC communication transcripts, along with a post-prediction heuristic rule override process to further boost recognition performance. The performance improvement of such post-recognition processing technique on intent learning is also highlighted in \Citep{chen2015closed}. We examine the usage of various token-level embedding models with specifications such as multilingual support and generative power. Finally, we match the extracted entities with the node names from the NASA FACET node-link graph, based on embedding similarities.

\subsection{Surface Movement Collision Risk Assessment \label{subsec: risk-review}}

Aviation systems are highly complex cyber-physical systems, where risk is related to the probability of failure when either a sub-module or a whole system is making inappropriate decisions while exposed to hazards \citep{netjasov2008review}. Risk assessment is defined as the systematic identification and evaluation of the risks posed by possible accident scenarios. It is a tool that supports decision making and therefore, risk management. Risk management is optimization of safety of a system, the verification process and risk acceptance, which support airport operations \citep{icao2009safety, icao2007runway}. Causal methods provide the theoretical framework for aviation risk assessment but are mostly data-driven and rely on data quality. Casual methods such as fault tree analysis \citep{kumamoto2000probablistic}, event tree analysis \citep{huang2001fuzzy}, and common case analysis \citep{ford1999aviation} are used to quantify the statistical probability of an accident or system component failure, while Bow-Tie analysis \citep{acarbay2020risk}, Petri nets \citep{blom1999accident}, and Bayesian Belief Networks \citep{luxhoj2006modeling} are employed to assess the anticipated risk associated with changes to the system. 

Collision risk between two aircraft seeks to quantify the probability of conflicts between aircraft by evaluating random deviations in position and speed, thereby informing separation standards and safety measures. It is a spatiotemporal problem where each aircraft location is uncertain, and researchers emphasize that ignoring the spatiotemporal uncertainty in motion estimation can lead to inaccurate risk assessments \citep{xin2021probabilistic}. Recent work attempts to tackle this challenge by modeling aircraft ground movement as a Markov Decision Process within a digital twin framework, integrating a context-aware speed model trained via imitation learning to improve real-time position estimation from noisy A-SMGCS data \citep{tran2025probabilistic}, and the improved tracking accuracy directly benefits collision risk assessment by providing more reliable temporal predictions for potential conflict nodes. It is quantified by integrating the joint probability of both agents occupying the same space at the same time, given their uncertain trajectories. Overall, on the tactical level, the civil aviation risk assessment model can be divided into midair risk assessment and surface risk assessment.  

Midair collision models, such as the Reich–Marks model \citep{reich1966analysis, shortle2004simulating}, represent aircraft as three-dimensional boxes and calculate the likelihood of collision by assessing the probability of aircraft proximity and the conditional probability of collision given that proximity. The Machol–Reich model \citep{machol1975aircraft, machol1995thirty} refines this approach by incorporating empirical data on lateral position errors, enabling more accurate predictions of vertical, horizontal, and longitudinal collision risks. Simpler intersection models \citep{siddiqee1973mathematical, geisinger1985airspace,barnett2000free} estimate collision probabilities at predetermined crossing points using traffic flow intensities, while geometric conflict models \citep{paielli1997conflict, paielli1999conflict, irvine2002geometrical} define conflict regions based on the extrapolation of aircraft trajectories. More recent advancements involve the generalized Reich model, which utilizes hybrid-state Markov processes and Monte Carlo simulation techniques \citep{bakker2001geometric, bakker1993air} to provide safety feedback for system redesign and to evaluate modifications in separation minima, as further adopted by the FAA \citep{kos2000probabilistic, blom2005study, blom2006safety}.

On the surface level, the collision risk assessment expands to other transportation domains. For instance, \cite{chen2024dynamic} propose a probabilistic conflict detection model for vessels that uses probability density functions (PDFs) of predicted positions to quantify the \textit{conflict criticality} between two trajectories. Each vehicle has a travel time distribution (TTD) over each road segment (link), which is one of the major sources of uncertainties. Link travel times are nonnegative and often skewed with a long tail (i.e., occasional heavy delays), making log-normal a natural choice. The route travel time is usually modeled as the sum of link travel times, so its distribution is the convolution of the link-level distributions. \cite{cardieri2000statistics} analyzed empirical travel times and found that a shifted log-normal distribution fit well for road link travel times. Similar approach is adopted for correlated link delays, where the link times are assumed to be log-normal and an approximate analytic form for the travel time density is derived by applying the Fenton-Wilkinson (FW) approximation \citep{chen2024dynamic}. Other location-scale distributions such as the log-logistic \citep{chu2011empirical}, log-normal\&normal \citep{kieu2015public} have been used for link travel time modeling. 

Airport surface movement refers to aircraft movement on the ground, including taxiing, takeoff, landing, and aircraft operations on runways, taxiways, taxilanes, and aprons. \cite{watnick1992airport} introduced the Airport Movement Area Safety System (AMASS) to prevent runway incursions from escalating into collisions. \cite{wang2021aircraft} evaluated several aircraft taxi time prediction models, identifying key features that enhance modeling accuracy. At both mesoscopic and macroscopic levels, \cite{yang2017fundamental} characterized airport surface flow and developed a cell transmission-based model to replicate its spatial–temporal dynamics. \cite{waldron2013quantifying} analyzed surface movement data to quantify discrete interactions during taxi operations, conceptualizing these as stages of increasing collision risk, while \cite{ford2014relating} used ASDE‑X surveillance data to assess the frequency and characteristics of potentially hazardous interactions linked to taxiway geometries and traffic flow constraints. Analysis and classification of runaway incursions based strictly on the risk of scenarios associated with the state at the start of the incursion have been investigated \citep{stroeve2016strengthening, stroeve2015risk, distefano2014risk}. 

Notably, in this work, we adopt the similar spatiotemporal risk modeling approach, where the total link travel time is modeled as the sum of log-normal distributions. Our primary interest is to estimate the collision probability based on the time reaching the given potential collision node for each aircraft, the spatial uncertainties of trajectory prediction model \citep{pang2021data, zhang2022airport, pang2022bayesian} are simplified as the neighborhood of the given node. 

\section{Methodology \label{sec: methodology}}

\begin{figure}
    \centering
    \includegraphics[width=0.95\textwidth]
    {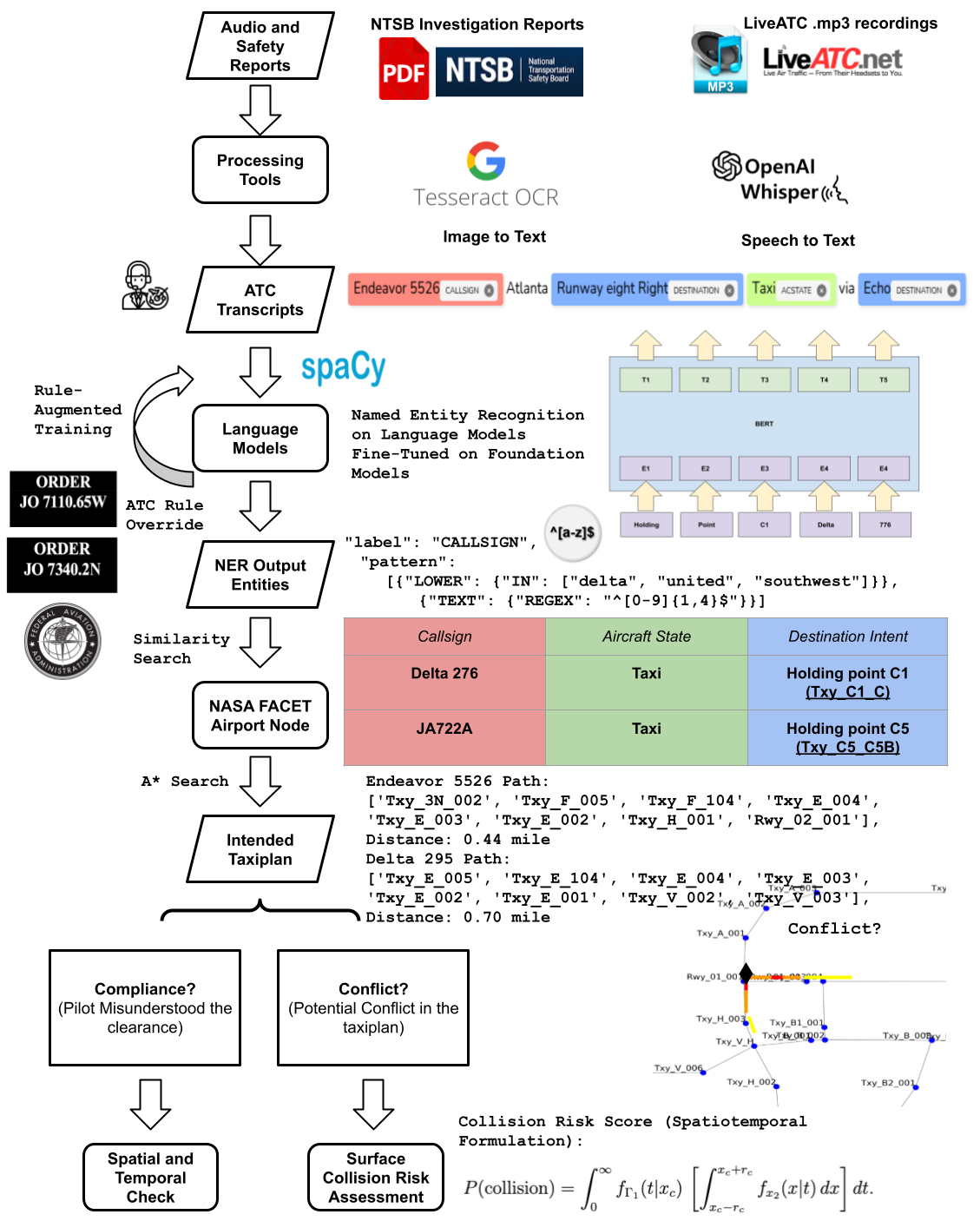}
    \caption{The proposed workflow of ATC communication transcript understanding and surface movement risk assessment.}
    \label{fig: flowchart}
\end{figure}

In this section, we introduce the detailed methodology proposed for language AI-powered pilot-ATC communication transcripts understanding for airport surface movement collision risk assessment. We propose the end-to-end pipeline that either goes from speech audio or processed communication transcripts to surface collision risk at a potential collision spot on the airport node-graph layout. \Cref{fig: flowchart} provides an overview of the workflow, where the training and risk assessment modules are shown in detail. For the learning function, we collect and obtain our ATC transcripts dataset and propose the rule-enhanced training pipeline by integrating ATC domain-specific rules. The surface movement collision risk assessment model takes the distribution of link travel time parameters as inputs and provides an estimation of conflict probability.

\subsection{ATC Rule-Enhanced NER \label{subsec: ner-learning}}
\subsubsection{Data Description}
We collect our data for the model training and testing from many sources. \href{https://www.liveatc.net/}{LiveATC.net} is a website providing live and recorded air traffic control (ATC) audio streams from airports worldwide. It offers both web-based streaming and mobile applications for real-time access to ATC audio, making it a valuable resource for research purposes. We adopt the recently developed, state-of-the-art speech-to-text engine, Whisper, to transcribe the audio files into text format and use them as the training set \citep{radford2023robust}. Besides that, we also download the public released accident investigation reports from NTSB and use them as our validation set. These transcripts are records of communications between pilots and air traffic controllers during specific flights or incidents. They are crucial for accident investigations, as they provide a timeline of communications, instructions, and responses that occurred during the flight. We adopt the Tesseract Optical Character Recognition (OCR) to extract text from these scanned documents \citep{smith2007overview}, which offers advanced multilingual support as well. Lastly, we obtain the communication transcripts published online for the test scenarios of our case studies. 

\begin{figure}[H]
\centering
    \begin{subfigure}[t]{0.45\textwidth}
        \centering
        \includegraphics[width=\textwidth]
        {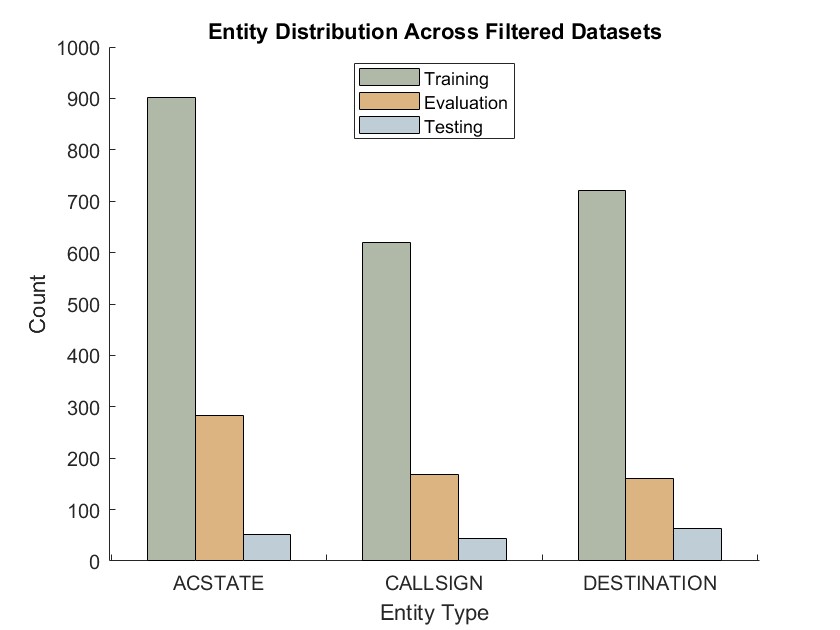}
        \caption{Entity count distribution in training, validation, and testing datasets.}
        \label{fig: data-dist-1}
    \end{subfigure}
    \begin{subfigure}[t]{0.45\textwidth}
        \centering
        \includegraphics[width=\textwidth]
        {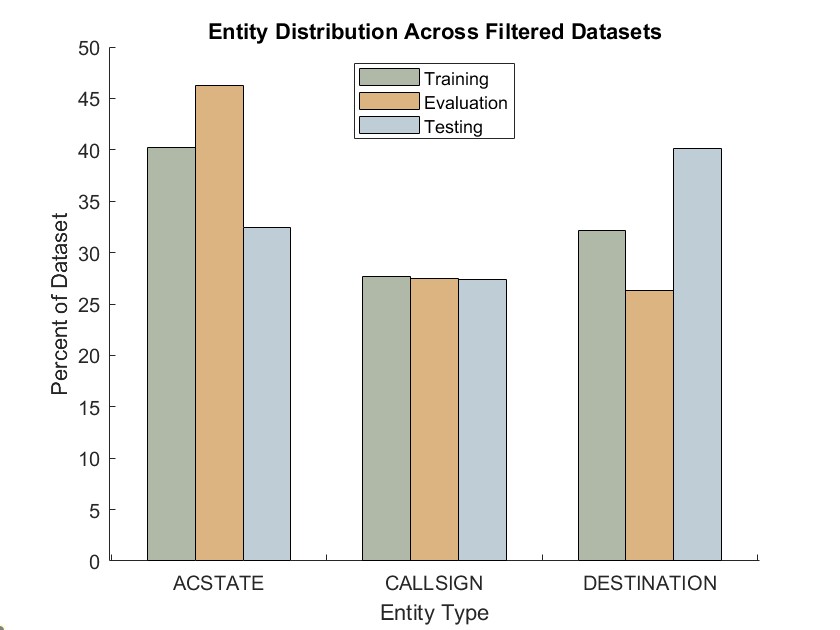}
        \caption{Entity percentage distribution in training, validation, and testing datasets.}
        \label{fig: data-dist-2}
    \end{subfigure}
\caption{The distribution of entity types of the dataset used in the ATC communication transcript information retrieval work.}
\label{fig: data-description}
\end{figure}

The online NER annotation tool is used for manual annotation of our defined entity types. In this study, we are interested in three types of entities, (a) callsign: the unique ID of each flight; (b) aircraft state: the state intention of the aircraft, which can be any of holding, taxiing, approaching, etc; (c) destination intent: the intended destination of the aircraft, which can be a runway name, or a taxiway name, or a gate. \Cref{fig: data-description} shows the count and share of different entity types in the training, validation, and testing sets. The annotation tool directly exports a json file for NER training. 

\subsubsection{Experiment Setup}

\begin{figure}
    \centering
    \includegraphics[width=0.75\textwidth]
    {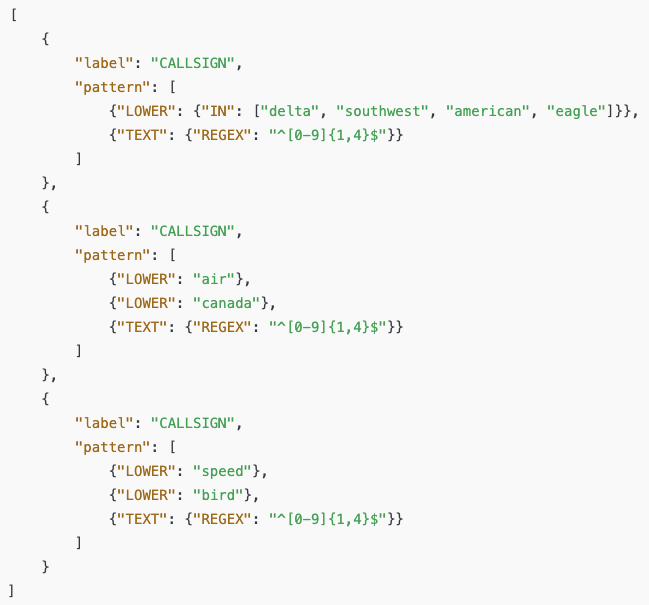}
    \caption{Three examples of entity rules of flight callsigns. Based on FAA Order JO 7340.2N, \textit{speed bird} is the \textit{nickname} of British Airlines \citep{FAA_7340.2N_2024}. }
    \label{fig: entity-rules}
\end{figure}

SpaCy is a fast and efficient open-source NLP toolbox designed for production-ready applications involving text analysis. It’s particularly known for its efficiency, accuracy, and ease of use with NER tasks \citep{vasiliev2020natural}. In this work, we are specifically interested in EntityRuler, a flexible SpaCy component that allows us to add their domain-specific rules to identify entities that the pretrained models might miss, which we call \textit{rule override}. The benefits of EntityRuler are manifold, (a) domain-specific customization makes it easy to adapt models to industry-specific terminology; (b) unlike black-box models, rules are clear and easy to debug, making it transparent and controllable; (c) rules can seamlessly integrate into existing models to boost the overall accuracy without retraining. \Cref{fig: entity-rules} shows several examples of flight callsign rules in regular expression format. We define these rules by reading the related regulations from the ATC manual defined by FAA Order JO 7110.65 and FAA Order JO 7340.2N. 

Importantly, our approach remains adaptable beyond the U.S. context. We evaluated multilingual embeddings (mBERT and mRoBERTa) to ensure coverage across non-English and mixed-language transcripts. And while FAA orders served as our initial heuristic base, we recognize that ICAO’s global radiotelephony standards (e.g., ICAO Doc 4444 (PANS‑ATM), ICAO Doc 9432 (the Manual of Radiotelephony), provide widely accepted phraseology that forms a common communication foundation in international operations. ICAO also mandates minimum English language proficiency (Operational Level 4) for pilots and controllers in international airspace to ensure a baseline of clarity across diverse linguistic and accent backgrounds. While regional adaptations do occur, for instance, the FAA’s use of TRACON, ARTCC, ramp, and center versus ICAO terms like FIR, apron, and ACC, the EntityRuler’s regex structure allows effortless extension to incorporate these localized terms. Broadening global rule coverage requires substantive engineering effort and ongoing maintenance of the heuristics class.

Embeddings serve as dense vector representations of tokens, words, or characters. Classical word embeddings, such as Word2Vec \citep{rong2014word2vec} and GloVe \citep{pennington2014glove}, assign a single, context-independent vector to each word, failing to capture the nuances of polysemy and contextual usage. In contrast, contextual embeddings generate dynamic representations for tokens, influenced by the surrounding text, thereby encapsulating both syntactic and semantic properties across diverse contexts. This advancement has significantly enhanced performance in various NLP tasks, including text classification, question answering, and text summarization \citep{liu2020survey}. Token-level embeddings refer to the \textit{fine-grained representation of individual tokens} (words or characters) within a sequence. Mathematically, Consider a sequence $S = (t_1, t_2, ..., t_N)$ composed of $N$ tokens, each token \( t_i \) is mapped to a dense vector representation \( h_{t_i} \) as,
\begin{equation}
    h_{t_i} = f(e_{t_1}, e_{t_2}, ..., e_{t_N})
\end{equation}
\noindent where \( e_{t_j} \) represents the non-contextualized embedding of token \( t_j \), and \( f \) is a function modeling dependencies between tokens \citep{liu2018generating, devlin2018bert}. Unlike traditional word embeddings that assign the same vector to a word regardless of context, token embeddings vary depending on surrounding words. This contextual nature allows models to capture words that have multiple meanings based on their unique usage in a particular sentence or phrase, which can capture syntactic dependencies and interactions between words \citep{liu2018generating, hewitt2019designing, artetxe2019massively}. 

In this work, we investigate and compare seven different token-level contextual embeddings models as,
\begin{itemize}
    \item Tok2Vec is considered local contextual token embedding, in contrary to Word2Vec \citep{rong2014word2vec}. It uses a CNN-based model for context-sensitive embeddings, making it more efficient and suitable for syntactic and semantic tasks.
    \item BERT (Bidirectional Encoder Representations from Transformers) is one of the most well-known embedding models due to its superior performance \citep{devlin2018bert}. Unlike traditional models that processed text in one direction, BERT remarkably introduces bidirectional context understanding by training on masked language modeling (MLM) and next sentence prediction (NSP) tasks. 
    \item RoBERTa (Robustly Optimized BERT Pretraining Approach) is the modified version of BERT by removing the NSP task, training on larger datasets, and using dynamic masking. These improvements result in enhanced performance across multiple benchmarks \citep{liu2019roberta}.
    \item Multilingual BERT is a variant of BERT trained on Wikipedia in 104 languages using a shared vocabulary. It enables cross-lingual transfer and zero-shot learning for various languages.
    \item Multilingual RoBERTa (XLM-R) builds on RoBERTa with multilingual capabilities. It is trained on 2.5TB of CommonCrawl data across over a hundred languages, surpassing multilingual BERT in performance \citep{goyal2021larger}.
    \item DistilBERT applies knowledge distillation to BERT, reducing model size by 40\% while maintaining 97\% of its performance. It is designed for applications with limited computational resources \citep{sanh2019distilbert}.
    \item BART (Bidirectional and Auto-Regressive Transformers) is a denoising autoencoder combining BERT’s bidirectional encoder with GPT’s autoregressive decoder. It is particularly effective in text generation tasks like summarization and dialogue generation \citep{lewis2019bart}. 
\end{itemize}

\subsubsection{NER Performance Comparison}
Depending on whether the data is augmented with ATC rules and whether an ATC rule override is applied to the model predictions, four distinct experimental setups are generated. We compare these models across the four setups by evaluating: (a) NER classification accuracy, which reflects the model's predictive performance; and (b) time and space complexities, which indicate the computational resources required for model deployment, a critical factor for real-world applications. 

The metrics used to quantify the classification accuracy are precision, recall, and micro F1 score. Precision tells about how much we can trust the model's positive predictions. Recall informs how well the model is able to find all the actual positive samples. Micro F1 score is used to measure how good the model's capability is to balance between precision and recall. 

\begin{table}[!ht]
\centering
\small
\caption{Experiment Results of the ATC Communication Transcript Retrieval Model. NER performance across four setups: 
\emph{\textbf{S1}} = Train w/o ATC rules, no rule override; 
\emph{\textbf{S2}} = Train w/o ATC rules, override w/ ATC rules; 
\emph{\textbf{S3}} = Train w/ ATC rules, no rule override; 
\emph{\textbf{S4}} = Train w/ ATC rules, override w/ ATC rules.}
\label{tab: ner-results}
\resizebox{\textwidth}{!}{%
\begin{tabular}{l|ccc|ccc|ccc|ccc}
\hline
\multirow{2}{*}{\textbf{Embedding Model}} 
 & \multicolumn{3}{c|}{\textbf{S1}} 
 & \multicolumn{3}{c|}{\textbf{S2}} 
 & \multicolumn{3}{c|}{\textbf{S3}} 
 & \multicolumn{3}{c}{\textbf{S4}} \\ 
 & Precision & Recall & microF1 
 & Precision & Recall & microF1 
 & Precision & Recall & microF1 
 & Precision & Recall & microF1 \\
\hline
Tok2Vec (Local-Contextual)  & 0.869 & 0.566 & 0.685 & 0.800 & 0.684 & 0.738 & 0.839 & 0.684 & 0.754 & 0.841 & 0.763 & 0.800 \\
BERT (Contextual)   & 0.847 & 0.546 & 0.664 & 0.781 & 0.658 & 0.714 & 0.838 & 0.750 & 0.792 & 0.839 & 0.822 & 0.831 \\
RoBERTa (Contextual)  & 0.859 & 0.559 & 0.677 & 0.805 & 0.678 & 0.736 & 0.853 & 0.724 & 0.783 & 0.855 & 0.816 & 0.835 \\
mBERT (Multilingual)  & 0.871 & 0.533 & 0.661 & 0.817 & 0.678 & 0.741 & 0.869 & 0.783 & 0.824 & 0.866 & 0.809 & 0.837 \\
mRoBERTa (Multilingual)  & 0.872 & 0.539 & 0.667 & 0.820 & 0.691 & 0.750 & 0.850 & 0.711 & 0.774 & 0.856 & 0.822 & 0.839 \\
DistilBERT (Distilled) & 0.856 & 0.546 & 0.667 & 0.800 & 0.711 & 0.753 & 0.831 & 0.711 & 0.766 & 0.840 & 0.796 & 0.818 \\
BART (Generative)   & 0.840 & 0.691 & 0.758 & 0.787 & 0.730 & 0.758 & 0.845 & 0.822 & 0.833 & 0.842 & 0.842 & 0.842 \\
\hline
\end{tabular}
}
\end{table}

\begin{figure}
\centering
    \begin{subfigure}[t]{0.45\textwidth}
        \centering
        \includegraphics[width=\textwidth,height=5cm]
        {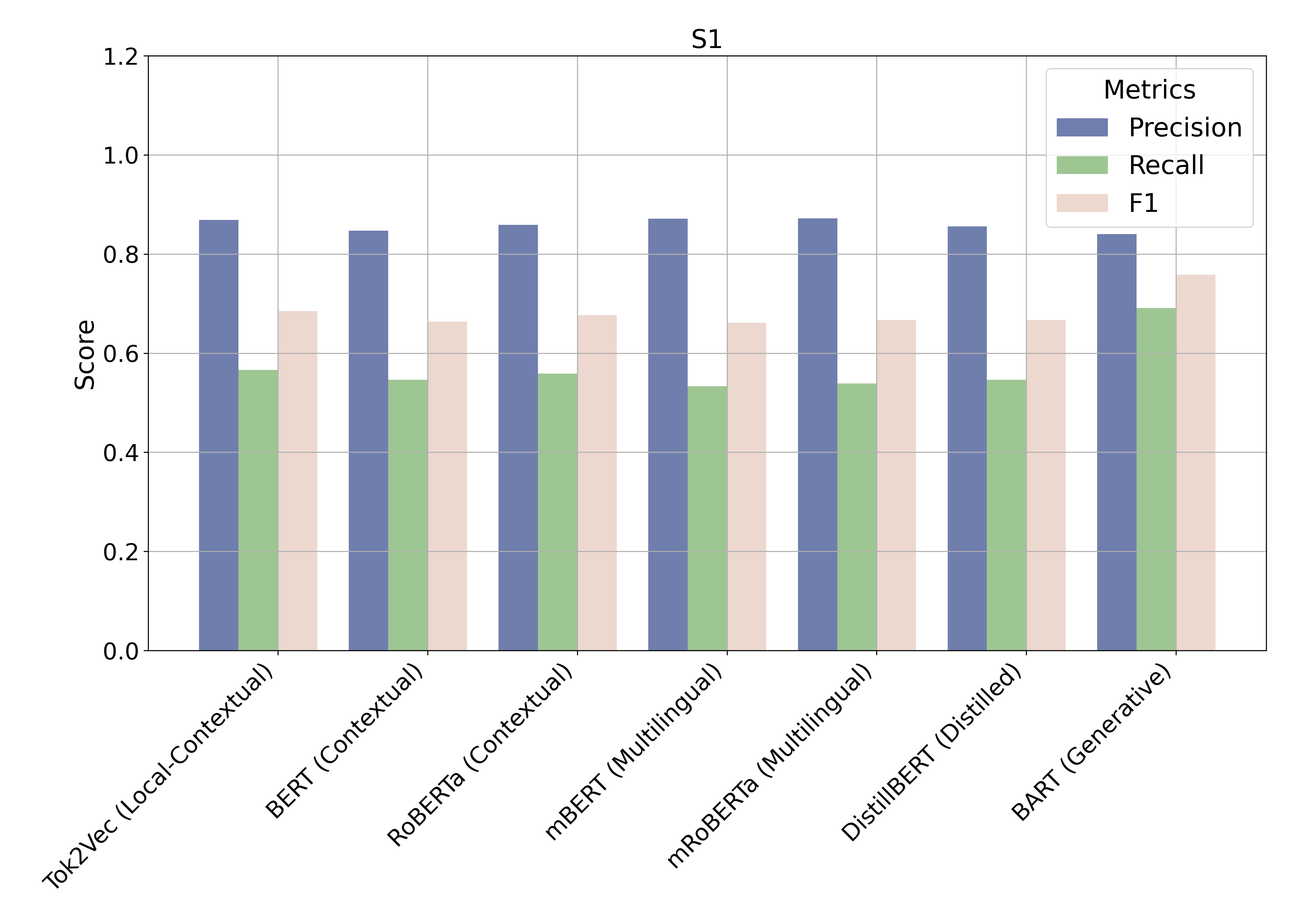}
        \caption{Performance of model trained without using ATC Rules and no prediction override. }
        \label{fig: score-set1}
    \end{subfigure}
    \begin{subfigure}[t]{0.45\textwidth}
        \centering
        \includegraphics[width=\textwidth,height=5cm]
        {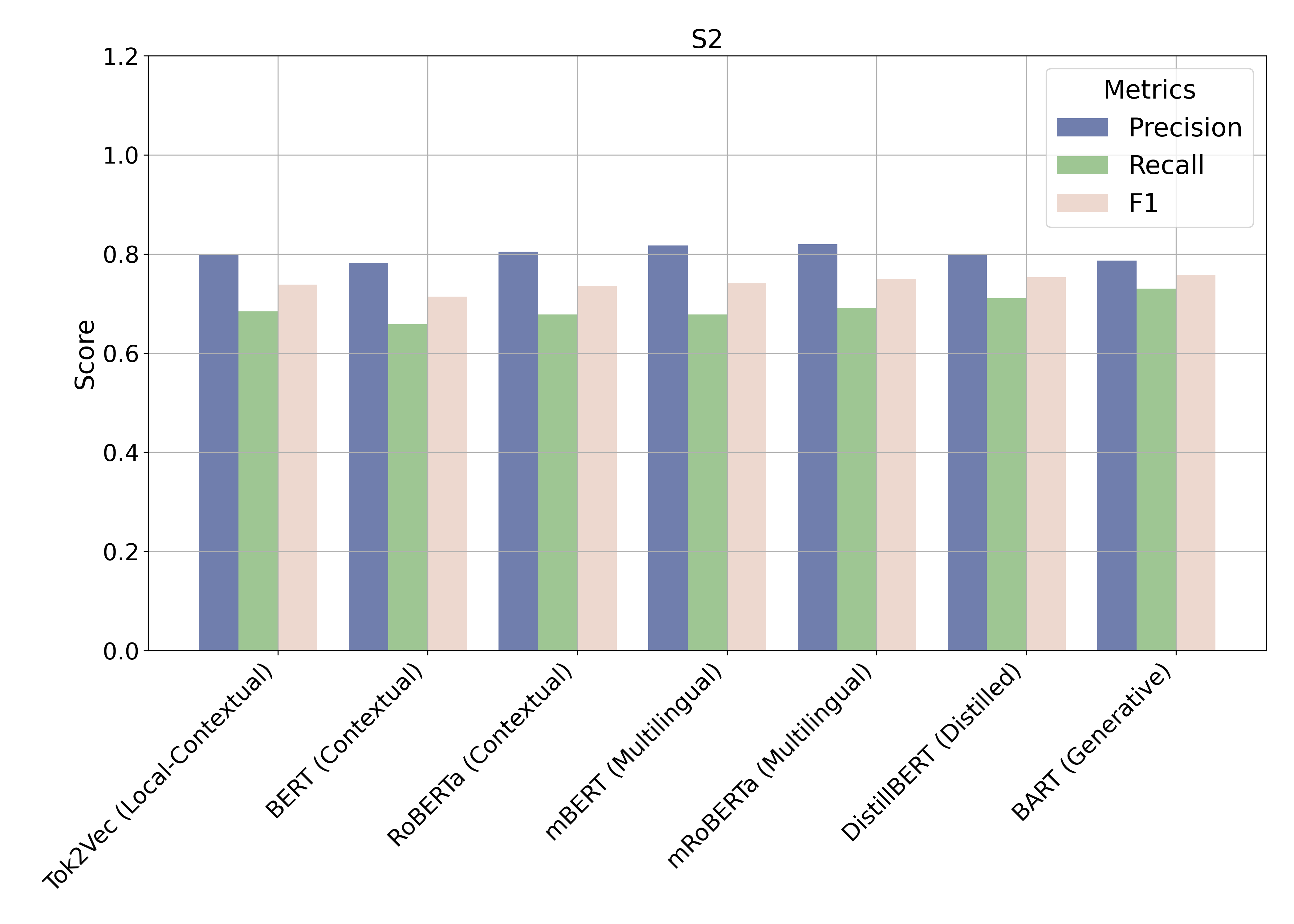}
        \caption{Performance of model trained without using ATC Rules but with prediction override. }
        \label{fig: score-set2}
    \end{subfigure}
    \\~\\
    \begin{subfigure}[t]{0.45\textwidth}
        \centering
        \includegraphics[width=\textwidth,height=5cm]
        {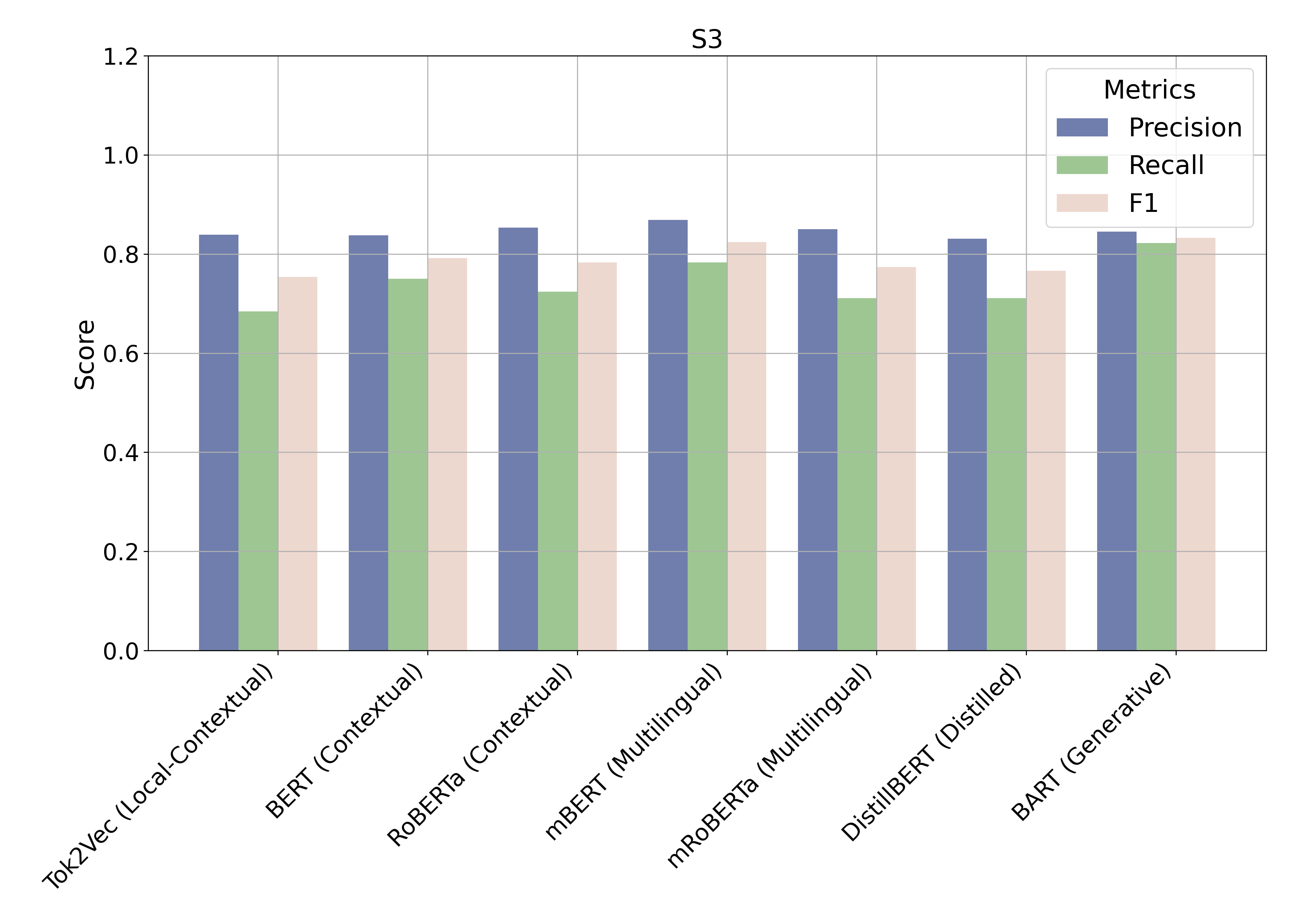}
        \caption{Performance of model trained with ATC Rules and no prediction override. }
        \label{fig: score-set3}
    \end{subfigure}
    \begin{subfigure}[t]{0.45\textwidth}
        \centering
        \includegraphics[width=\textwidth,height=5cm]
        {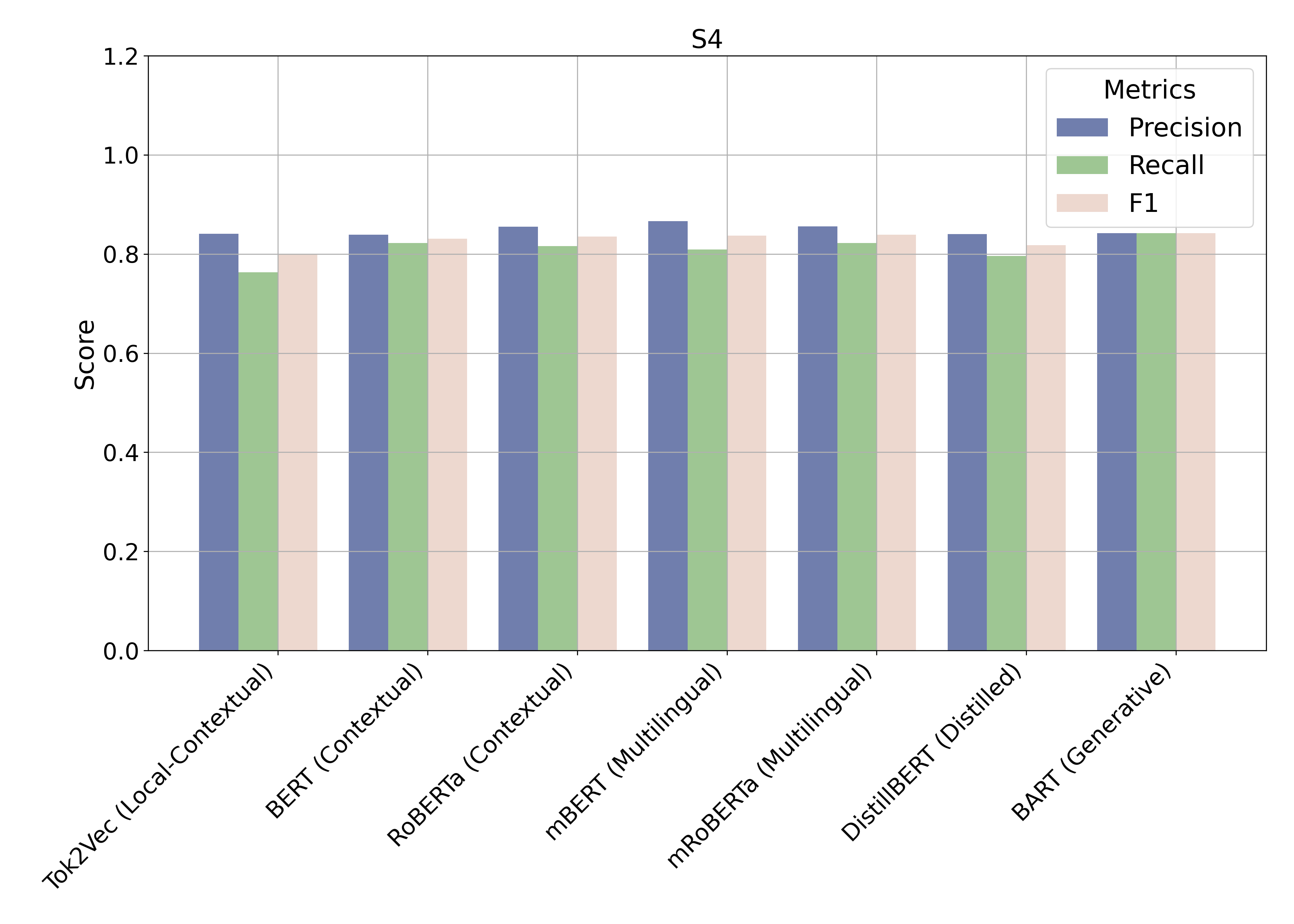}
        \caption{Performance of model trained with ATC Rules and with prediction override. }
        \label{fig: score-set4}
    \end{subfigure}
\caption{Performance comparison between different embedding models under various setups. The last setup clearly gives better overall performance for any contextual embeddings. }
\label{fig: scores}
\end{figure}

\Cref{tab: ner-results} lists the performance of seven embedding models with four different setups, and \Cref{fig: scores} visualizes the same results as histograms. The results obviously indicate that incorporating ATC rules into the model pipeline, both during training and as a post-prediction override, leads to consistent improvements across various embedding models. In particular, the use of ATC rules tends to boost recall significantly, even though it may sometimes lead to a slight reduction in precision. This trade-off is significant, as evidenced by the increased F1 scores, which balance the contributions of both precision and recall. For instance, the metrics of Tok2Vec move from training without ATC rules and no override (Precision: 0.869, Recall: 0.566, F1: 0.685) to training with ATC rules and override (Precision: 0.841, Recall: 0.763, F1: 0.800) show a marked increase in recall and F1.

When comparing traditional models like Tok2Vec with transformer-based models such as BERT, RoBERTa, mBERT, mRoBERTa, DistilBERT, and BART, the transformer architectures generally demonstrate superior performance. The advanced contextual representations learned by these models are further enhanced when combined with ATC rules. For instance, in several cases, transformer models achieve higher F1 scores when ATC rules are applied during training and as a post-prediction screening mechanism. This effect is particularly notable in multilingual models like mBERT and mRoBERTa, where integrating ATC rules helps to overcome lower recall rates observed when these rules are not applied.


One of the most compelling observations is that the optimal performance is almost always reached when ATC rules are integrated into both the training phase and the post-prediction process. For instance, BART achieves balanced performance with equal precision and recall (0.842) when both training and override rules are applied. This balanced improvement is a recurring theme across the board and underscores the value of combining data augmentation with rule-based adjustments to achieve a robust model performance.

For practitioners, these results indicate that integrating ATC rules both during model training and as a post-processing step can lead to better prediction performance, especially when working with complex data where pure statistical models might miss subtle patterns. This is especially useful in scenarios where high recall is critical. 

\subsubsection{Embedding Model Complexity Study}
\begin{figure}
    \centering
    \begin{subfigure}[t]{0.75\textwidth}
        \centering
        \includegraphics[width=\textwidth]
        {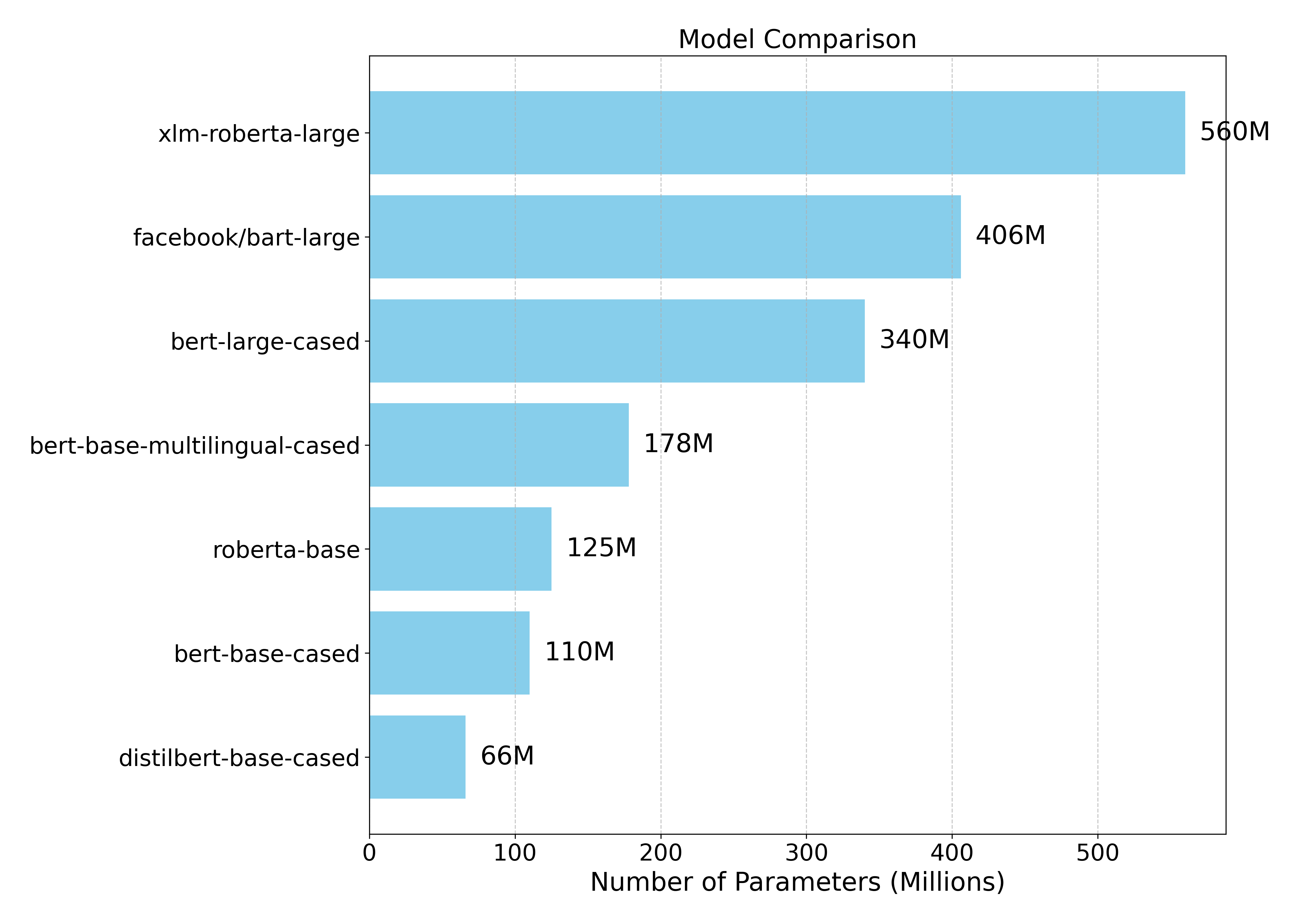}
        \caption{Number of parameters in each model.}
        \label{fig: n_parameters}
    \end{subfigure}
    \begin{subfigure}[t]{0.75\textwidth}
        \centering
        \includegraphics[width=\textwidth]
        {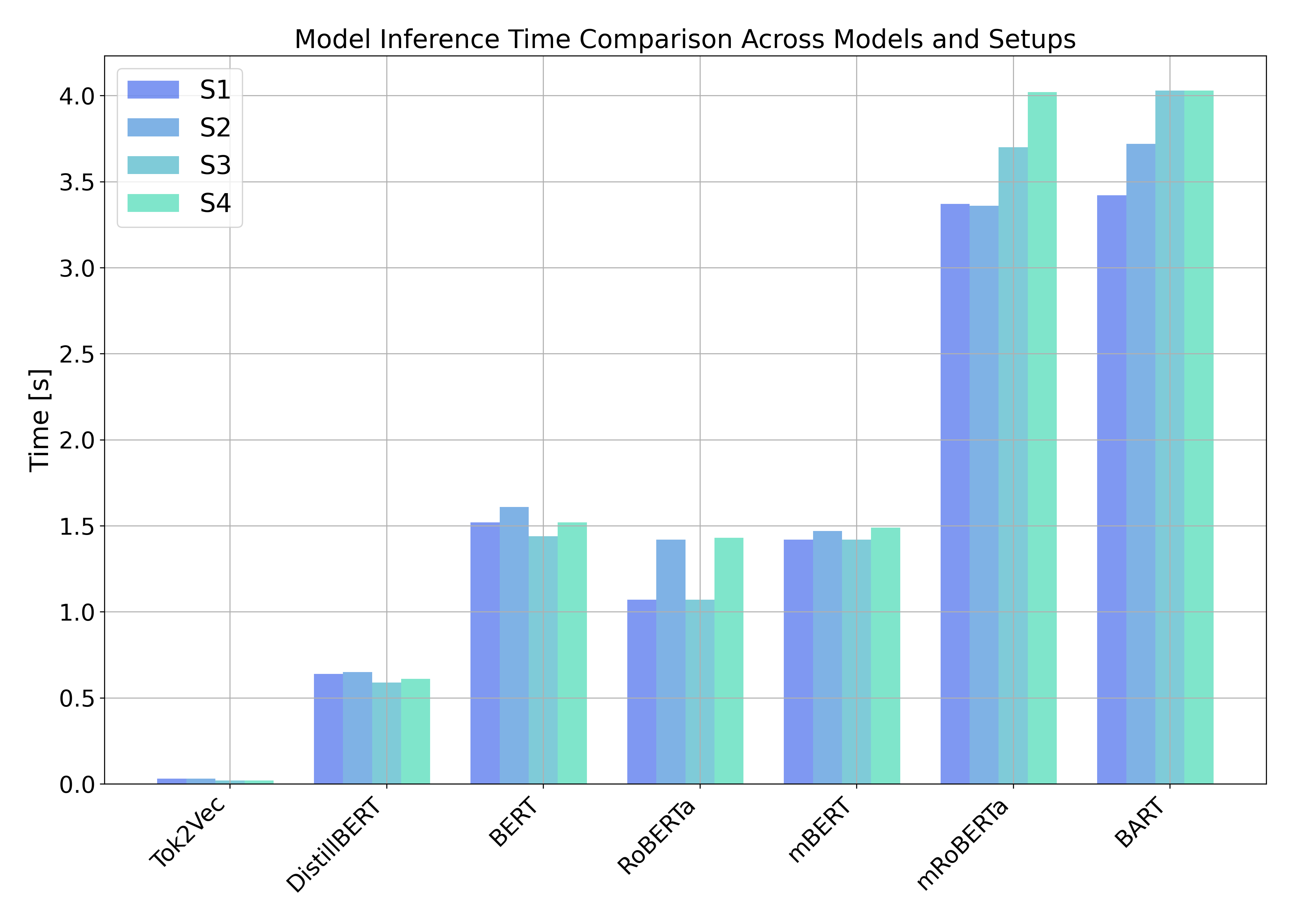}
        \caption{Time complexity comparison between different models. }
        \label{fig: time_compare}
    \end{subfigure}
\caption{The space and time complexity comparison between different models. For \Cref{fig: time_compare}, a sample communication transcript of \texttt{"Japan Air 179, Tokyo Tower, good evening, number 3, taxi to holding point C1."} is used for testing.}
\label{fig: space-time-compare}
\end{figure}

\Cref{fig: space-time-compare} lists the comparison of time and space complexities of our experiments. The metrics to evaluate the space complexity are the number of parameters in the model, which serves as a proxy for its space complexity and also correlates with its capacity to learn complex representations. For instance, mRoBERTa and BART have parameter counts of 560 and 406 million, respectively, suggesting a significant capacity to capture intricate patterns in the data. This increased capacity, as observed in previous accuracy results, translates into higher performance metrics (i.e., F1 scores) when the models are properly fine-tuned. However, this clearly comes at the cost of increased memory requirements and computational overhead during inference.

In contrast, models like DistilBERT (i.e., 66 million parameters) offer a lightweight alternative. The inference time data also reflects this efficiency. While Tok2Vec models are extremely fast (i.e., 0.02 or 0.03 seconds per inference), they tend to lag in accuracy compared to transformer-based models. DistilBERT strikes a balance, with inference times of approximately 0.6 seconds, making it a viable option when computational resources or real-time constraints are a concern. However, models such as BERT and RoBERTa require 1.5 and 1.1–1.4 seconds, respectively, indicating a trade-off between increased model complexity (and hence better accuracy) and slower processing speeds.

The inference time increases significantly for larger models like mRoBERTa and BART, which require approximately 3.4–4 seconds per inference. This substantial increase in processing time is a direct consequence of their higher parameter counts and more sophisticated architectures. As a result, while these models tend to yield superior accuracy, especially when augmented with ATC rules, their heavier computational demands are not suitable for real-world applications at all. The numbers here underscore the trade-off between time and space complexity when choosing the appropriate embedding model with desired performance.

\begin{figure}
    \centering
    \begin{subfigure}[t]{0.45\textwidth}
        \centering
        \includegraphics[width=\textwidth]
        {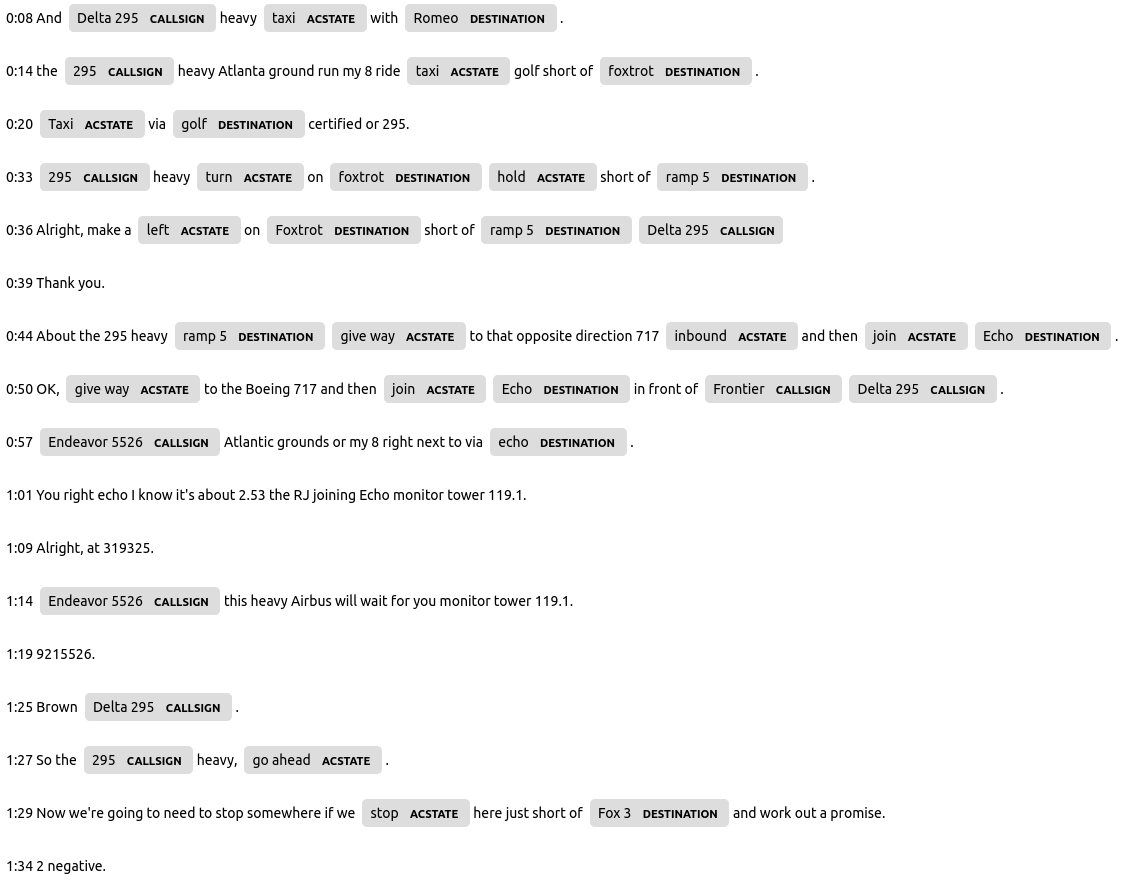}
        \caption{2024 KATL taxiway collision where no ATC rule override applied.}
        \label{fig: example-1-norules}
    \end{subfigure}
    \begin{subfigure}[t]{0.45\textwidth}
        \centering
        \includegraphics[width=\textwidth]
        {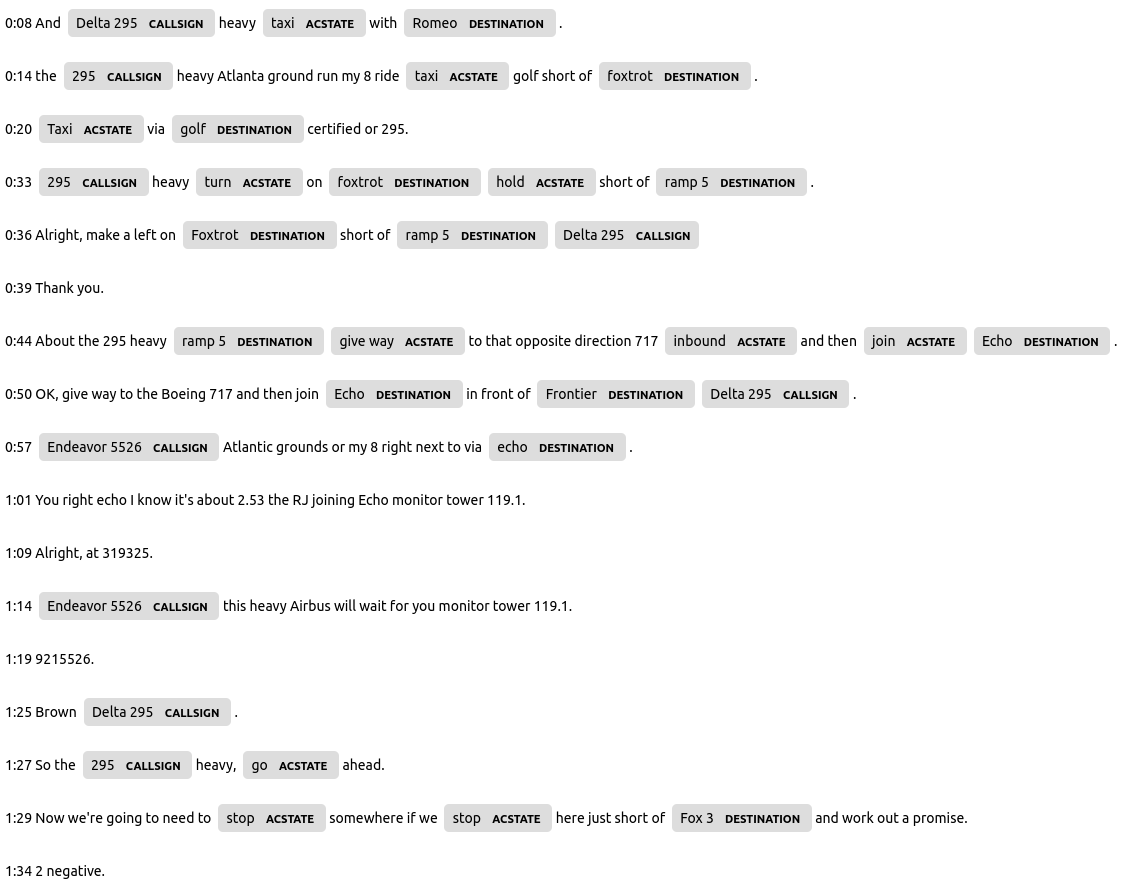}
        \caption{2024 KATL taxiway collision with ATC rule override.}
        \label{fig: example-1-rules}
    \end{subfigure}
    \\~\\~\\
    \begin{subfigure}[t]{0.45\textwidth}
        \centering
        \includegraphics[width=\textwidth]
        {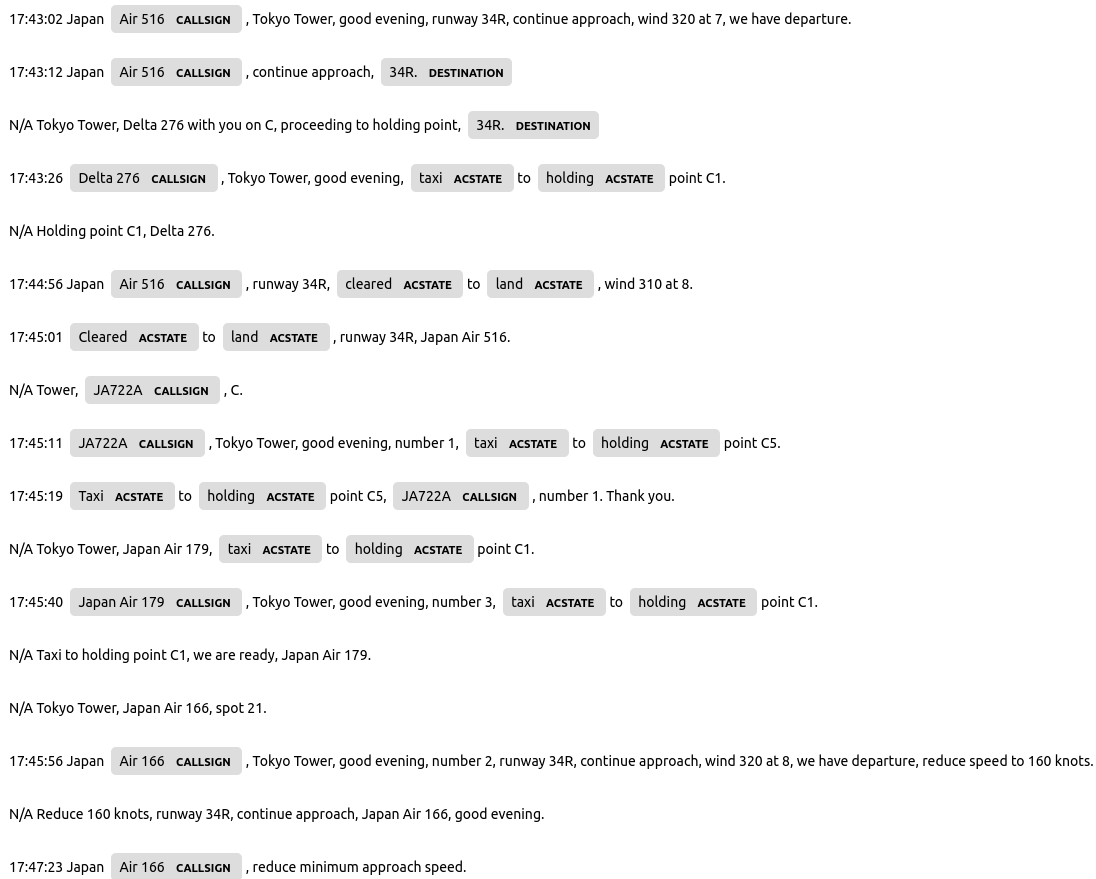}
        \caption{2024 Haneda Airport runway collision where no ATC rule override applied.}
        \label{fig: example-2-norules}
    \end{subfigure}
    \begin{subfigure}[t]{0.45\textwidth}
        \centering
        \includegraphics[width=\textwidth]
        {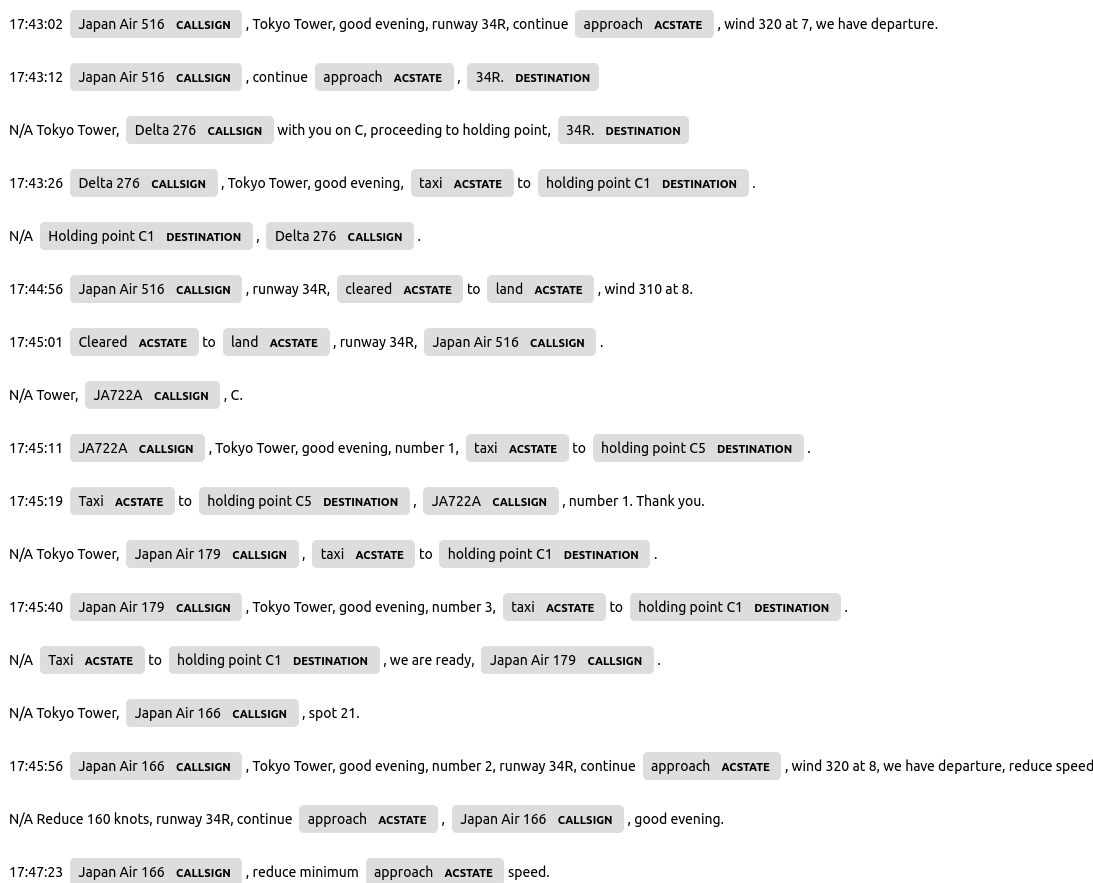}
        \caption{2024 Haneda Airport runway collision with ATC rule override.}
        \label{fig: example-2-rules}
    \end{subfigure}
    \\~\\~\\
    \begin{subfigure}[t]{0.45\textwidth}
        \centering
        \includegraphics[width=\textwidth]
        {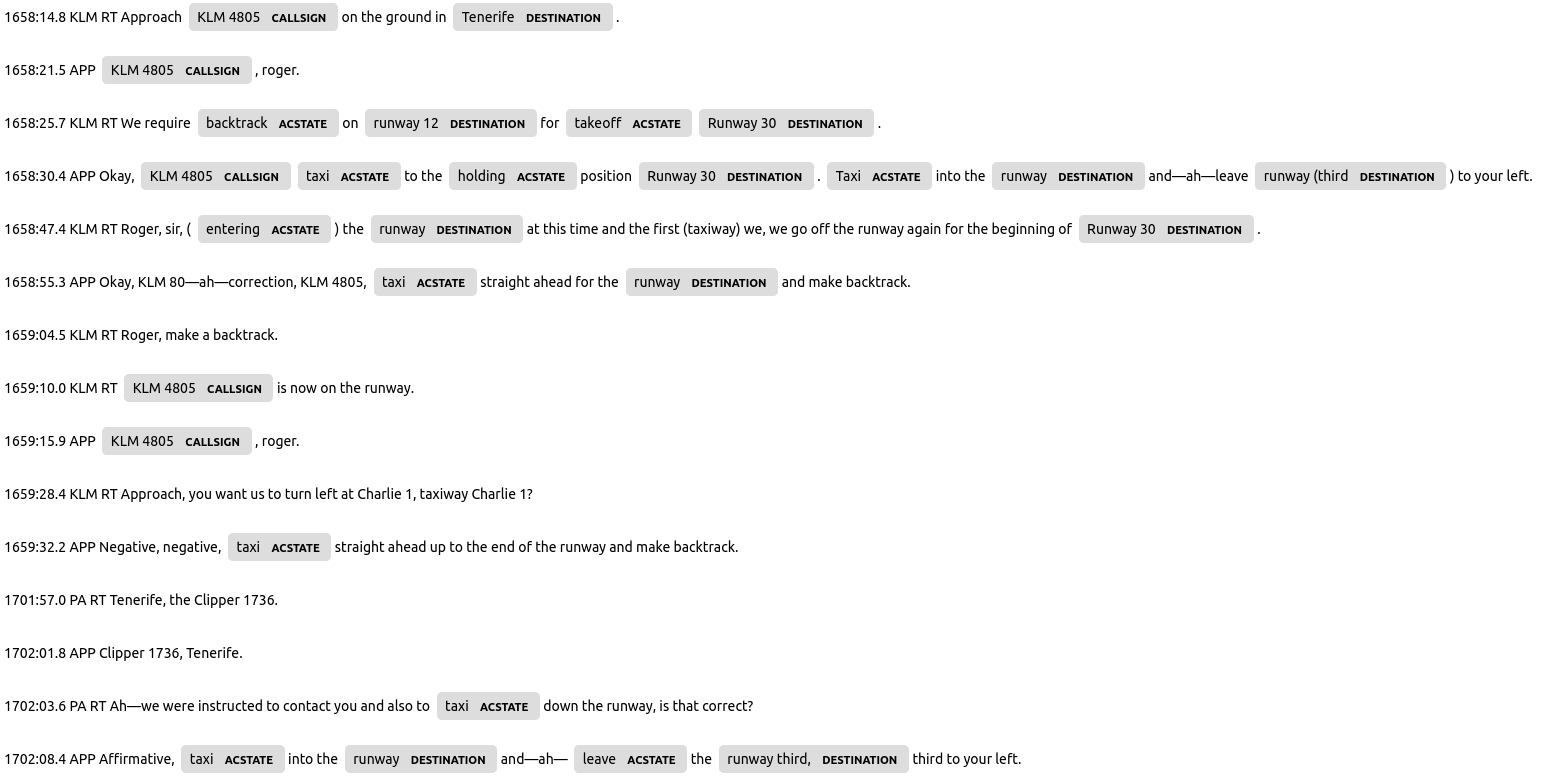}
        \caption{1977 Los Rodeos Airport runway collision where no ATC rule override applied.}
        \label{fig: example-3-norules}
    \end{subfigure}
    \begin{subfigure}[t]{0.45\textwidth}
        \centering
        \includegraphics[width=\textwidth]
        {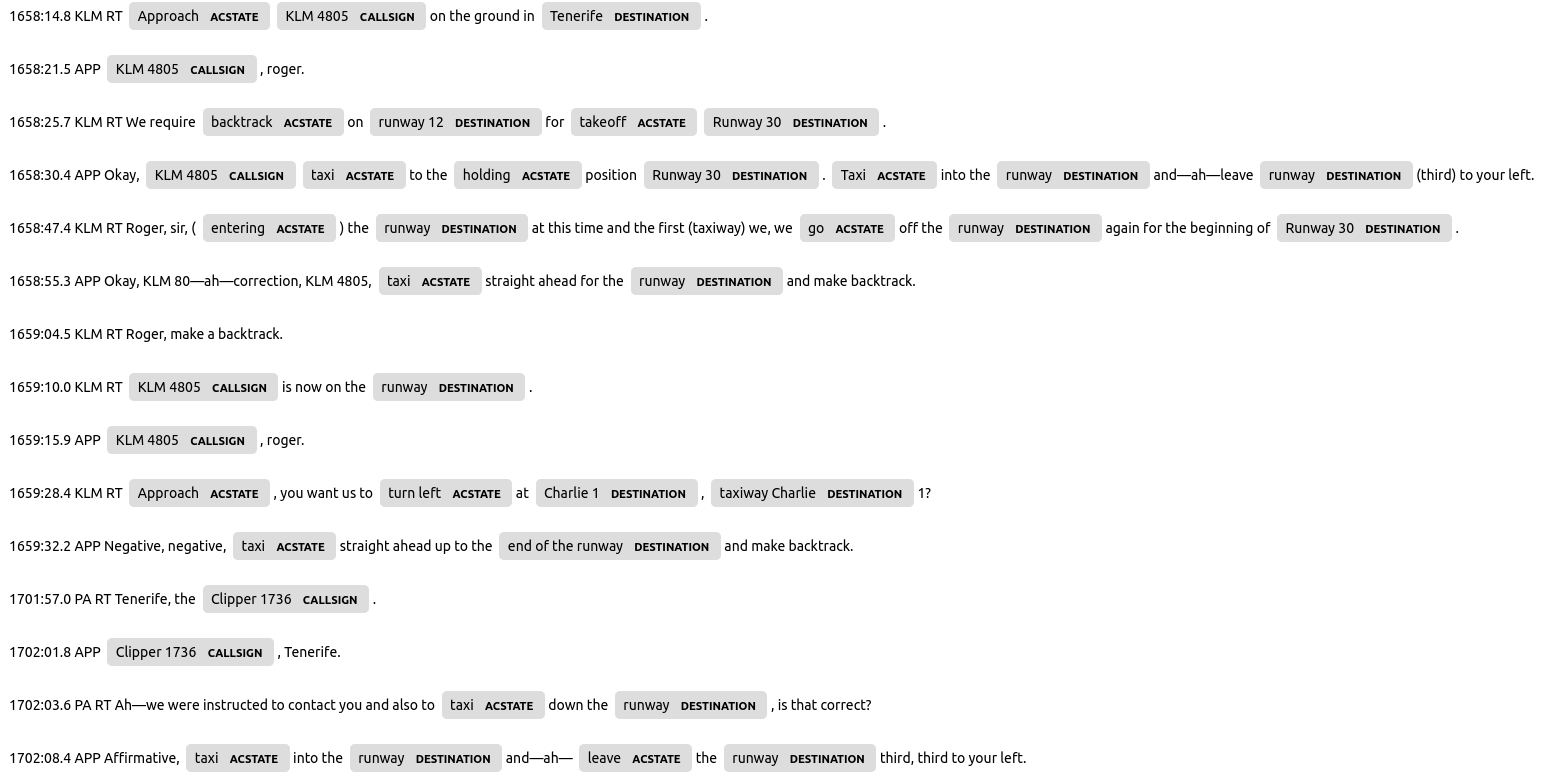}
        \caption{1977 Los Rodeos Airport runway collision with ATC rule override.}
        \label{fig: example-3-rules}
    \end{subfigure}
\caption{Performance comparison of three case studies. The first case study is the 2024 KATL taxiway collision happened on September 10th. The second case study is the 2024 Haneda Airport runway collision on January 2nd. The third case study is the Tenerife airport disaster in 1977 (only a portion of the communication transcript). It is obvious that the entity recognition accuracy greatly improved after the ATC rule override.}
\label{fig: example-scenarios}
\end{figure}

\subsubsection{Embedding Sensitivity}
\begin{table}[!ht]
\centering
\small
\caption{Cross-embedding sensitivity statistics. 
\emph{\textbf{S1}} = Train w/o ATC rules, no rule override; 
\emph{\textbf{S2}} = Train w/o ATC rules, override w/ ATC rules; 
\emph{\textbf{S3}} = Train w/ ATC rules, no rule override; 
\emph{\textbf{S4}} = Train w/ ATC rules, override w/ ATC rules.}
\label{tab: ner-sensitivity}
\begin{tabular}{lccc}
\toprule
\textbf{Setup} & \textbf{Mean microF1} & \textbf{SD across embeddings} & \textbf{MicroF1 Range} \\
\midrule
S1 & 0.683 & 0.032 & 0.097 \\
S2 & 0.741 & 0.014 & 0.044 \\
S3 & 0.789 & 0.027 & 0.079 \\
S4 & 0.829 & 0.014 & 0.042 \\
\bottomrule
\end{tabular}
\vspace{2pt}
\footnotesize
Means/SDs computed across the seven embeddings using microF1; SD is population SD; \textit{Range} is maximum microF1 minus minimum microF1.
\end{table}

We quantify how NER performance depends on the embedding choice by pooling microF1 across seven encoders under the four pipelines in \Cref{tab: ner-results} and summarizing cross-embedding variability in \Cref{tab: ner-sensitivity}. Without rules (S1: train w/o rules, no override), F1 spans 0.661–0.758 (mean 0.683, SD 0.032, range 0.097), indicating substantial sensitivity to the embedding. Adding only a post-prediction override (S2) already compresses variability (mean 0.741, SD 0.014, range 0.044). Training with rules (S3) improves the mean to 0.789 but retains a wider spread (SD 0.027, range 0.079). The most accurate and stable configuration is S4 (train +rules, +override), where F1 concentrates in 0.800–0.842 (mean 0.829, SD 0.014, range 0.042). 

Overall, integrating ATC rules reduces sensitivity to the embedding method by roughly 2.3$\times$ (range 0.097 in S1 vs. 0.042 in S4) while increasing mean F1 by 0.146. This pattern is consistent with domain adaptation: ATC-specific heuristics both raise recall and constrain error modes across encoders, thereby mitigating latent-space mismatch (e.g., numerals, callsigns, clipped phraseology, mixed accents). Practically, S4 enables competitive performance even with lighter models (e.g., DistilBERT) for low-latency settings, while larger encoders (e.g., BART, mRoBERTa) yield the highest F1 when compute permits.

\subsection{Airport Surface Collision Risk Model \label{subsec: riskmodel}}

Mathematically, the collision risk is calculated based on the joint probability that each aircraft reaches the same area at the same timestamp of the node-link airport layout graph. The location of a potential collision can be the intersection of runways, taxiways, taxilanes, or the combination of any two. The collision risk model is defined as the joint distribution of the time overlap for two aircraft occupying the same area around a node in the airport layout map. In this section, we provide the full probabilistic formulation of collision risk at a certain node in the node-link graph.

\subsubsection{Total Travel Time Modeling \label{subsubsec: travel-time-model}}
The $k$-th aircraft travels a total of $n$ taxiway links until reaching the area of interest (i.e., potential collision spot), where the total travel time is given by $\Gamma_k$. We assume each taxiway link has an associated distance $d_{k,i}$ and a taxi speed $v_{k,i}$ that is log-normally distributed with parameters $\mu_{k,i}$ and $\sigma_{k,i}^2$, which is
\begin{align}
    v_{k,i} &\sim \mathsf{Lognormal}(\mu_{k,i}, \sigma^2_{k,i}).
\end{align}
or,
\begin{align}
 f_{v_{k,i}}(v_{k,i})&= \frac{1}{v_{k,i}\, \sigma_{k,i} \sqrt{2 \pi}} \exp\!\Bigl(-\frac{(\ln v_{k,i} - \mu_{k,i})^2}{2 \sigma_{k,i}^2}\Bigr), \quad \forall v_{k,i}>0.
\end{align}

It is obvious that $\Gamma_k = \sum_{i=0}^n \tau_{k,i}$ where $\tau_{k,i} = \frac{d_{k,i}}{v_{k,i}}$ is the distribution of the $k$-th aircraft travel time duration for the $i$-th node link. We also have,
\begin{align}
\label{eq: lognormal}
    \tau_{k,i} \sim \mathsf{Lognormal}(\ln d_{k,i} - \mu_{k,i}, \sigma^2_{k,i}).
\end{align}

\begin{lemma}
    We can prove \Cref{eq: lognormal} by the standard formula for transformations of random variables, if $\tau_{k,i} = g(v_{k,i}) = \frac{d_{k,i}}{v_{k,i}}$, then,  
\begin{equation}
    f_{\tau_{k,i}}(\tau_{k,i}) = 
    f_{v_{k,i}}\bigl(g^{-1}(\tau_{k,i})\bigr)\;\left|\frac{d}{d\tau_{k,i}}\,g^{-1}(\tau_{k,i})\right|. 
\end{equation}

\noindent where $g^{-1}(\tau_{k,i}) = \frac{d_{k,i}}{\tau_{k,i}}$ gives us,
\begin{equation}
    \begin{aligned}
    f_{\tau_{k,i}}(\tau_{k,i}) &= f_{v_{k,i}}\bigl(\frac{d_{k,i}}{\tau_{k,i}}\bigr)\;\cdot \left|\frac{d}{d\tau_{k,i}}(\frac{d_{k,i}}{\tau_{k,i}})\right| \\
    &= f_{v_{k,i}}\bigl(\frac{d_{k,i}}{\tau_{k,i}}\bigr)\ \cdot \frac{d_{k,i}}{\tau_{k,i}^2} \\
    &= \frac{1}{\bigl(\tfrac{d_{k,i}}{\tau_{k,i}}\bigr)\;\sigma_{k,i} \sqrt{2 \pi}} \exp\!\Bigl[-\tfrac{\bigl[\ln(\tfrac{d_{k,i}}{\tau_{k,i}}) - \mu_{k,i} \bigr]^2}{2\,\sigma_{k,i}^2}\Bigr] \cdot\ \frac{d_{k,i}}{\tau_{k,i}^2} \\
    &= \frac{1}{\sigma_{k,i} \sqrt{2 \pi}} \;\frac{1}{\tau_{k,i}} \exp\!\Bigl[\!-\,\tfrac{1}{2\sigma_{k,i}^2}\,\bigl[\ln\!\bigl(\tfrac{d_{k,i}}{\tau_{k,i}}\bigr) - \mu_{k,i} \bigr]^2\Bigr] \\
    &= \frac{1}{\tau_{k,i}\,\sigma_{k,i}\,\sqrt{2\pi}} \exp\!\Bigl[-\,\frac{\bigl(\ln \tau_{k,i} - [\ln d_{k,i} - \mu_{k,i}] \bigr)^2}{2\,\sigma_{k,i}^2} \Bigr],\quad \forall \tau_{k,i}>0.
\end{aligned}
\end{equation}

That is, each $\tau_{k,i}$ is a log-normal‐type variable, with parameters shifted by $\ln d_{k,i}$ with,
\begin{align}
    \mathbb{E}[\tau_{k,i}] &= d_{k,i}\,\exp\!\Bigl[-\mu_{k,i} + \tfrac{\sigma_{k,i}^2}{2}\Bigr].\\
    \operatorname{Var}[\tau_{k,i}] &= d_{k,i}^2\,\exp\!\Bigl(-2\mu_{k,i} + \sigma_{k,i}^2\Bigr)\Bigl[\exp\!\bigl(\sigma_{k,i}^2\bigr)-1\Bigr].  
\end{align}  
\qed
\end{lemma}

The total travel time for the $k$-th aircraft, $\Gamma_k$, is the $n$-fold convolution of each individual link distributions as,
\begin{equation}
    \begin{aligned}
    f_{\Gamma_k}(t_k) &= [f_{\tau_{k,1}}(\tau_{k,1}) \circledast f_{\tau_{k,2}}(\tau_{k,2}) \circledast \cdots \circledast f_{\tau_{k,n}}(\tau_{k,n})](t_k) \\
    &= \int_{0}^{t_k} \!\int_{0}^{t_k - x_1} \!\cdots \int_{0}^{t_k - x_1 - \cdots - x_{n-2}} f_{\tau_{k,1}}(x_1)\;f_{\tau_{k,2}}(x_2)\ \\ &\cdots\ f_{\tau_{k,n-1}}(x_{n-1}) f_{\tau_{k,n}}\bigl(t_k - (x_1 + \dots + x_{n-1})\bigr) dx_{n-1}\dots dx_1.
    \end{aligned}
\end{equation}

\noindent where $\circledast$ is the distribution convolution symbol. 

In practice, we approximate $f_{\Gamma_k}(t_k)$ for any time $t_k>0$ by either Monte Carlo Simulations or Moment-Matching Approximations. For the convolution of log-normal distributions with moderate variance and $n_k$, the Fenton-Wilkinson approach provides a feasible solution to directly match the first two moments, and is widely-adopted as the approximated analytical solution of log-normal sums in various fields \citep{fenton1960sum, mehta2007approximating, cobb2012approximating}. Fenton found that the sum of several independent log-normal variables can be reasonably approximated by another log-normal. Under this approximation, the mean of the route travel time equals the sum of link means, and the variance equals the sum of all link variances and covariance terms. These matched moments define the parameters of an approximate log-normal for the route \citep{fenton1960sum}. Furthermore, \cite{cardieri2000statistics} note that this log-normal approximation is computationally efficient compared to brute-force convolution or simulation, with only a modest loss of accuracy. That is, we are looking for the parameters of an approximate distribution of $\Gamma_k \approx X^*_k$, where $X_k^* \sim \mathsf{Lognormal}(\mu_k^*, \sigma_k^{*2})$. That is,
\begin{align}
\label{eq: fw-approximation}
    f_{\Gamma_k}(t_k) \approx \frac{1}{t_k\,\sigma_k^*\sqrt{2\pi}}\exp\!\Bigl[-\frac{\bigl(\ln t_k - \mu_k^*\bigr)^2}{2\,\sigma_k^{*2}}\Bigr].
\end{align}

\noindent and the associated cumulative density function (CDF) is as,
\begin{equation}
    F_{\Gamma_k}(t_k) \approx \Phi( \frac{\ln t_k - \mu_k^*}{\sigma_k^*}).
\end{equation}

\noindent where,
\begin{align}
    \mu_k^* &= \ln M_k - \frac12 \ln\Bigl(1 + \frac{V_k}{M_k^2}\Bigr), \quad \sigma_k^{*2} = \ln\Bigl(1 + \frac{V_k}{M_k^2}\Bigr).
\end{align}
\noindent with,
\begin{align}
    M_k &= \sum_{i=1}^{n_k} \mathbb{E}[\tau_{k,i}] , \quad V_k = \sum_{i=1}^{n_k} \mathrm{Var}[\tau_{k,i}].
\end{align}

\subsubsection{Spatiotemporal Risk Formulation \label{subsubsec: risk-model}}

As reviewed in \Cref{subsec: risk-review}, the collision between two moving aircraft on the surface is quantified by the probability of conflicts between aircraft by evaluating random deviations in position and speed. This also guides our spatiotemporal risk formulation. In our formulation, we assume that a collision occurs when two interchangeable aircraft satisfy the following two conditions,

\begin{itemize}
    \item The first aircraft arrives at the collision point \(x_c\) at time $t$
    \item The second aircraft is positioned within a spatial collision radius \(r_c\) of the collision point \(x_c\) at the same time.
\end{itemize}

We derive an expression to approximate the probability that a collision occurs at \(x_c\) at any time $t$. We start with a joint probability distribution $f_{X_i,\Gamma_i}(x,t)$ for the spatiotemporal state of the aircraft. Define \(f_{\Gamma_1}(t|x)\) be the conditional PDF of aircraft 1’s arrival time at location $x$, \(f_{\Gamma_2}(t|x)\) be the PDF of aircraft 2’s arrival time at location $x$, and \(f_{X_2}(x|t)\) be the PDF describing aircraft 2’s spatial position at time \(t\).  

If we take aircraft 1 at the time \(t\) it reaches \(x_c\), then a collision requires that aircraft 2 is located in the interval \([x_c-r_c,\, x_c+r_c]\) at the same time \(t\), where $r_c$ is viewed as the averaged wingspan of two aircraft to extend the point mass formulation in the simplest way. A fully coupled expression for the probability of a collision at any time is then given as,
\begin{equation}
\label{eq: fw-formulation}
    P_{FW}(x_c) = \int_0^\infty f_{\Gamma_1}(t|x_c)\,  \left[ \int_{x_c-r_c}^{x_c+r_c} f_{X_2}(x|t)\, dx \right] dt.
\end{equation}

For \(r_c\), it is valid to assume that \(f_{x_2}(x,t)\) is nearly constant over \([x_c-r_c,\, x_c+r_c]\), when wingspans are small compared to the distance traveled. Thus, we can approximate with,
\begin{equation}
\int_{x_c-r_c}^{x_c+r_c} f_{X_2}(x|t)\, dx \approx 2r_c\, f_{X_2}(x_c|t),
\label{eq: rcintegral}
\end{equation}


Approximating the velocity at collision as independent of $x$ and $t$, a change of variables from space to time near $x_c$ can be made using the expected inverse of the speed at the point of collision,
\begin{equation}
f_{X_2}(x|t) = \mathbb{E}\!\left[\frac{1}{v_2}\right] f_{\Gamma_2}(t|x).
\label{eq: speed}
\end{equation}

\begin{lemma}
We prove \Cref{eq: speed} using the joint density of the position, arrival time, and velocity $f_{X_2,\Gamma_2,V_2}(x_2,t_2,v_2)$ and the relation $\frac{dx}{dt} = v_2$. We assume that the aircraft velocity distribution is independent and that the time distribution is independent and constant in time such that $f_{\Gamma_2}(t_2)=f_{\Gamma_2}(\cdot)$. The position distribution is dependent on the time and velocity.

\begin{align}
f_{X_2,\Gamma_2,V_2}(x_2,t_2,v_2) = f_{X_2}(x_2|t_2,v_2)f_{\Gamma_2}(t_2)f_{V_2}(v_2).
\end{align}

We first use Bayes' rule for conditional distributions to relate the conditional position-velocity distribution to the conditional arrival time distribution.
\begin{equation}
\begin{aligned}
f_{X_2}(x_2|t_2) &=  f_{\Gamma_2}(t_2|x_2)\frac{{f_{X_2}(x_2)}}{f_{\Gamma_2}(t_2)}\\
\end{aligned}
\label{eq: bayes}\end{equation}



We compute the $f_{X_2}$ distribution using the integral of joint probability over time and velocity. The $\frac{dx}{dt} = v$ relation is utilized for a change of variables from position to time. We also utilize the fact that $f_{\Gamma_2}$ is constant. The remaining integral results in the $\mathbb{E}[{v_2^{-1}}]$ term.

\begin{equation}
\begin{aligned}
f_{X_2}(x_2) &= \iint f_{X_2}(x_2|t,v)f_{\Gamma_2}(t)f_{V_2}(v) \, dt \, dv \\
&= f_{\Gamma_2}(\cdot)\iint f_{X_2}(x_2|t,v)f_{V_2}(v) \, dt \, dv \\
&= f_{\Gamma_2}(\cdot)\iint \frac{f_{\Gamma_2}(t)}{v}f_{V}(v) \, dt \, dv \\
&=  f_{\Gamma_2}(\cdot)\mathbb{E}\bigg[\frac{1}{v_2}\bigg] \\
\end{aligned}
\end{equation}

The position distribution is plugged into the \eqref{eq: bayes} and the constant time distribution terms cancel out.
\begin{equation}
\begin{aligned}
f_{X_2}(x_2|t_2) &=  f_{\Gamma_2}(t_2|x_2)\frac{{f_{X_2}(x_2)}}{f_{\Gamma_2}(t_2)}
\\&= f_{\Gamma_2}(t_2|x_2)\frac{f_{\Gamma_2}(\cdot)\mathbb{E}[v^{-1}_2] }{f_{\Gamma_2}(\cdot)}
\\&= \mathbb{E}\left[\frac{1}{v_2}\!\right]f_{\Gamma_2}(t_2|x_2)
\end{aligned}   
\end{equation}
Thus we have recovered \Cref{eq: speed}. \qed
\end{lemma}

The approximated probability of aircraft 2 occupying \([x_c-r_c,\, x_c+r_c]\) at time $t$ is obtained by substituting \Cref{eq: speed} into \Cref{eq: rcintegral},
\begin{equation}
\int_{x_c-r_c}^{x_c+r_c} f_{X_2}(x|t)\, dx \approx 2r_c\, \mathbb{E}\!\left[\frac{1}{v_2}\right] f_{\Gamma_2}(t|x_c).
\end{equation}

Substitute this spatial approximation into the collision probability formulation,
\begin{equation}
\begin{aligned}
P_{FW}(x_c) 
&\approx \int_0^\infty f_{\Gamma_1}(t|x_c)\,  \left[ 2r_c\, \mathbb{E}\!\left[\frac{1}{v}\right]\,f_{\Gamma_2}(t|x_c) \right] dt\\[1mm]
&= 2r_c\, \mathbb{E}\!\left[\frac{1}{v}\right] \int_0^\infty f_{\Gamma_1}(t|x_c)\, f_{\Gamma_2}(t|x_c)\, dt.
\end{aligned}
\end{equation}

Note the integral $\int_0^\infty f_{\Gamma_1}(t|x_c)\, f_{\Gamma_2}(t|x_c)\, dt$ is viewed as the temporal overlap collision density. To keep the notation consistent with \Cref{subsubsec: travel-time-model}, we remove the spatial condition of $f_{\Gamma_k}(t|x_c)$ and use $f_{\Gamma_k}(t)$, which is,
\begin{equation}
f_\digamma(\digamma = 0) = \int_0^\infty f_{\Gamma_1}(t)\, f_{\Gamma_2}(t)\, dt.\end{equation}
\noindent where $\digamma$ represents the arrival time difference between two aircraft,
\begin{equation}
    \digamma = \Gamma_1 - \Gamma_2.
\end{equation}

We can obtain the compact expression as,
\begin{equation}
P_{FW}(x_c) \approx 2r_c\, \mathbb{E}\!\left[\frac{1}{v}\right] \, f_\digamma(\digamma = 0).
\end{equation}

Further substituting the F-W approximated PDFs, we have $f_\digamma(\digamma = 0)$ as,
\begin{equation}
\begin{aligned}
    &f_\digamma(\digamma = 0) \\ &= \int_0^\infty \frac{1}{t\,\sigma_1^*\sqrt{2\pi}}\exp\!\Biggl[-\,\frac{\Bigl(\ln t - \mu_1^*\Bigr)^2}{2\,\sigma_1^{*2}}\Biggr] \cdot \frac{1}{t\,\sigma_2^*\sqrt{2\pi}}\exp\!\Biggl[-\,\frac{\Bigl(\ln t - \mu_2^*\Bigr)^2}{2\,\sigma_2^{*2}}\Biggr] dt \\
    &= \frac{1}{2\pi\,\sigma_1^*\sigma_2^*} \int_0^\infty \frac{1}{t^2}\,\exp\!\Biggl[-\frac{\bigl(\ln t - \mu_1^*\bigr)^2}{2\sigma_1^{*2}} - \frac{\bigl(\ln t - \mu_2^*\bigr)^2}{2\sigma_2^{*2}}\Biggr] dt.
\end{aligned}
\end{equation}

This spatiotemporal collision risk formulation provides the probability score at certain pre-defined potential collision spot $x_c$ with a collision radius $r_c$, and depends on the link travel time distributions of the two aircraft. 

Additionally, the approximation error analysis of the proposed formulation is given in \Cref{sec: error-proof}.

\subsubsection{Petri-Net Formulation \label{subsubsec: petri-net}}
As an alternative to the proposed spatiotemporal risk formulation, we also provide a Petri-Net formulation for aircraft surface risk calculation. We follow the flow of \cite{sun2024collision} and propose the following formulation.

Following the previously defined notation, let $k\in\{1,2\}$ be the aircraft index. Let $\Gamma_k(x_c)$ denote the random arrival time of aircraft $k$ at the node $x_c$ (i.e., collision spot). Similarly, a graph-theoretic node is a point set of measure zero. Hence instantaneous occupancy at $x_c$ is zero in continuous time unless we introduce a small temporal window. We therefore define an \textit{operational} coincidence window $\varepsilon>0$ around the instant of arrival and the corresponding node-time occupancy indicator as,
\begin{equation}
\mathrm{Occ}^{(\varepsilon)}_i(t; s_k)
:= \mathbf{1}\{|t-\Gamma_i(s_k)|<\varepsilon\}.
\end{equation}

\noindent Then the Petri-net node co-occupancy probability (within small $\varepsilon$) is,
\begin{equation}
\label{eq: pn-formulation}
\begin{aligned}
P_{\text{PN,node}}^{(\varepsilon)}(x_c)
& = \mathbb{E}\!\left[\int_0^\infty 
\mathrm{Occ}^{(\varepsilon)}_1(t; s_k)\,
\mathrm{Occ}^{(\varepsilon)}_2(t; s_k)\,dt\right] \\
& = 2\varepsilon\,\underbrace{\int_0^\infty f_{\Gamma_1}(t)f_{\Gamma_2}(t)\,dt}_{f_\digamma(\digamma=0)} \;+\; o(\varepsilon),
\end{aligned}
\end{equation}

\begin{lemma}
\label{lem: pn-small-window}
We realize \Cref{eq: pn-formulation} here. Let $\Gamma_k(x_c)$ denote the random arrival time of aircraft $k\in\{1,2\}$ at node $x_c$ and define
$\mathrm{Occ}^{(\varepsilon)}_k(t;x_c):=\mathbf{1}\{|t-\Gamma_k(x_c)|<\varepsilon\}$.
Write $f_{\Gamma_k}(t)\equiv f_{\Gamma_k}(t\mid x_c)$ and $F_k(t)\equiv F_{\Gamma_k}(t\mid x_c)$.
\begin{equation}
\label{eq:pn_chain}
\begin{aligned}
P_{\mathrm{PN,node}}^{(\varepsilon)}(x_c)
&:= \Pr\!\big(|\Gamma_1(x_c)-\Gamma_2(x_c)|<\varepsilon\big) \\[2pt]
&= \int_0^\infty \Pr\!\big(|\Gamma_1-\Gamma_2|<\varepsilon \,\big|\, \Gamma_1=t\big)\,f_1(t)\,dt 
\quad &&\text{(conditioning on $\Gamma_1$)}\\[2pt]
&= \int_0^\infty \big[F_2(t+\varepsilon)-F_2(t-\varepsilon)\big]\,f_1(t)\,dt 
\quad &&\text{(definition of $F_2$)}\\[2pt]
&= \int_0^\infty \big[2\varepsilon\,f_{\Gamma_2}(t)+o(\varepsilon)\big]\,f_1(t)\,dt 
\quad &&\text{(mean value theorem)}\\[2pt]
&= 2\varepsilon \int_0^\infty f_{\Gamma_1}(t)f_{\Gamma_2}(t)\,dt \;+\; o(\varepsilon) \\[2pt]
&= 2\varepsilon\, f_\digamma(\digamma=0) \;+\; o(\varepsilon),
\end{aligned}
\end{equation}
where $f_\digamma(\digamma=0)=\int_0^\infty f_{\Gamma_1}(t)f_{\Gamma_2}(t)\,dt$.

\noindent Equivalently, the overlap-integral form satisfies,
\begin{equation}
\begin{aligned}
P_{\mathrm{PN,node}}^{(\varepsilon)}(x_c)
& =  \frac{1}{2\varepsilon}\,
\mathbb{E}\!\left[\int_0^\infty \mathrm{Occ}^{(\varepsilon)}_1(t;x_c)\,\mathrm{Occ}^{(\varepsilon)}_2(t;x_c)\,dt\right] \\
& =  2\varepsilon\, f_\digamma(\digamma=0) + o(\varepsilon).
\end{aligned}
\end{equation}
Thus, we have recovered \Cref{eq: pn-formulation}.  \qed
\end{lemma}

\noindent where $f_\digamma(\digamma=0)$ is the value at zero of the PDF of $\Gamma_1-\Gamma_2$ in the FW formulation. $P_{\text{PN,node}}^{(\varepsilon)}$ is the probability the two arrivals fall within a $\pm\varepsilon$ coincidence window at the node. Further recall the FW-based risk probability at the same node $x_c$ with spatial capture radius $r_c$ is,

\begin{equation}
    P_{\text{FW}}(x_c)\;\approx\; 2r_c\,\mathbb{E}\!\left[\tfrac{1}{v}\right]\; f_\digamma(\digamma=0)
\end{equation}

This is derived by converting the temporal overlap into a spatial hit via the (small) window $[x_c-r_c, x_c+r_c]$ and the expected inverse speed near $x_c$. In the link model,
$\mathbb{E}[1/v]=e^{-\mu+\tfrac{1}{2}\sigma^2}$ using the incoming link’s $(\mu,\sigma)$.

Comparing the two expressions shows they share the same term $f_\digamma(\digamma=0)$. Eliminating $f_\digamma(\digamma=0)$ gives the proportionality,

\begin{equation}
P_{\text{FW}}(x_c)
\;\approx\;
\frac{r_c}{\varepsilon}\;\mathbb{E}\!\left[\tfrac{1}{v}\right]\;
P_{\text{PN,node}}^{(\varepsilon)}(x_c)
\end{equation}

If we choose $\varepsilon$ to match a temporal resolution (e.g., surveillance update or controller time bin) and pick $r_c$ as the spatial resolution (e.g., the half wingspan of the aircraft with a safety buffer), then the factor $\frac{r_c}{\varepsilon}\mathbb{E}[1/v]$ converts the PN node coincidence probability into the FW risk probability.

In summary, Petri-Net and FW formulations are tightly linked through the common overlap term $f_\Delta(0)$ where PN gives a temporal probability at the node. FW multiplies by a space–time factor to yield a collision risk probability at the same node. They are both probabilities. $P_{\text{PN,node}}^{(\varepsilon)}(x_c)\in[0,1]$ is a probability of temporal coincidence within $\pm\varepsilon$ at the node, which answers the question of \textit{Do the two arrivals occur within $\varepsilon$ seconds of each other at the node?} $P_{\text{FW}}(x_c)\in[0,1]$ is a probability of a collision event at the node given a spatial capture radius $r_c$.

\subsubsection{Real-Time Risk Assessment \label{subsubsec: real-time-risk}}
We enable real-time risk assessment capability of the node-based framework by inserting, at uniform temporal intervals $t=m\,\Delta t$ (i.e., $\Delta t=1$\,s), a \emph{virtual node} $\mathcal{V}$ at the instantaneous position of each aircraft along its current link. Let aircraft $k\in\{1,2\}$ be on link $i$ at time~$t$, with link length $d_{k,i}$ and lognormal speed $v_{k,i}\sim \text{Lognormal}(\mu_{k,i},\sigma_{k,i}^2)$ as in \Cref{subsubsec: travel-time-model}. Denote the residual distance on the current link as the remaining arc-length from the aircraft’s instantaneous position to the next topological node as,
\begin{equation}
    \tilde d_{k,i}(t)\in(0,d_{k,i}].
\end{equation}

Per the transformation in \Cref{eq: lognormal}, the residual time on the current link is,
\begin{equation}
\tilde\tau_{k,i}(t)\;=\;\frac{\tilde d_{k,i}(t)}{v_{k,i}}
\;\sim\;
\text{Lognormal}\!\bigl(\ln \tilde d_{k,i}(t)-\mu_{k,i},\,\sigma_{k,i}^2\bigr),
\end{equation}

\noindent with $\mathbb{E}[\tilde\tau_{k,i}(t)]$ and $\text{Var}[\tilde\tau_{k,i}(t)]$ obtained from the same moment formulas cited below \Cref{eq: lognormal} by replacing $d_{k,i}$ with $\tilde d_{k,i}(t)$.

For any overlapping node $x_c$ downstream of the current position, define the residual route at time $t$ as the concatenation of the residual portion $(\tilde d_{k,i}(t),\mu_{k,i},\sigma_{k,i})$ at $\mathcal{V}$ and all full links from the next node up to $x_c$. The residual travel time to $x_c$ is then a sum of link-times of the lognormal type in \Cref{eq: lognormal}. Its mean $M_k(t,x_c)$ and variance $V_k(t,x_c)$ are computed by summing the corresponding link moments (using the residual moments for the partial link and the original moments for full links). The route-level lognormal approximation,
\begin{equation}
T_k(t,x_c)\;\approx\;\text{Lognormal}\bigl(\mu_k^*(t,x_c),\,\sigma_k^{*2}(t,x_c)\bigr)
\end{equation}

\noindent is again obtained by the Fenton–Wilkinson moment-matching as in \Cref{eq: fw-approximation}. Moreover, the calculations of $P_{\text{FW}}(x_c)$ and $P_{\text{PN,node}}^{(\varepsilon)}(x_c)$ remain the same as in \Cref{subsubsec: risk-model}.

The above construction is repeated for every $t=m\,\Delta t$ and for every overlapping node $x_c$. The per-node risk time series remain negligible until the aircraft converge near the terminal conflict node, at which point the overlap density and the associated risk rise sharply.

\section{Case Studies \label{sec: case-study}}
In this section, we present three illustrative case studies to demonstrate our proposed framework. The first case study examines the 2024 Haneda Airport runway collision in \Cref{subsec: case-1}, the second focuses on the 2024 KATL taxiway collision accident in \Cref{subsec: case-2}, and a reconstruction of the 1977 Tenerife airport disaster in \Cref{subsec: case-3}. These accidents underscore the complex interplay of human error, communication misunderstandings, and operational failures. However, we use the assumed values for the link travel time speed distribution parameters of the first case study, based on recommended values mentioned in the FAA airport design recommendations, 150/5300-13B, as in \cite{FAA_AC_150_5300_13B_2024}. For the second study, we obtain the ground movement data (i.e., ASDE-X) from Sherlock and use the link travel time speed distribution parameters from the real-world data. Details are given in \Cref{subsubsec: linkparameters}. For the last one, we adopt the inferred taxi speed parameters based on the timestamp given from the literature \citep{weick1990vulnerable}. 

\subsection{Case I: Haneda Airport Runway Incursion \label{subsec: case-1}}

\begin{figure}[hbt!]
\centering
\includegraphics[width=0.85\textwidth]
{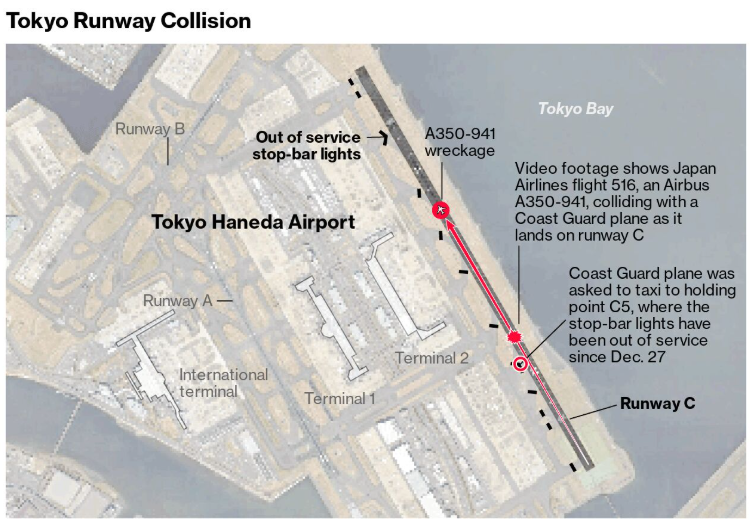}
\caption{The debriefing of the Haneda airport runway collision with the airport layout \citep{Bloomberg2024HanedaClearance}}
\label{fig: case1-overview}
\end{figure}

The Haneda Airport collision occurred on January 2, 2024, when Japan Airlines Flight 516 (Airbus A350) collided with a Japan Coast Guard DHC-8-315Q aircraft on the runway. The Coast Guard aircraft was stationed on the runway to deliver relief supplies following the Noto Peninsula earthquake. Despite the dramatic collision and ensuing fire, all 379 occupants aboard the A350 were successfully evacuated, while five of the six crew members on the smaller aircraft lost their lives. Investigations have attributed the accident primarily to miscommunication and human error where the Coast Guard pilot misinterpreted air traffic control instructions and mistakenly believed he had clearance to enter the runway. The communication transcripts ahead of the occurrence of the accident were released and tested against our developed learning model as in \Cref{fig: example-1-norules} and \Cref{fig: example-1-rules}. It is clear that JA722A was given clearance to hold at C5 and should wait for departure clearance. However, the JA722A goes to the runway and misunderstood as the first to take off at runway 34R. Our ATC Rule-Enhanced Learning model processed the timeline of clearance given from the controller and the state of each aircraft and listed in \Cref{tab: case-1}. 

\begin{table}
\centering
\caption{Key ATC communication transcript extracted with the knowledge-enhanced hybrid learning model for the 2024 Haneda airport runway incursion case study.}
\label{tab: case-1}
\resizebox{0.9\textwidth}{!}{%
\begin{tabular}{c|c|c|c|c}
\hline
\textbf{TIME} & \textbf{CALLSIGN} & \textbf{ACSTATE}   & \textbf{DEST\_RUNWAY} & \textbf{DESTINATION} \\ \hline
17:43:02      & Japan Air 516     & approach,departure & 34R                   & Rwy\_03\_001         \\ \hline
17:43:12      & Japan Air 516     & approach           & 34R                   & Rwy\_03\_001         \\ \hline
17:43:26      & Delta 276         & taxi               & 34R                   & holding point C1(Txy\_C1\_C)     \\ \hline
17:44:56      & Japan Air 516     & cleared,land       & 34R                   & Rwy\_03\_001         \\ \hline
17:45:01      & Japan Air 516     & cleared,land       & 34R                   & Rwy\_03\_001         \\ \hline
17:45:11      & JA722A            & taxi               &                       & holding point C5(Txy\_C5\_C5B)    \\ \hline
17:45:19      & JA722A            & taxi               &                       & holding point C5(Txy\_C5\_C5B)    \\ \hline
17:45:40      & Japan Air 179     & taxi               &                       & holding point C1(Txy\_C1\_C)      \\ \hline
17:45:56      & Japan Air 166     & approach           & 34R                   & Rwy\_03\_001         \\ \hline
17:47:23      & Japan Air 166     & approach           & 34R                   &                      \\ \hline
17:47:27      & Japan Air 166     &                    & 34R                   &                      \\ \hline
17:47:30      & Japan Air 516     & collision          &                       &                      \\ \hline
17:47:30      & JA722A            & collision          &                       &                      \\ \hline
\end{tabular}%
}
\end{table}

\begin{figure}[H]
    \centering
    \begin{subfigure}[t]{0.45\textwidth}
        \centering
        \includegraphics[width=\textwidth]
        {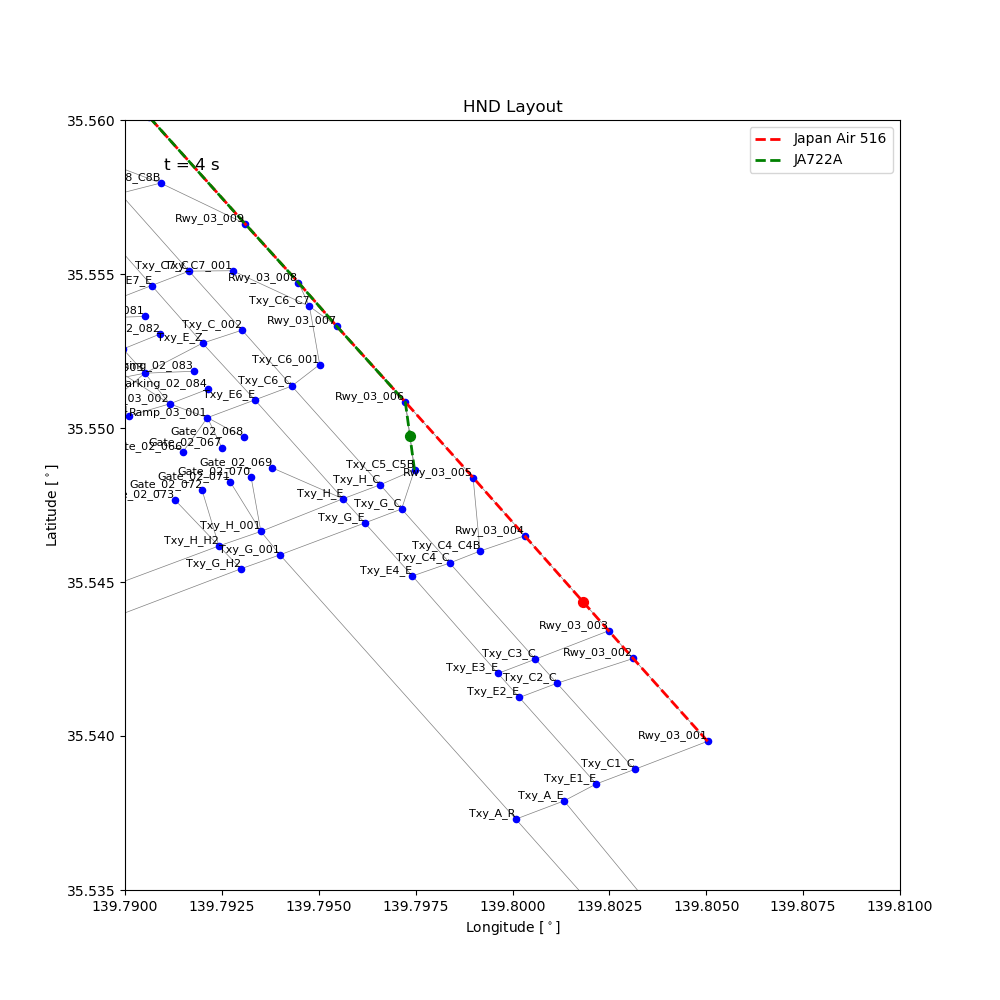}
    \end{subfigure}
    \begin{subfigure}[t]{0.45\textwidth}
        \centering
        \includegraphics[width=\textwidth]
        {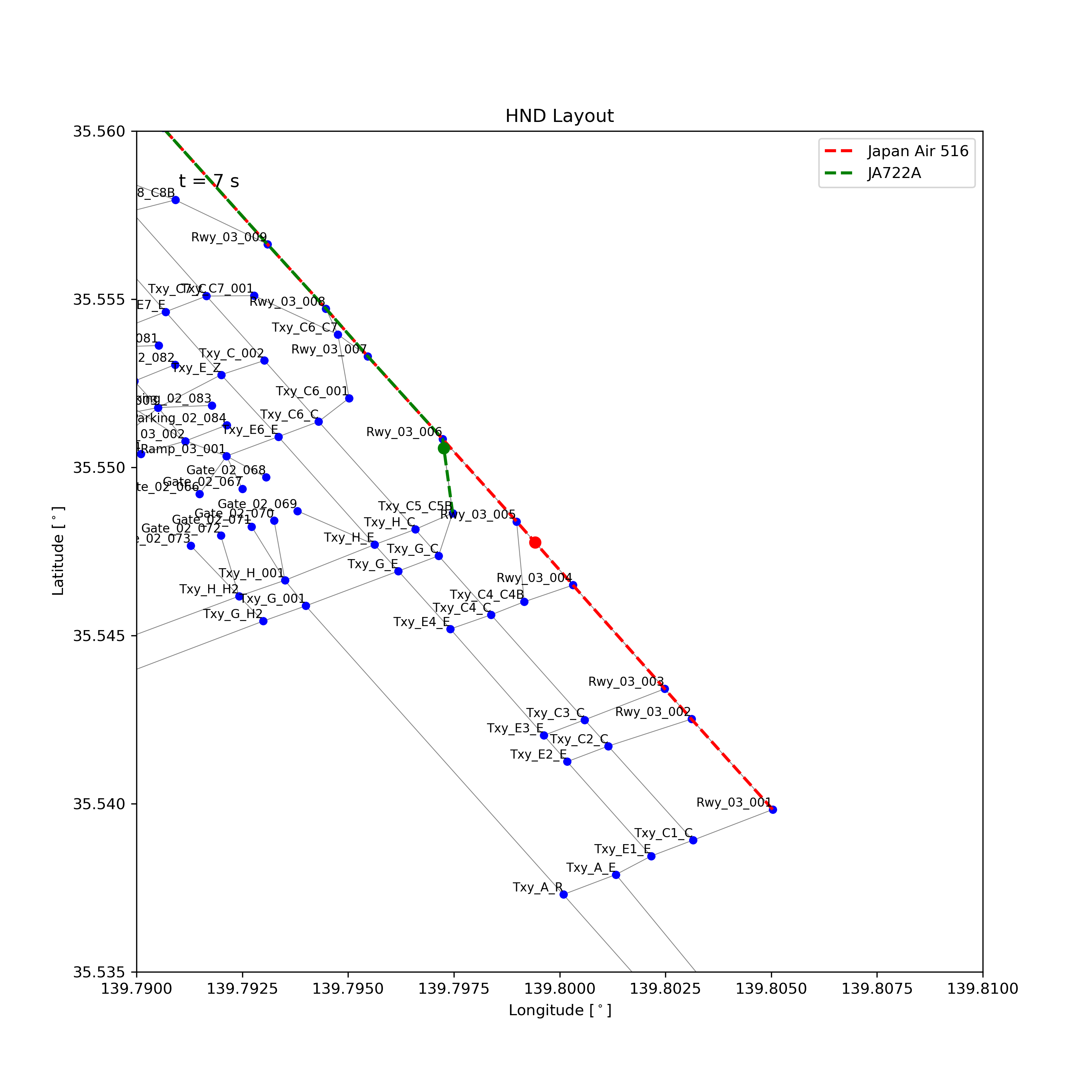}
    \end{subfigure}
    \\~\\
    \begin{subfigure}[t]{0.45\textwidth}
        \centering
        \includegraphics[width=\textwidth]
        {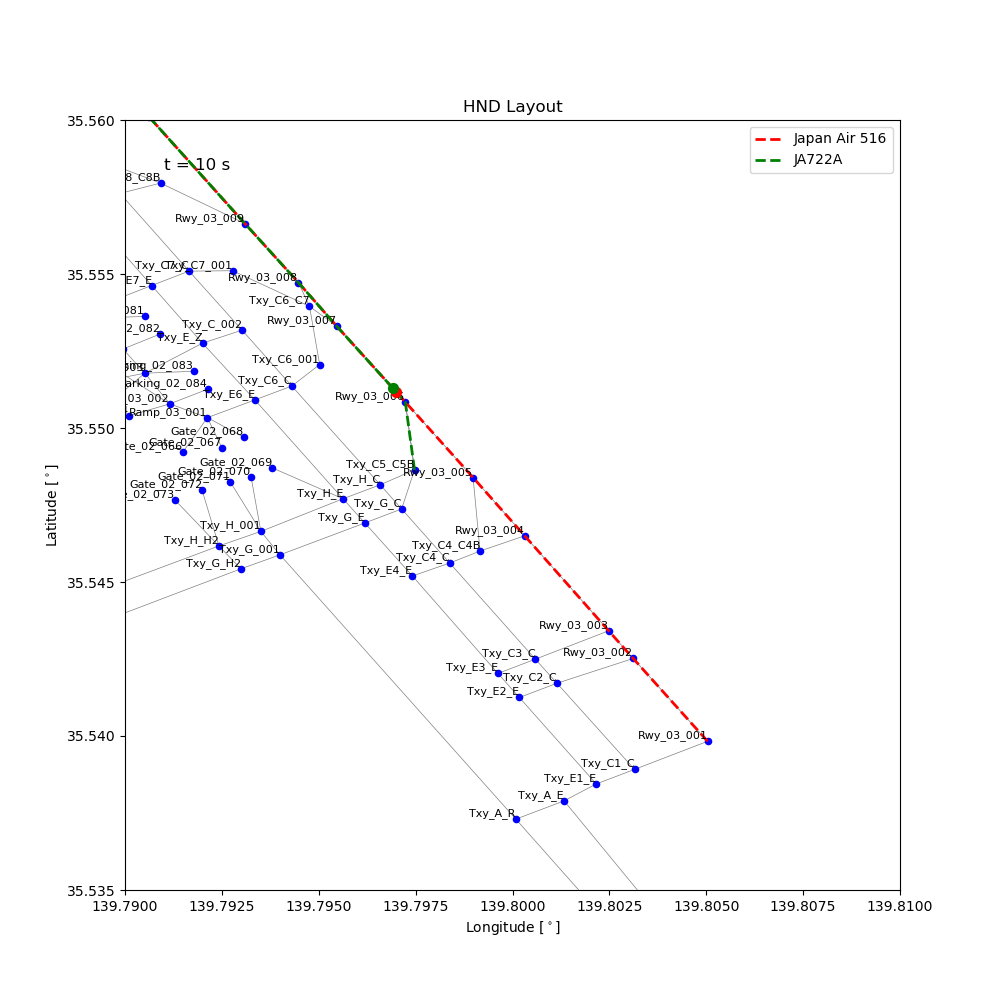}
    \end{subfigure}
    \begin{subfigure}[t]{0.45\textwidth}
        \centering
        \includegraphics[width=\textwidth]
        {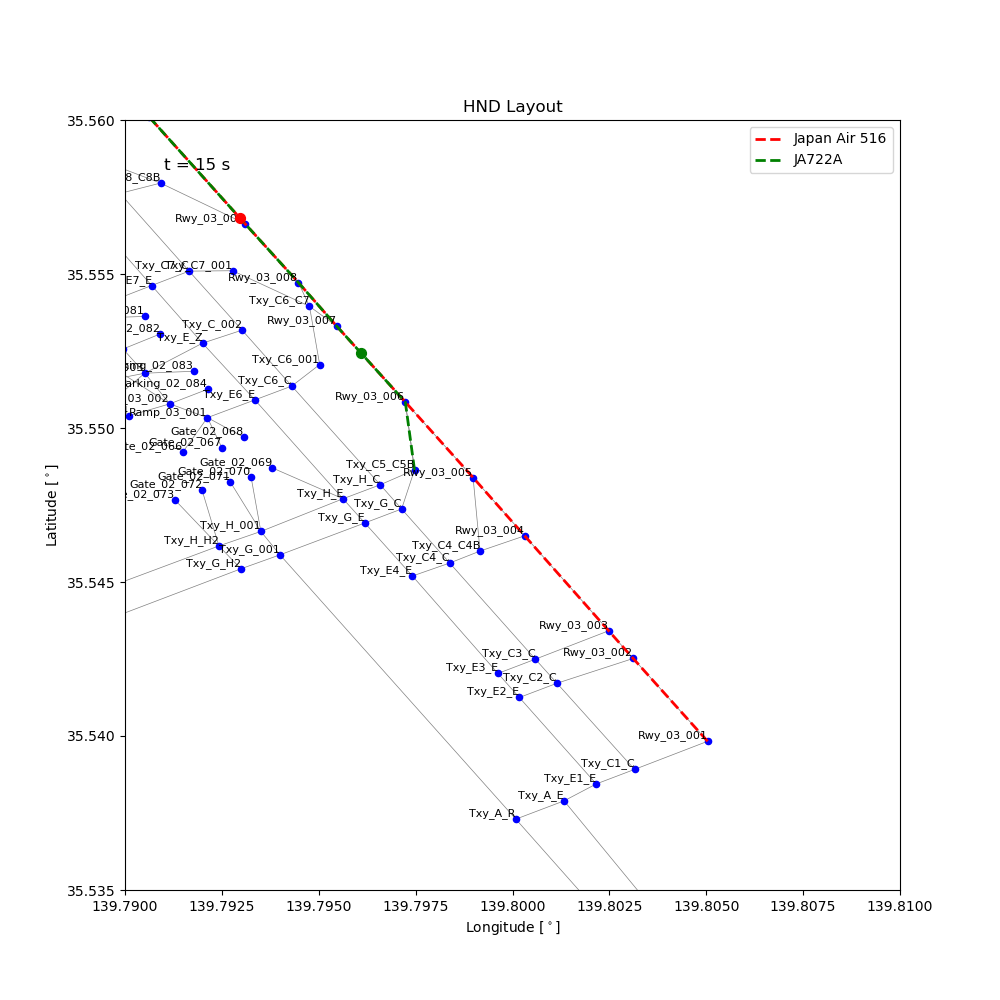}
    \end{subfigure}
\caption{Node-link simulation of the accident that happened at the Haneda in January 2024. }
\label{fig: case-2-pregress}
\end{figure}

Based on \Cref{tab: case-1}, we build the surface movement simulation environment with the airport node-link graph layout obtained from NASA FACET \citep{bilimoria2001facet}. We simulate the occurrence of the accident from the timestamp where the last clearance was given to JA722A from the ATCo in \Cref{fig: case-2-pregress}. The collision happened at the timestamp of the lower left figure. The simulation takes three inputs for each link, the link distance $d_{k,i}$, the link travel speed parameters $v_{k,i} \sim \text{Lognormal}(\mu_{k,i}, \sigma^2_{k,i})$. $d_{k,i}$ was calculated based on the location of the node on the node-link graph, with the speed parameters assumed to be $v_{k,i} \sim \text{Lognormal}(30, 10^2)$ if the link is between runway nodes, and $v_{k,i} \sim \text{Lognormal}(25, 5^2)$ between runway and taxiways. For links between two taxiway nodes, we assume the speed follows $v_{k,i} \sim \text{Lognormal}(20, 5^2)$, and $v_{k,i} \sim \text{Lognormal}(10, 5^2)$ for all other scenarios (taxilanes, ramps, etc). These speed assumptions are based on the recommended values mentioned in the FAA airport design recommendations, 150/5300-13B, as in \cite{FAA_AC_150_5300_13B_2024}. A detailed explanation of taxiways, runways, and terminology is given in \Cref{fig: airport-layout}. 

\begin{figure}
\centering
    \begin{subfigure}[t]{0.75\textwidth}
        \centering
        \includegraphics[width=\textwidth]
        {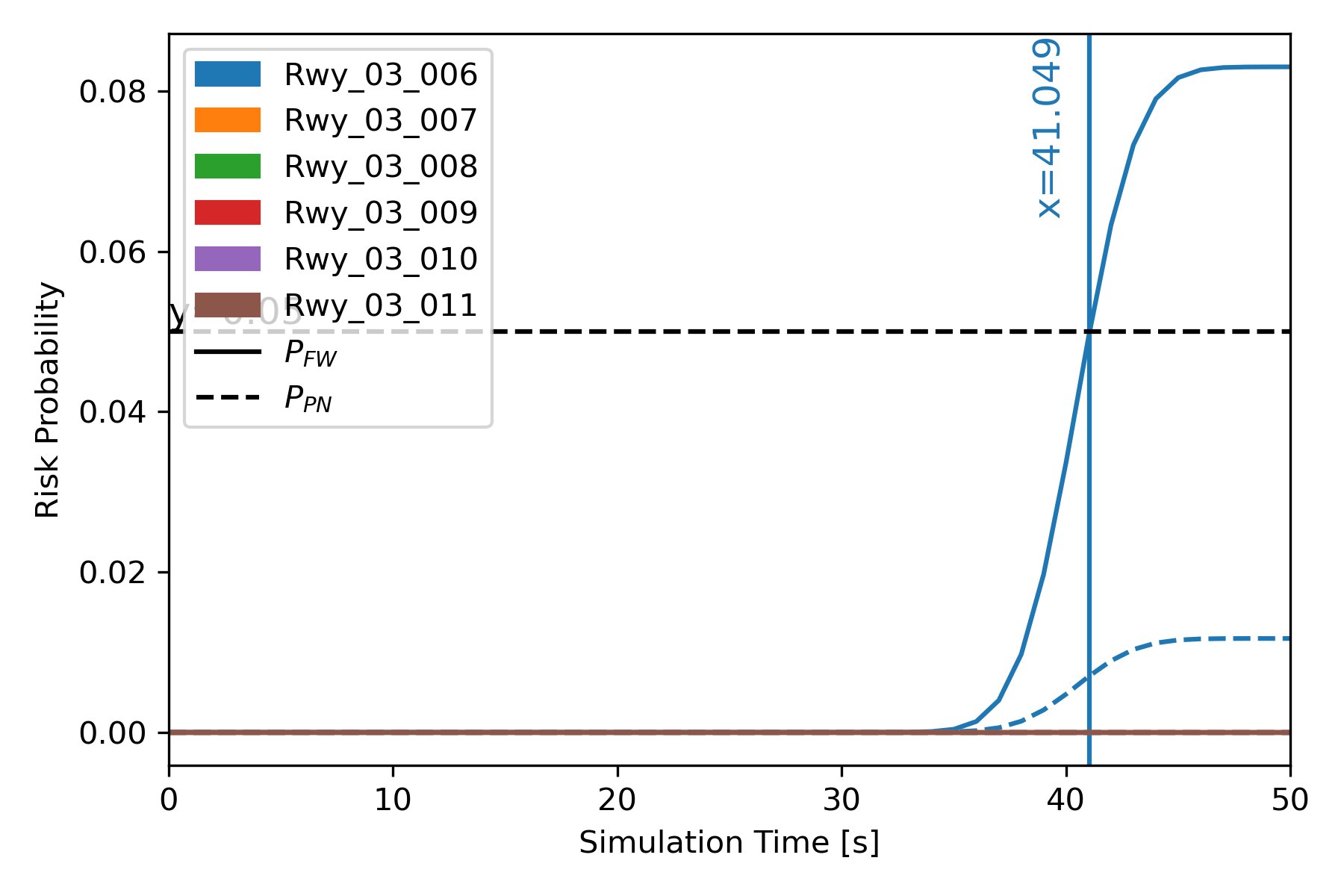}
        \caption{Haneda Simulation with safety threshold at 0.05. }                
        \label{fig: case1-0.05}
    \end{subfigure}
    \begin{subfigure}[t]{0.75\textwidth}
        \centering
        \includegraphics[width=\textwidth]
        {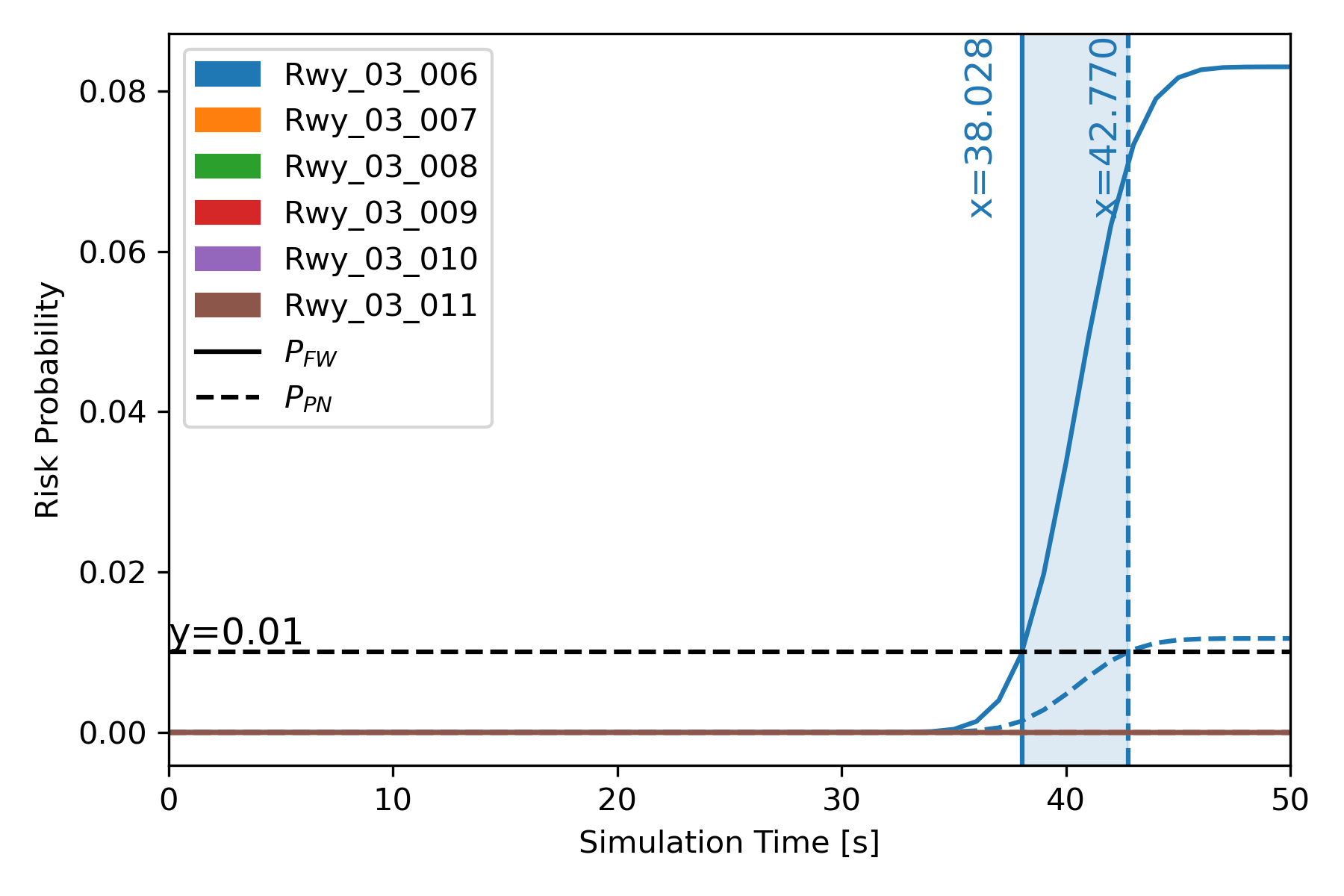}
        \caption{Haneda Simulation with safety threshold at 0.01. }
        \label{fig: case1-0.01}
    \end{subfigure}
\caption{The real-time risk probability calculated based on the Haneda airport disaster simulation. With different warning thresholds, we identify the lead times of the warning scheme from both the FW and PN formulation. We show that FW formulation provides larger lead time compared with the PN formulation. The shaded area highlights the lead time difference under each threshold.}
\label{fig: case1-risk}
\end{figure}

\Cref{tab: case-1} shows that the NER model is able to capture the destination runway for both Japan Air 516 and Delta 276, where the key information such as callsign, aircraft state and also retrieved. The destination node from the node-link graph are correlated by either, (a) the entry node of the given runway; (b) the similarity between the destination node name and the entities from pilot-ATC communication transcripts. \Cref{tab: case-1} visualizes several timestamps of the simulation, based on the assumed speed distribution parameters. Incorporating the information retrieved from the NER model, this simulation successfully replicates the occurrence of the Haneda runway incursion accident. 

We provide the real-time risk at each overlapping nodes across the entire simulation time in \Cref{fig: case1-risk}. In this scenario, these are nodes \texttt{Rwy\_03\_006} to \texttt{Rwy\_03\_011}. The collision happens at the first overlapping node while the collision at other nodes remain subtle. For risk probability tolerance threshold at 0.05, the FW formulation provides a warning at around 41 seconds while the PN formulation fails to show the warning as in \Cref{fig: case1-0.05}. For lower risk tolerance at 0.01, both methods show warning but FW shows larger lead time compared with the PN approach as in \Cref{fig: case1-0.01}.

\subsection{Case II: KATL Taxiway Collision\label{subsec: case-2}}
In the second case study, we investigate the taxiway collision occurred at KATL on September 10, 2024, where two Delta Air Lines aircraft collided while taxiing. An Airbus A350, preparing for an international departure to Tokyo, struck the tail of an Endeavor Air Bombardier CRJ-900, which was scheduled for a domestic flight to Lafayette, Louisiana. The collision, which occurred at the intersection of two taxiways, resulted in significant damage to the tail section of the smaller aircraft. Although no injuries were reported among the 277 passengers and crew, preliminary findings indicate that the pilot of the larger aircraft was momentarily distracted—likely due to efforts to monitor opposing traffic, thereby contributing to the mishap. Similar to case study I, this accident was caused by miscommunication between the tower controller and the pilot \citep{NTSB_DCA24FA299_2024}.

\subsubsection{Link Travel Speed Parameters \label{subsubsec: linkparameters}}

For this case study, we look for a more realistic simulation with link speed distribution parameters obtained from ASDE-X from the Sherlock Data Warehouse (SDW).
SDW is a comprehensive big data system developed specifically to support air traffic management (ATM) research, which collects raw data from multiple trusted sources such as the FAA and the National Oceanic and Atmospheric Administration (NOAA). Data sources include flight plans, flight tracks from Air Route Traffic Control Centers (ARTCCs) and TRACON facilities, as well as meteorological such as wind, temperature, pressure, and precipitation from NOAA’s Rapid Refresh (RR) system, along with convective weather details (e.g., echo tops) from the Convective Integrated Weather Service (CIWS) and FAA SWIM. The raw input data are parsed and processed to ensure that the data are reliable, queryable, and ready for further analysis. SDW is a central resource for researchers who need to analyze complex datasets and derive insights for improving air traffic operations and safety \citep{srivastava2011improving, pang2019recurrent, pang2020conditional, wang2019predicting}. 

\begin{figure}
\centering
\includegraphics[width=0.95\textwidth]
{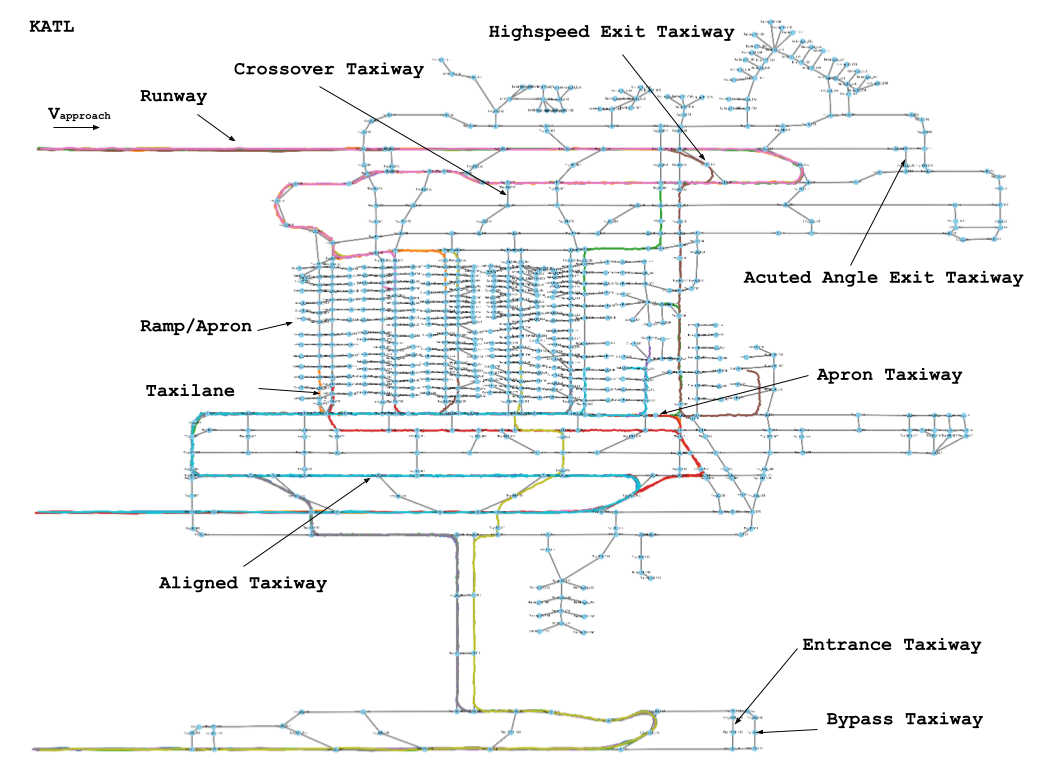}
\caption{KATL node-link graph of the airport layout and configurations of each line segment, with several samples of ASDE-X aircraft landing trajectories overlaid. The detailed classification of taxiways can be found in FAA Airport Design Manual \cite{FAA_AC_150_5300_13B_2024}.}
\label{fig: airport-layout}
\end{figure}

\Cref{fig: airport-layout} provides the node-link graph layout of KATL, where several arrival flight trajectories from SDW ASDE-X are layered on top of it. According to \citep{FAA_AC_150_5300_13B_2024}, taxiways are further divided to crossover taxiways, apron taxiways, bypass taxiways, etc. For the sake of simplicity, the detailed classification of different taxiway types is not given here. Similar to previous work \citep{pang2022bayesian}, we adopt the open source high performance geospatial data processing tool, Apache Sedona, to calculate the aircraft taxi speed at each node link. 

\begin{figure}
    \centering
    \begin{subfigure}[t]{0.45\textwidth}
        \centering
        \includegraphics[width=\textwidth]
        {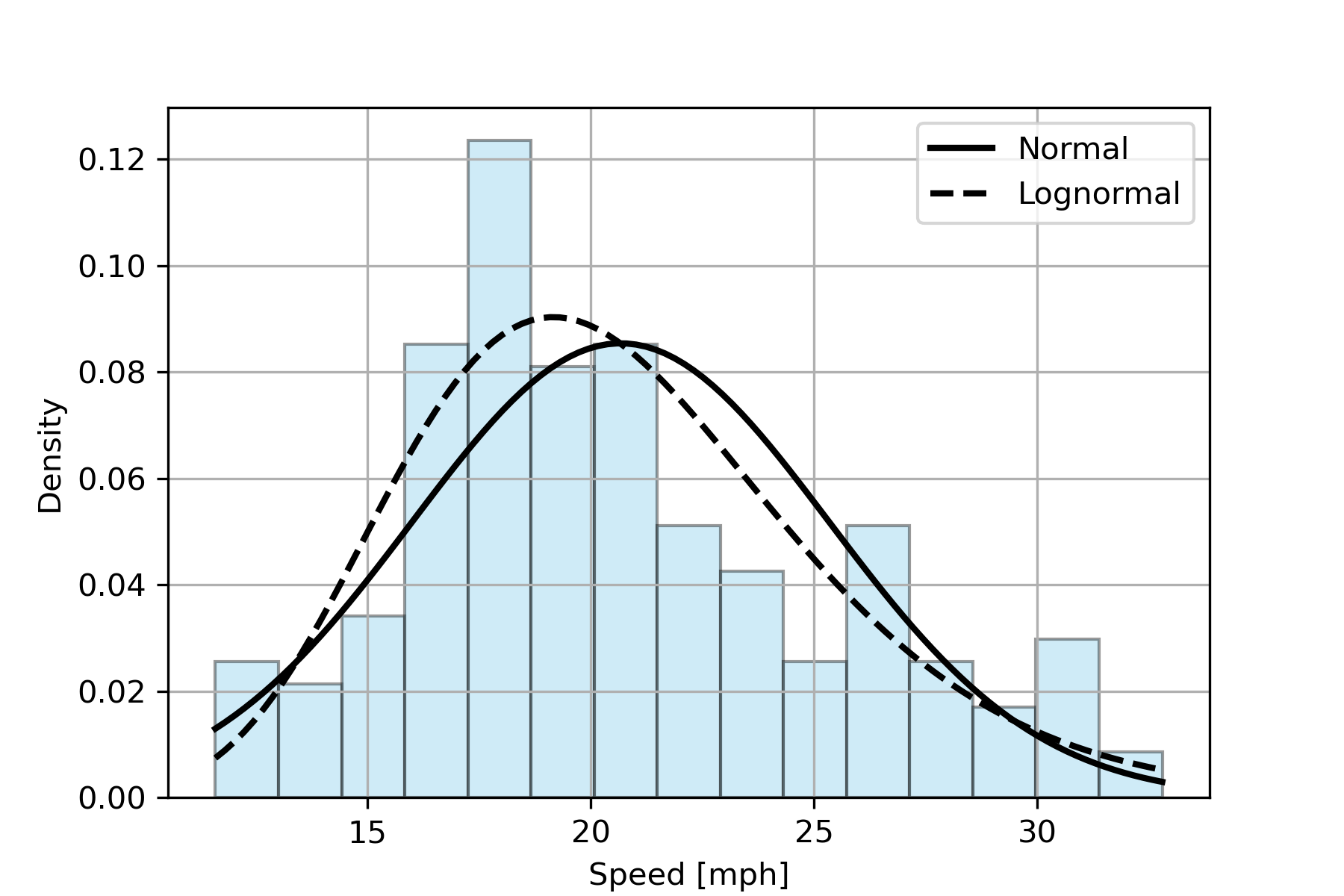}
        \caption{Link: $\texttt{Txy\_E\_004} \rightarrow \texttt{Txy\_E\_003}$. Data obtained from IFF ASDEX on 05/08/2023.}
        \label{fig: 0508-hist-392}
    \end{subfigure}
    \begin{subfigure}[t]{0.45\textwidth}
        \centering
        \includegraphics[width=\textwidth]
        {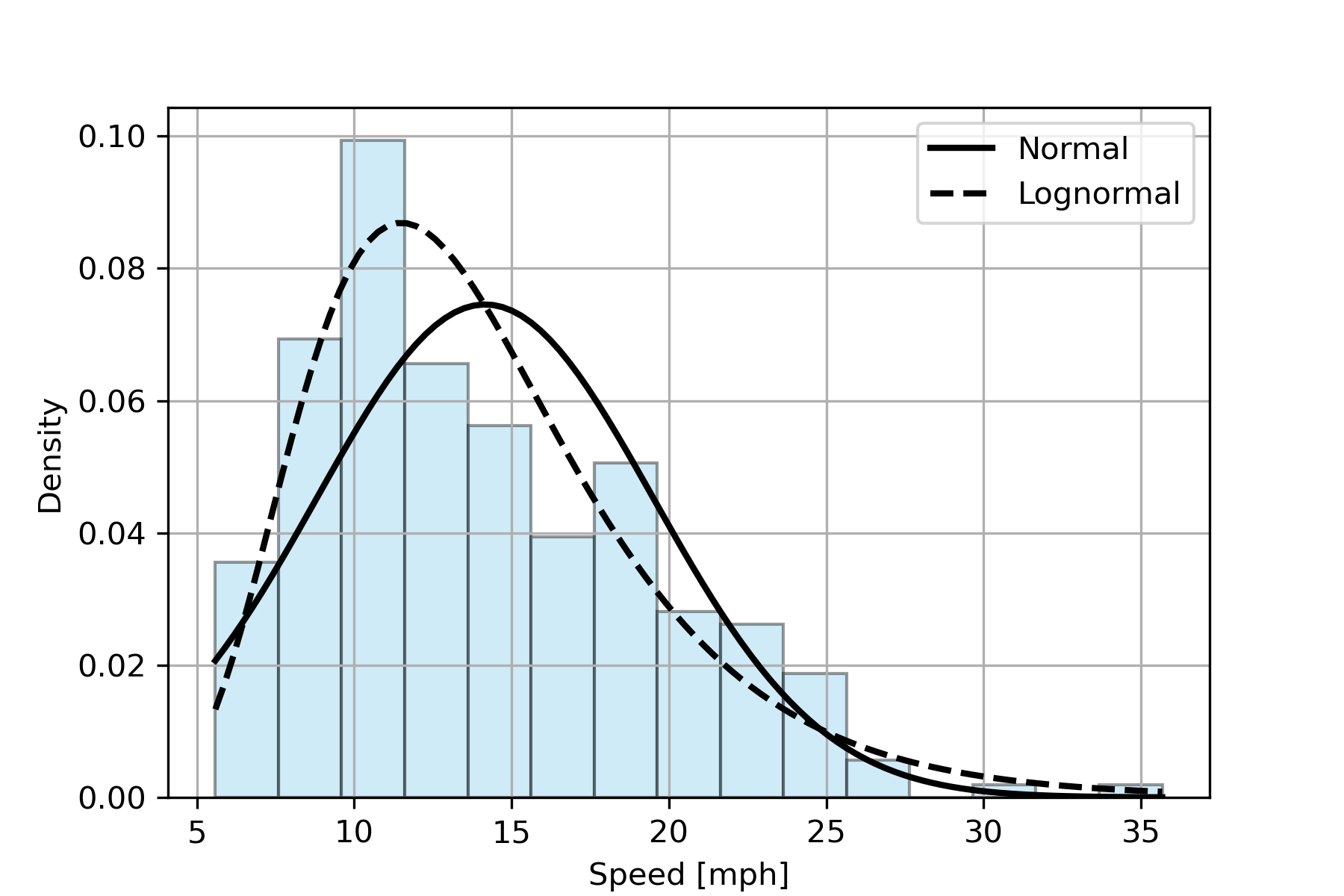}
        \caption{Link: $\texttt{Txy\_E\_004} \rightarrow \texttt{Txy\_E\_003}$. Data obtained from IFF ASDEX on 05/11/2023.}
        \label{fig: 0511-hist-392}
    \end{subfigure}
    \\~\\
    \begin{subfigure}[t]{0.45\textwidth}
        \centering
        \includegraphics[width=\textwidth]
        {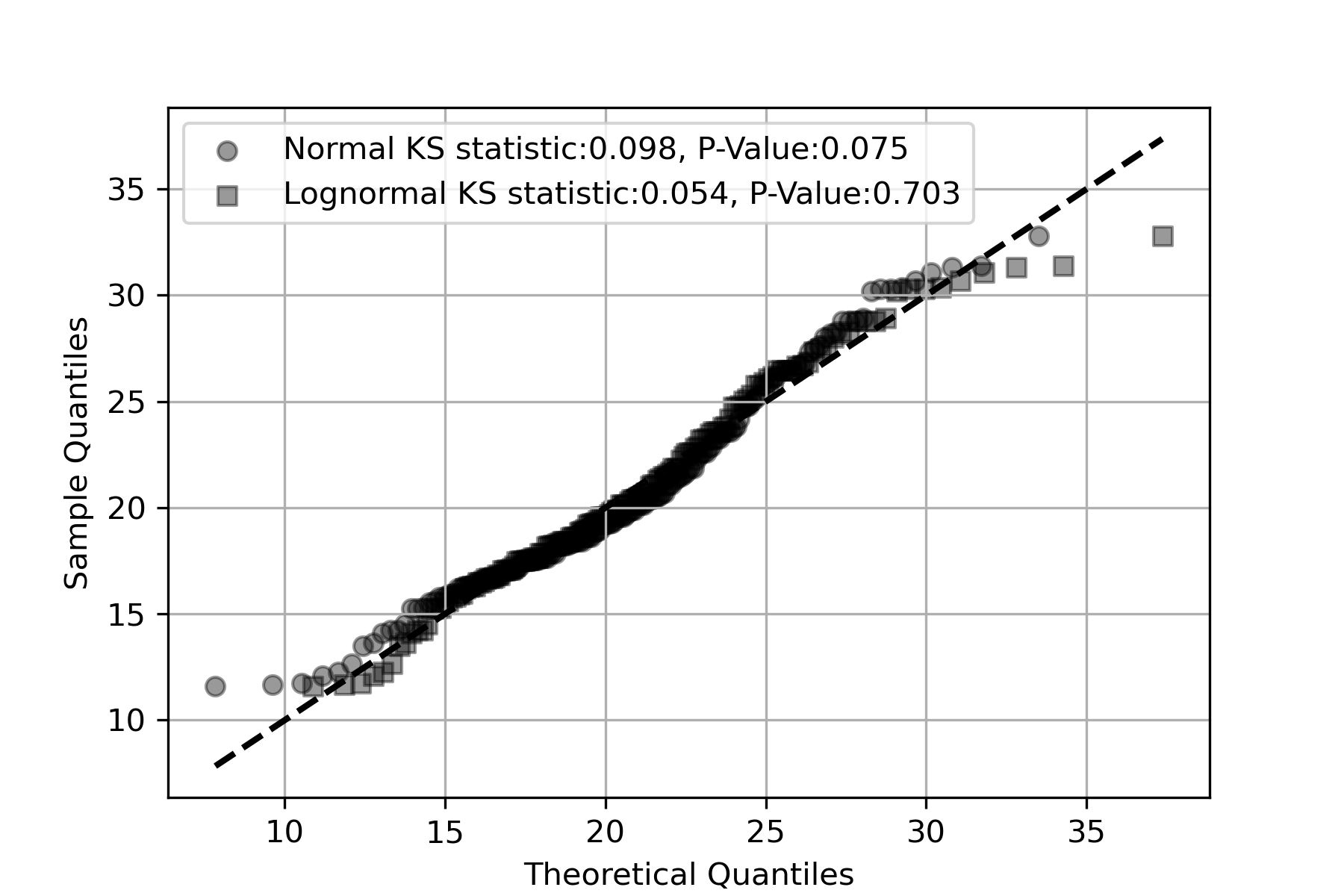}
        \caption{Link: $\texttt{Txy\_E\_004} \rightarrow \texttt{Txy\_E\_003}$. Data obtained from IFF ASDEX on 05/08/2023.}
        \label{fig: 0508-qq-392}
    \end{subfigure}
    \begin{subfigure}[t]{0.45\textwidth}
        \centering
        \includegraphics[width=\textwidth]
        {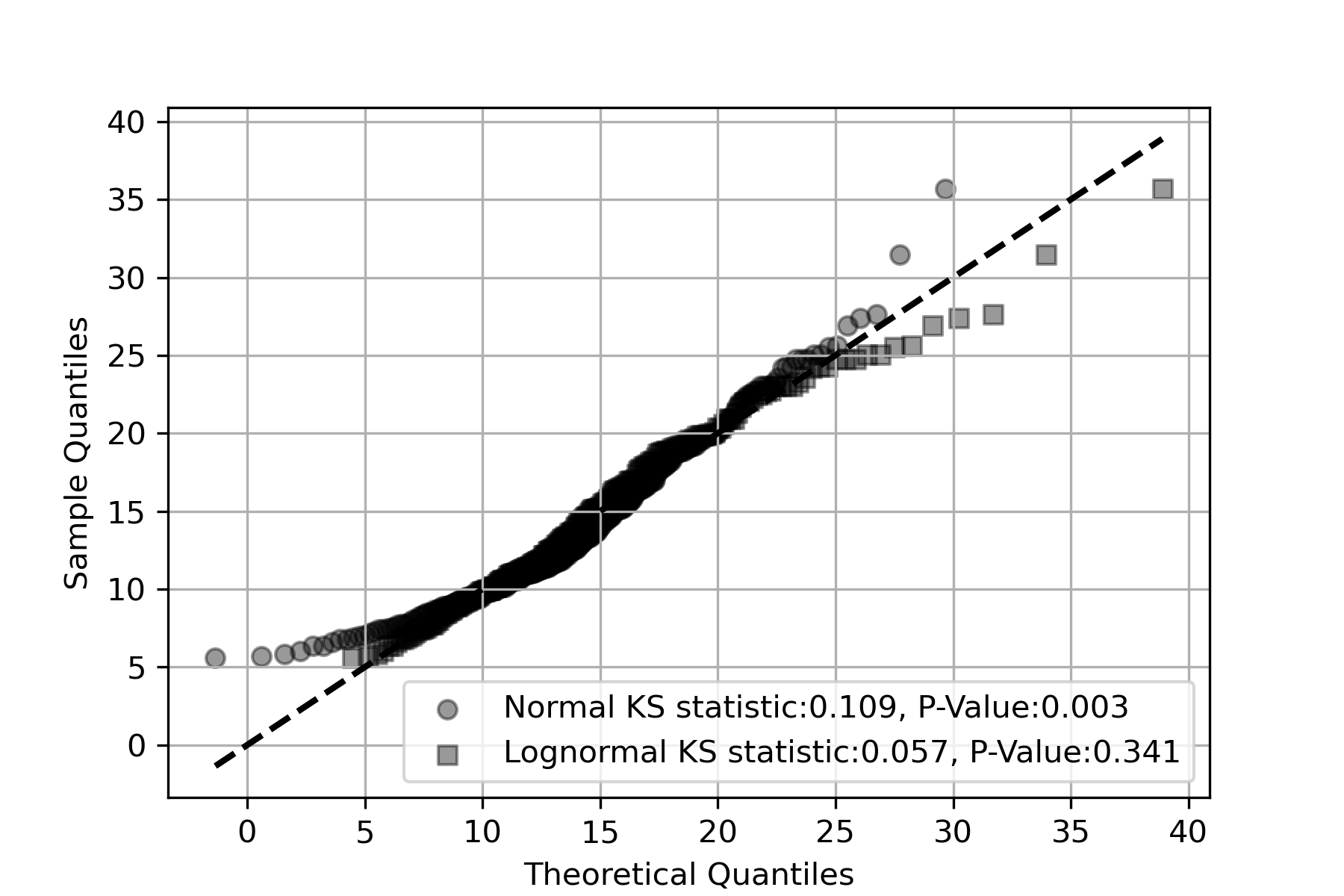}
        \caption{Link: $\texttt{Txy\_E\_004} \rightarrow \texttt{Txy\_E\_003}$. Data obtained from IFF ASDEX on 05/11/2023.}
        \label{fig: 0511-qq-392}
    \end{subfigure}
\caption{Data analysis and statistical tests on Link travel speed distributions. Based on K-S results, \textbf{Log-normal} distributions are better fits in these cases. }
\label{fig: stats-test-lognormal}
\end{figure}

\begin{figure}
    \centering
    \begin{subfigure}[t]{0.45\textwidth}
        \centering
        \includegraphics[width=\textwidth]
        {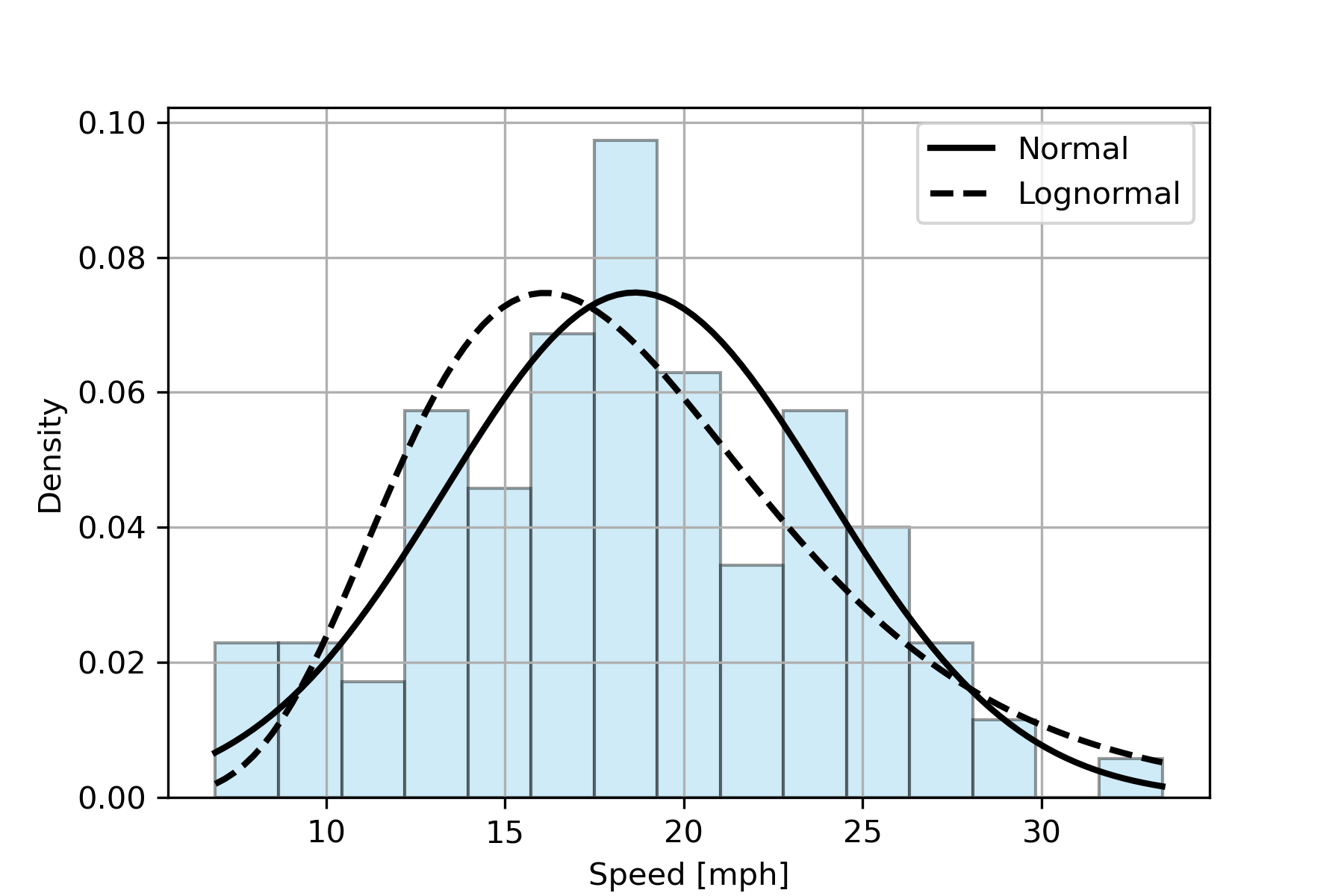}
        \caption{Link: $\texttt{Txy\_E\_003} \rightarrow \texttt{Txy\_E\_002}$. Data obtained from IFF ASDEX on 05/08/2023.}
        \label{fig: 0508-hist-391}
    \end{subfigure}
    \begin{subfigure}[t]{0.45\textwidth}
        \centering
        \includegraphics[width=\textwidth]
        {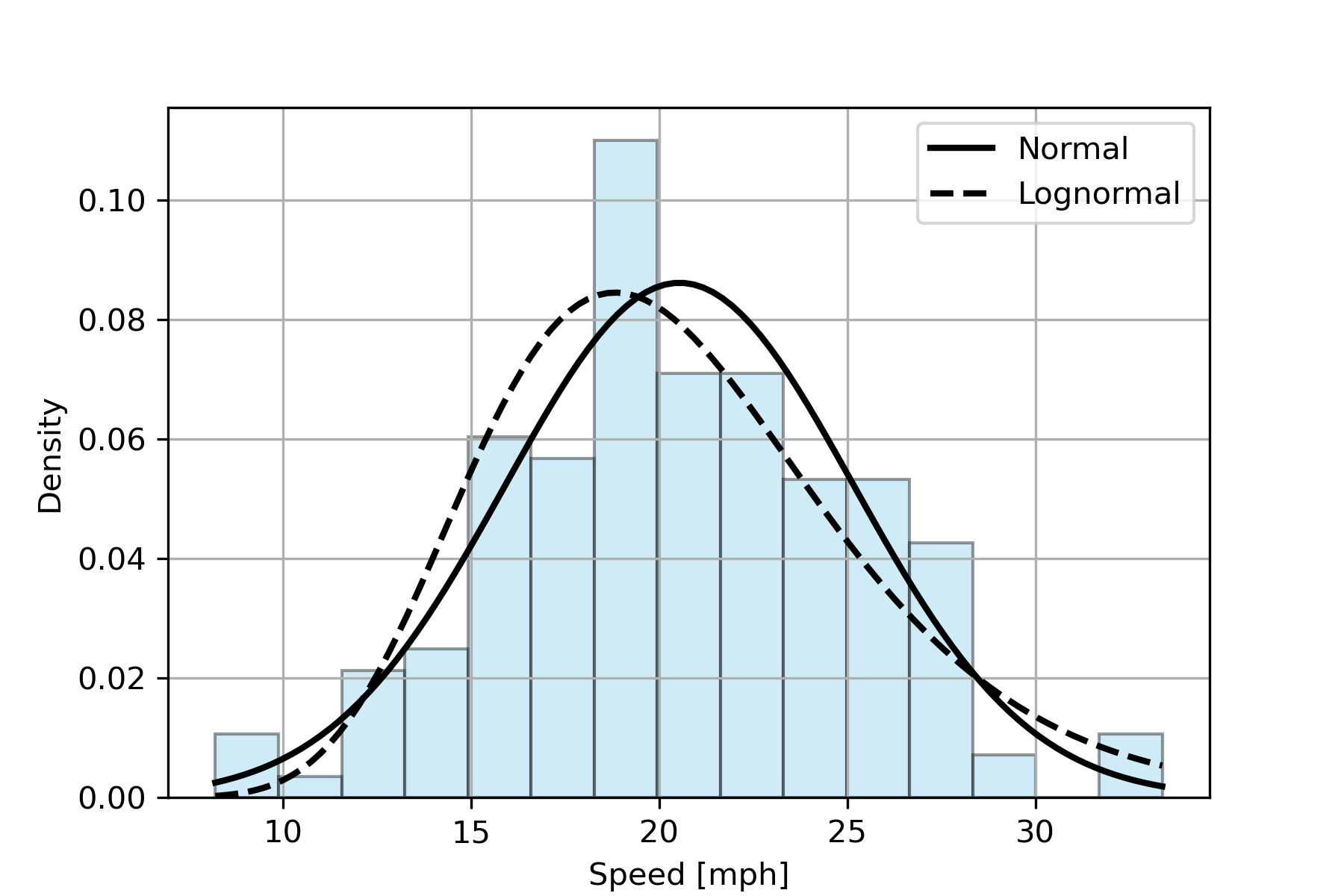}
        \caption{Link: $\texttt{Txy\_E\_004} \rightarrow \texttt{Txy\_E\_003}$. Data obtained from IFF ASDEX on 05/09/2023.}
        \label{fig: 0509-hist-392}
    \end{subfigure}
    \\~\\
    \begin{subfigure}[t]{0.45\textwidth}
        \centering
        \includegraphics[width=\textwidth]
        {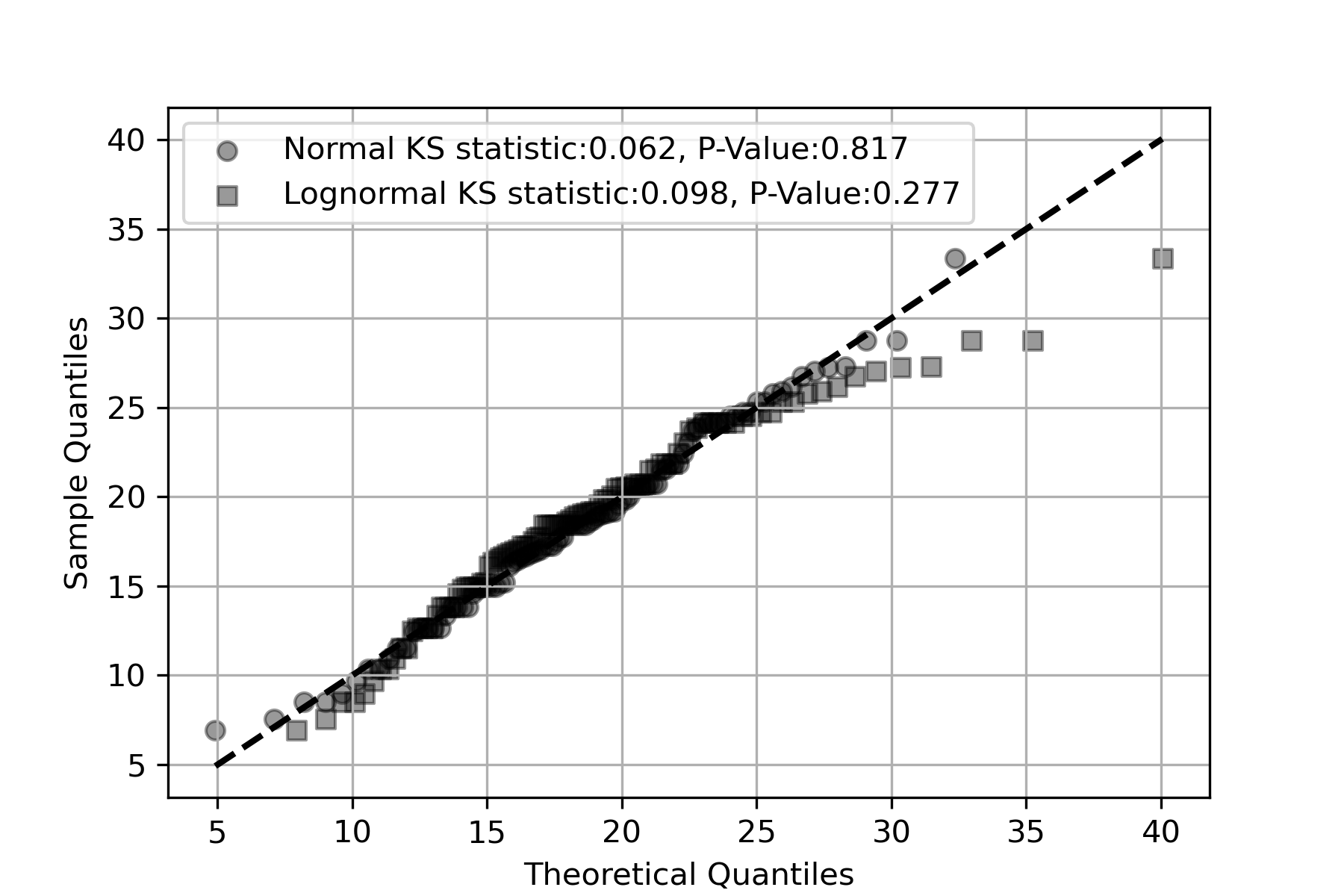}
        \caption{Link: $\texttt{Txy\_E\_003} \rightarrow \texttt{Txy\_E\_002}$. Data obtained from IFF ASDEX on 05/08/2023.}
        \label{fig: 0508-qq-391}
    \end{subfigure}
    \begin{subfigure}[t]{0.45\textwidth}
        \centering
        \includegraphics[width=\textwidth]
        {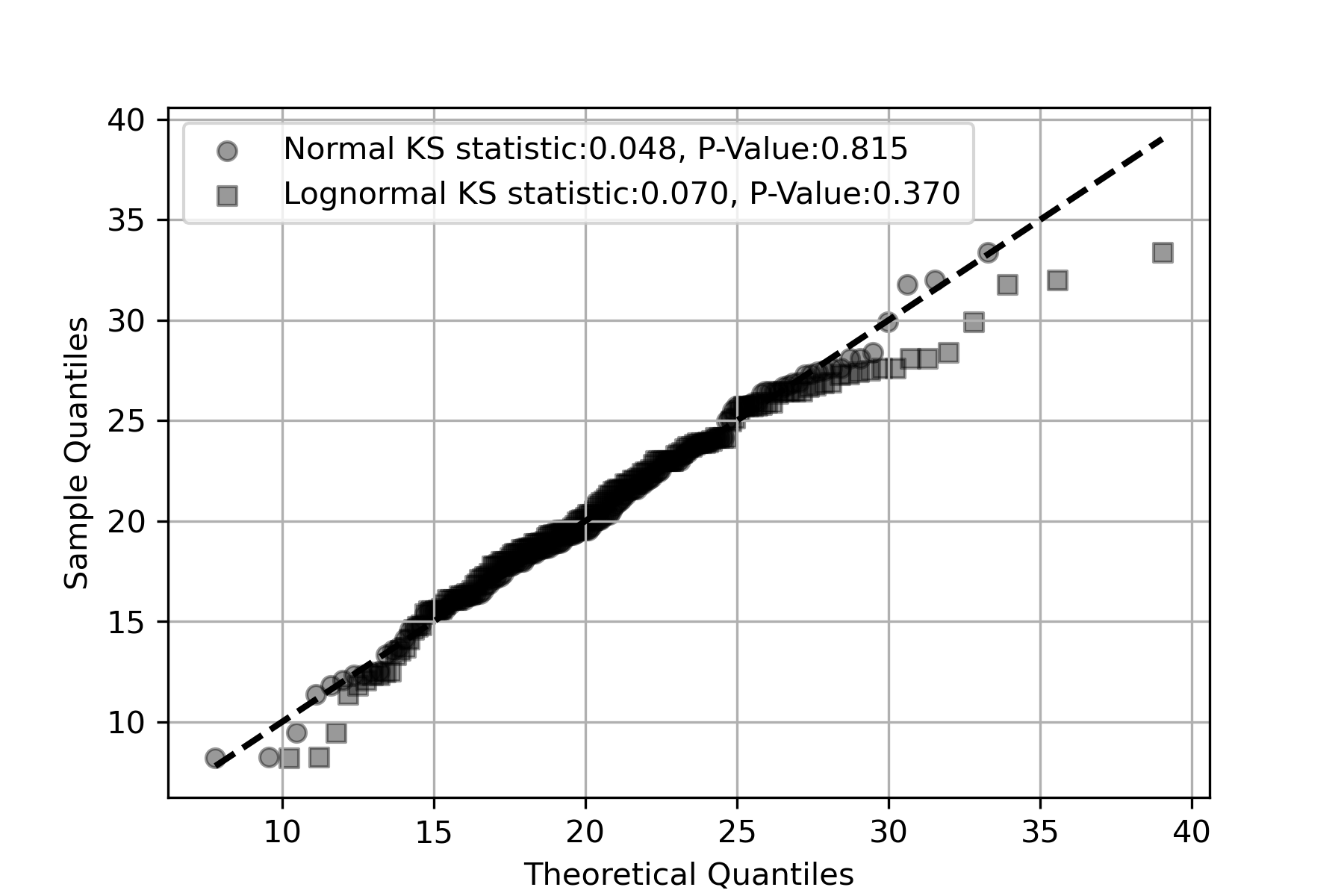}
        \caption{Link: $\texttt{Txy\_E\_004} \rightarrow \texttt{Txy\_E\_003}$. Data obtained from IFF ASDEX on 05/09/2023.}
        \label{fig: 0509-qq-392}
    \end{subfigure}
\caption{Data analysis and statistical tests on Link travel speed distributions. Based on K-S results, \textbf{Normal} distributions are better fits in these cases. }
\label{fig: stats-test-normal}
\end{figure}

\Cref{fig: stats-test-lognormal} and \Cref{fig: stats-test-normal} are the data analysis study results from the nodes of interest (i.e., overlapping nodes for both aircraft) from several days of ASDE-X. From \Cref{fig: stats-test-lognormal}(a) and \Cref{fig: stats-test-lognormal}(b), we show that Log-normal assumptions on link speed distributions are valid. \Cref{fig: stats-test-lognormal}(c) and \Cref{fig: stats-test-lognormal}(d) are the Quantile-Quantile (QQ) plot to compare the distribution of the data samples with assumed distributions. Moreover, the Kolmogorov-Smirnov (KS) test statistics are given, and the null hypothesis is that the sample follows a desired distribution, where a p-value greater or equal to $.05$ are considered significant since there is no significant evidence against it. The log-normal assumption is valid and convenient because it ensures positive speeds and provides tractable transformations for travel time. However, \Cref{fig: stats-test-normal}(a) and \Cref{fig: stats-test-normal}(b) show that at some links, the link speed distributions follow both normal and log-normal assumptions (i.e., p-value greater than 0.05), while normal distributions are better fits. To properly adapt to the risk formulation in \Cref{subsubsec: travel-time-model}, the normal distribution assumptions of link speed are better used as truncated normal distributions and modify the spatial integral accordingly, or simply use log-normal across all links.

\begin{table}
\centering
\caption{ANOVA and Kruskal-Wallis test on weight class impact to taxi speed. }
\label{tab: weight-tests}
\begin{tabular}{c|cc|cc}
\hline
\multirow{2}{*}{} & \multicolumn{2}{c|}{$\mathsf{Txy\_E\_004} \rightarrow \mathsf{Txy\_E\_003}$}          & \multicolumn{2}{c}{$\mathsf{Txy\_E\_003} \rightarrow \mathsf{Txy\_E\_002}$}             \\ \cline{2-5}
                  & \multicolumn{1}{c|}{F}     & p-value & \multicolumn{1}{c|}{$\chi^2$} & p-value \\ \hline
Anova             & \multicolumn{1}{c|}{4.406} & 0.004   & \multicolumn{1}{c|}{11.372}   & 0.010   \\ \hline
Kruskal-Wallis    & \multicolumn{1}{c|}{2.528} & 0.056   & \multicolumn{1}{c|}{3.803}    & 0.284   \\ \hline
\end{tabular}
\end{table}

Furthermore, a study on the impact of weight class on taxi speed is briefly conducted, to understand the impact of aircraft weight class to taxi speed. The weight class is derived based on the FAA Order JO 7360.1E \citep{faa_order_jo_7360.1e}. As shown in \Cref{tab: weight-tests}, Analysis of Variance (ANOVA) \citep{fisher1970statistical} and the Kruskal-Wallis \citep{kruskal1952use} test are considered to study the impact of aircraft weight class both parametrically and non-parametrically. ANOVA is a statistical test used to determine whether there are significant differences between the means of two or more independent groups. It assumes that the data follows a normal distribution and that variances are equal across groups. The Kruskal-Wallis test is a non-parametric alternative to ANOVA, used when data does not meet normality or equal variance assumptions. Instead of comparing means, it ranks the data and compares the distributions across groups. The results for Link $\texttt{Txy\_E\_004} \rightarrow \texttt{Txy\_E\_003}$ indicate that weight class influences taxi speed, while the test results from link $\texttt{Txy\_E\_003} \rightarrow \texttt{Txy\_E\_002}$ is saying no strong evidence that weight class affects taxi speed. This suggests that the effect of weight class on taxi speed might be location-dependent or influenced by other factors like taxiway geometry, congestion, or operational procedures. To confirm a general relationship, further analysis (e.g., post-hoc tests for specific weight class differences, additional links) is needed and is listed as a major future study.

\begin{table}
\centering
\caption{Key ATC communication transcript extracted with the knowledge-enhanced hybrid learning model for the 2024 KATL taxiway collision case study.}
\label{tab: case-2}
\resizebox{0.9\textwidth}{!}{%
\begin{tabular}{c|c|c|c|c}
\hline
\textbf{TIME} & \textbf{CALLSIGN} & \textbf{AC\_STATE}    & \textbf{DEST\_RUNWAY} & \textbf{DESTINATION} \\ \hline
0:08          & Delta 295         & taxi                  & 08R                   & Romeo                \\ \hline
0:14          & Delta 295         & taxi                  & 08R                   & Rwy\_02\_001         \\ \hline
0:20          & Delta 295         & Taxi                  & 08R                   & foxtrot              \\ \hline
0:33          & Delta 295         & continue,hold         & 08R                   & ramp 5               \\ \hline
0:44          & Delta 295         & give way,inbound,join & 08R                   & Echo(Txy\_E\_002)    \\ \hline
0:50          & Delta 295         & give way              & 08R                   &                      \\ \hline
0:57          & Endeavor 5526     & taxi                  & 08R                   & Rwy\_02\_001         \\ \hline
1:27          & Delta 295         & go                    & 08R                   &                      \\ \hline
1:35          & Delta 295         & continue,hold         & 08R                   &                      \\ \hline
1:45          & Delta 295         & holding               & 08R                   & Victor(Txy\_V\_003)  \\ \hline
1:54          & Endeavor 5526     & line up,wait          & 08R                   &                      \\ \hline
2:10          & Endeavor 5526     & collision             &                       &                      \\ \hline
2:10          & Delta 295         & collision             &                       &                      \\ \hline
\end{tabular}
}
\end{table}

\begin{figure}
\centering
    \begin{subfigure}[t]{0.75\textwidth}
        \centering
        \includegraphics[width=\textwidth]{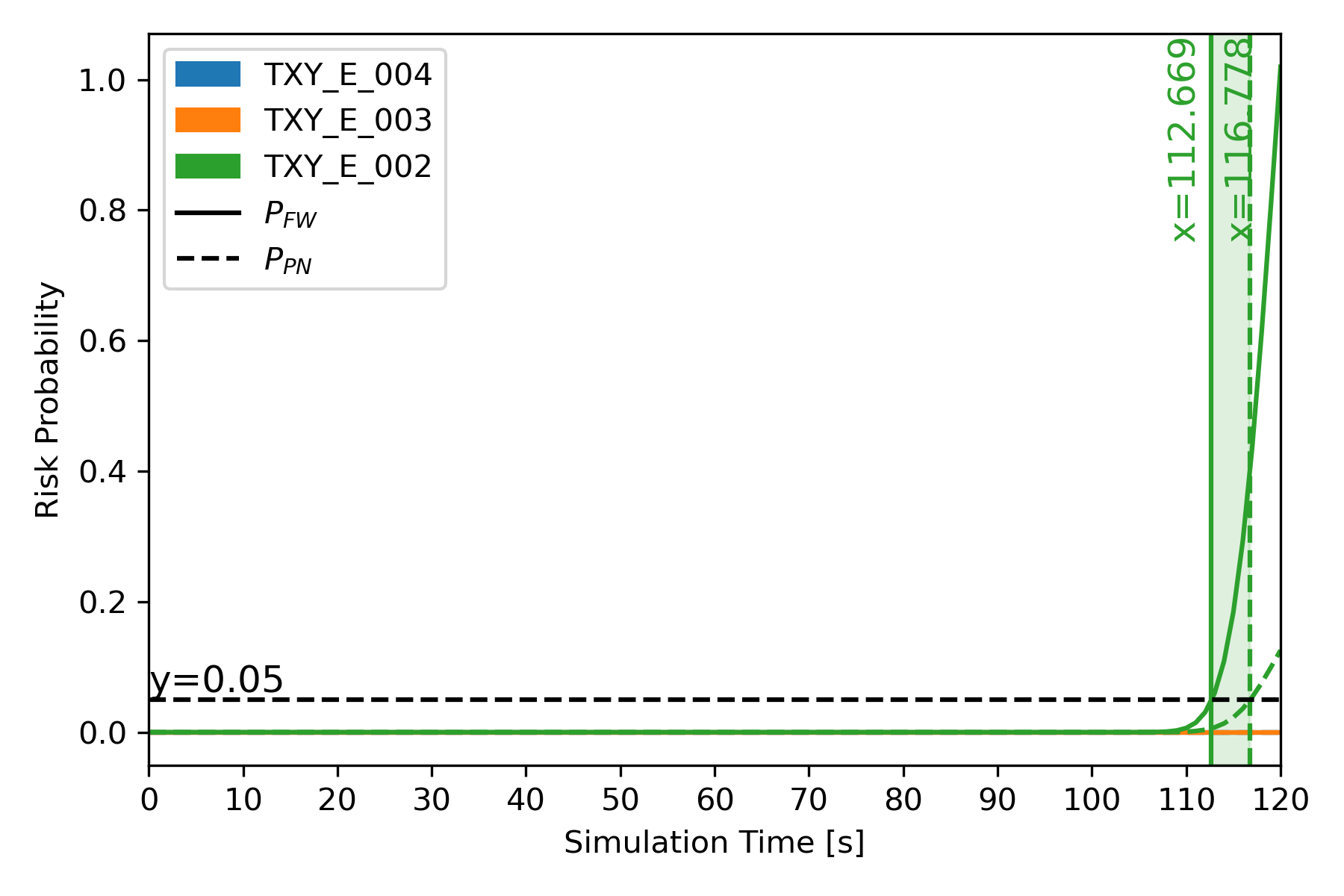}
        \caption{KATL Simulation with safety threshold at 0.05.}
        \label{fig: case2-0.05}
    \end{subfigure}
    \begin{subfigure}[t]{0.75\textwidth}
        \centering
        \includegraphics[width=\textwidth]{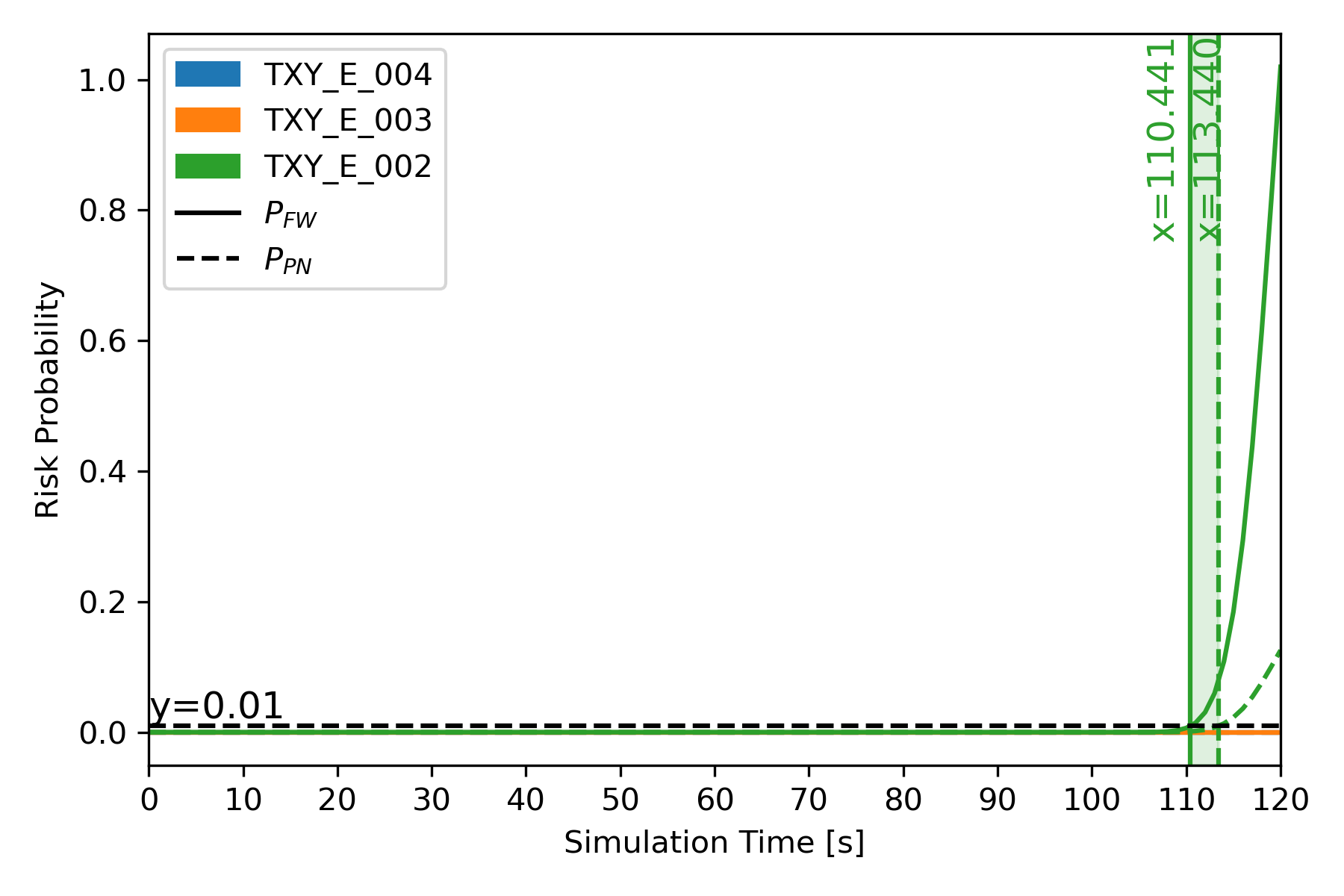}
        \caption{KATL Simulation with safety threshold at 0.01.}
        \label{fig: case2-0.01}
    \end{subfigure}
\caption{The real-time risk probability calculated based on the KATL taxiway collision simulation. FW and PN risk profiles are shown for each overlapping node.}
\label{fig: case2-risk}
\end{figure}

Once the real-world travel time parameters are obtained. We conduct the simulation and risk calculation as in \Cref{subsec: case-1}. Similarly, \Cref{tab: case-2} shows the NER model output as the guidance for taxiplan generation of the second simulation, where \Cref{fig: case-2-pregress} lists the progression of the taxiway collision case study. The associated risk score along each node and links are visualized in \Cref{fig: case2-risk}. It is worth pointing out that only the nodes that are the overlaps of the two generated taxiplans are considered as potential collision spots, with a risk score. 

\begin{figure}[H]
    \centering
    \begin{subfigure}[t]{0.45\textwidth}
        \centering
        \includegraphics[width=\textwidth]
        {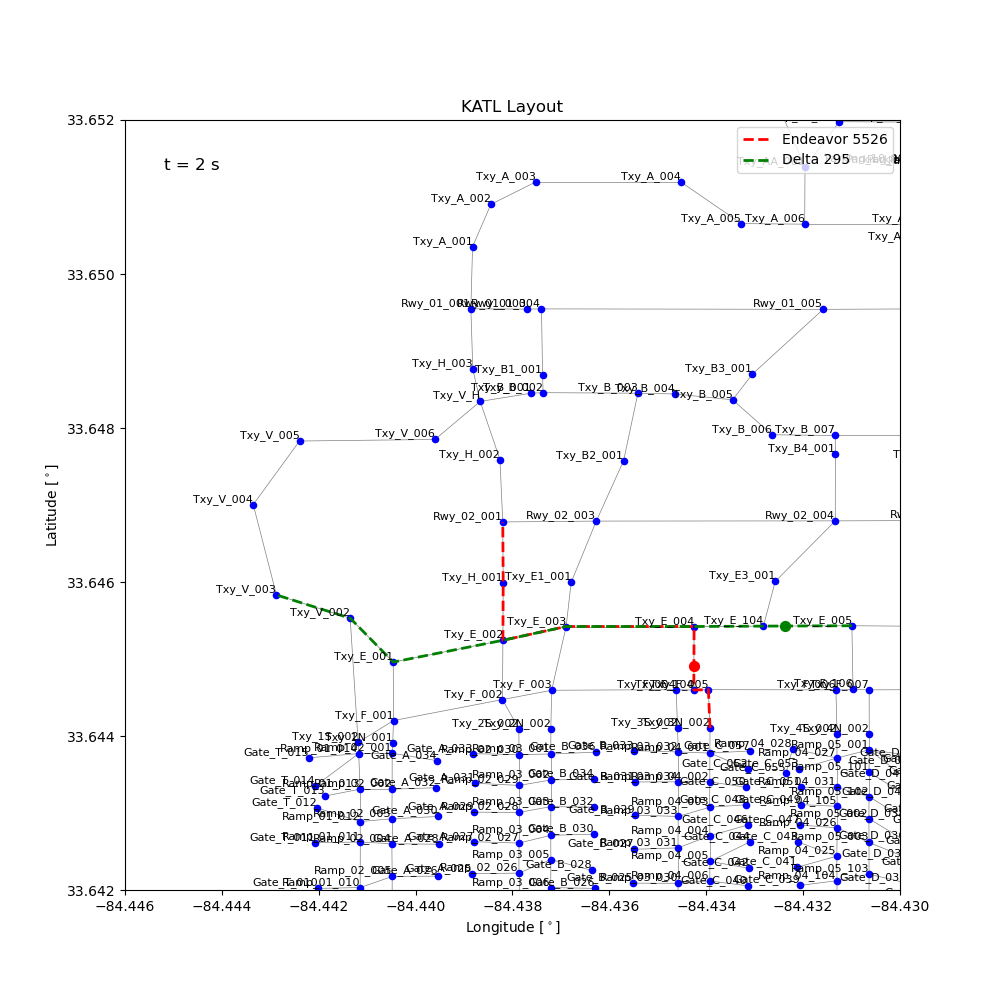}
    \end{subfigure}
    \begin{subfigure}[t]{0.45\textwidth}
        \centering
        \includegraphics[width=\textwidth]
        {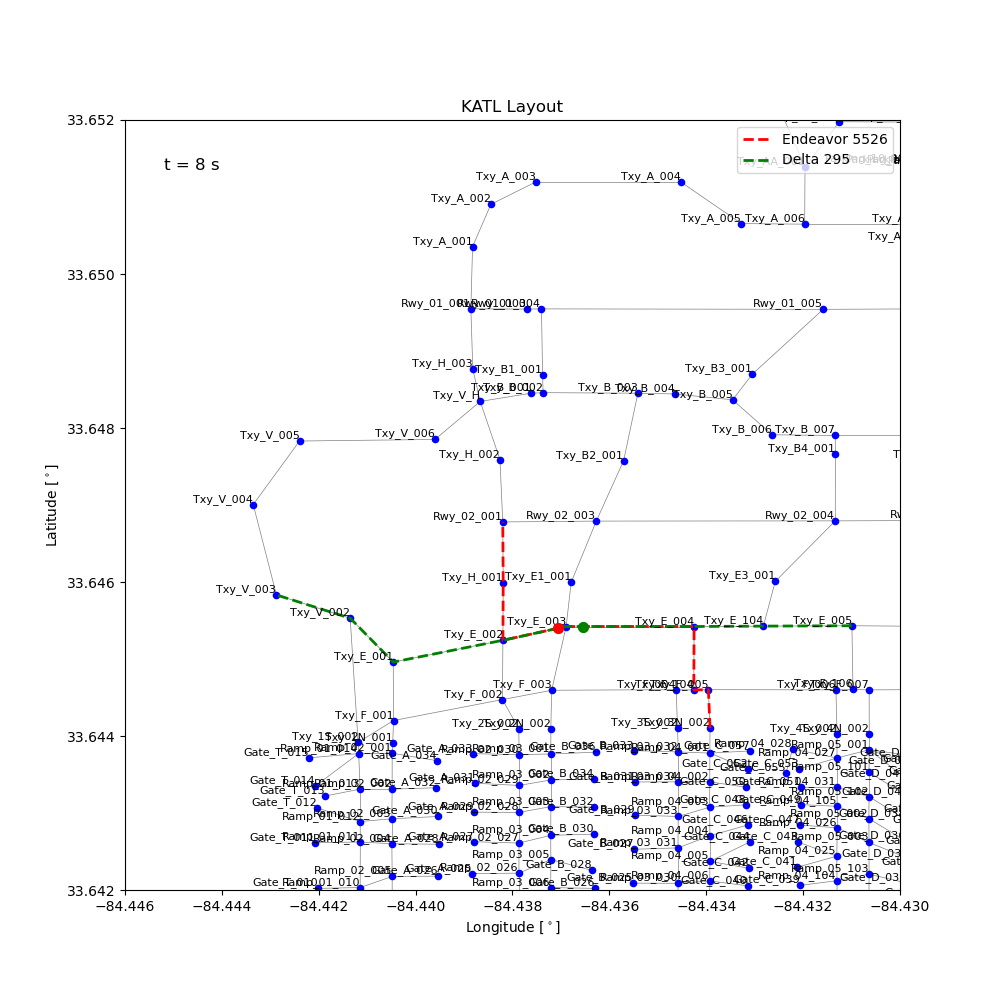}
    \end{subfigure}
    \\~\\
    \begin{subfigure}[t]{0.45\textwidth}
        \centering
        \includegraphics[width=\textwidth]
        {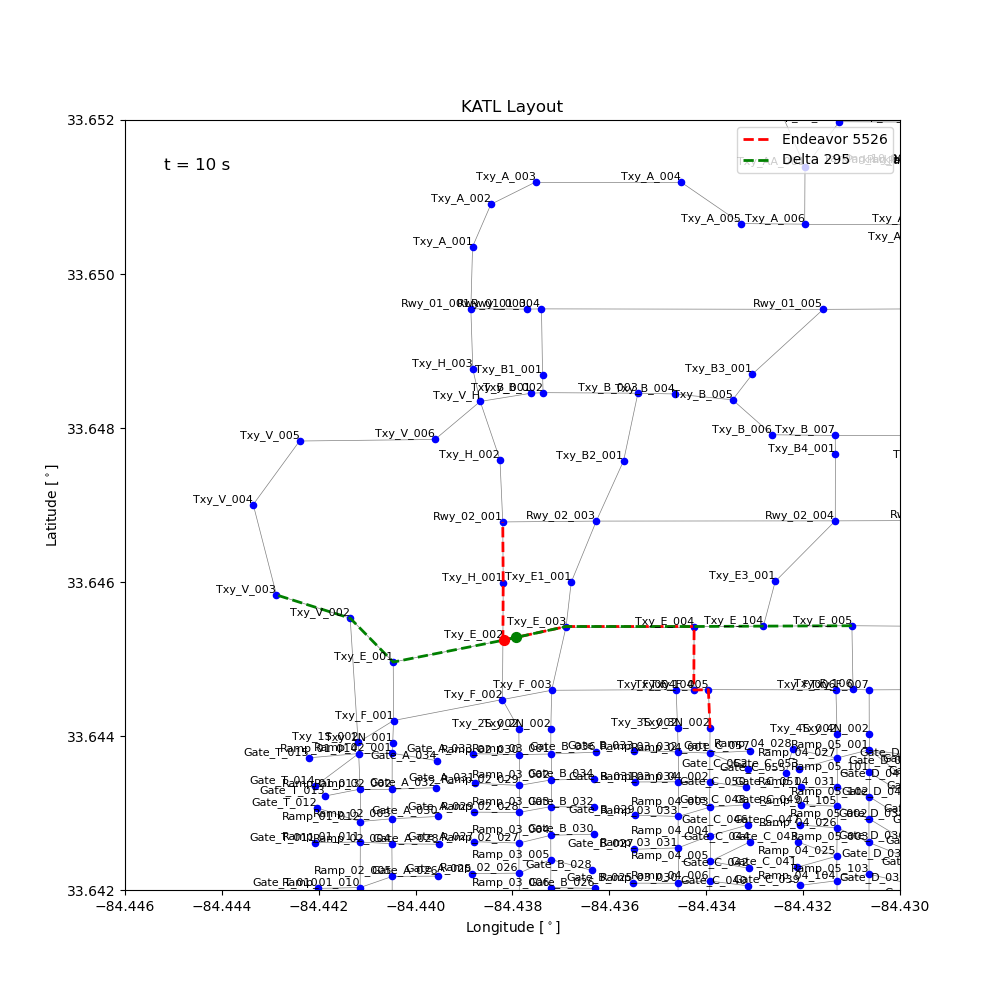}
    \end{subfigure}
    \begin{subfigure}[t]{0.45\textwidth}
        \centering
        \includegraphics[width=\textwidth]
        {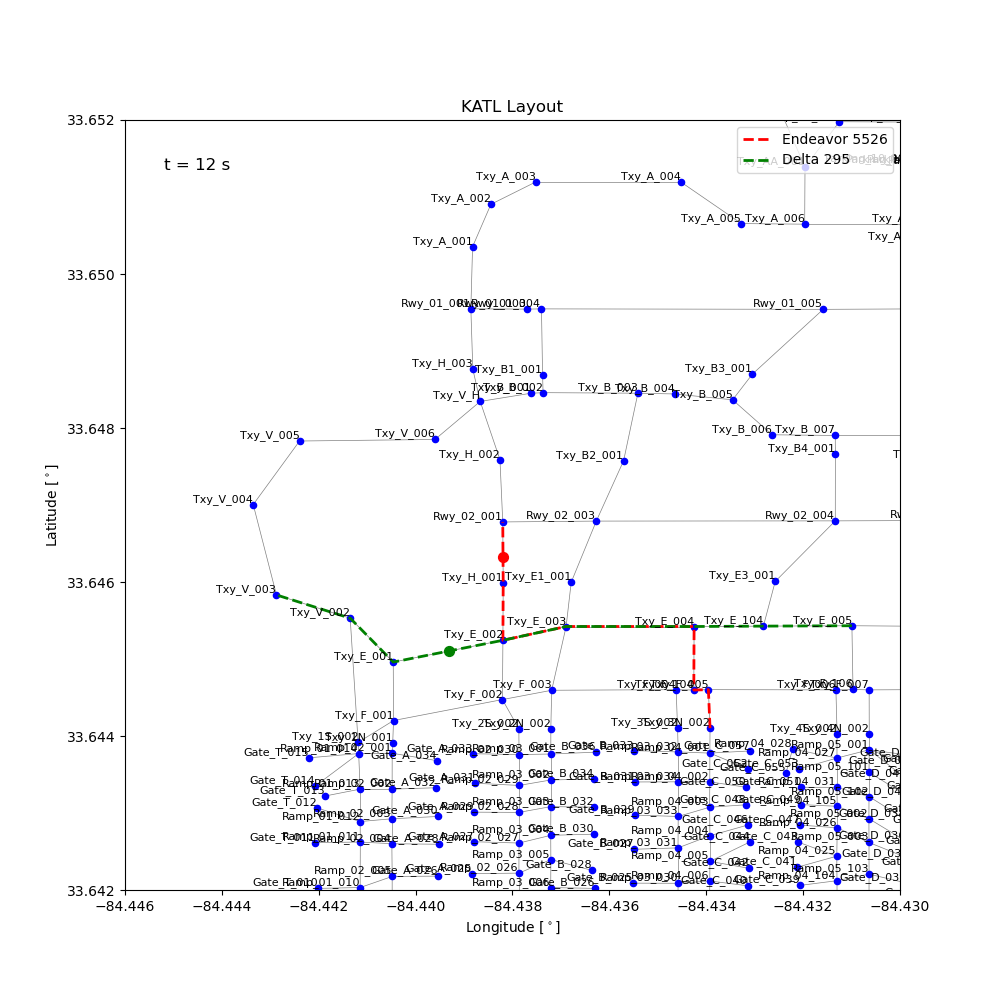}
    \end{subfigure}
\caption{Node-link simulation of the accident happened at KATL in September 2024.}
\label{fig: case-1-pregress}
\end{figure}

Again, we report the real-time risk curves for the overlapping nodes \texttt{Txy\_E\_004}, \texttt{Txy\_E\_003}, and \texttt{Txy\_E\_002} in \Cref{fig: case2-risk}. The simulation runs for around 120 seconds. Similar to case 1, only the terminal conflict node (\texttt{Txy\_E\_002}) exhibits significant risk, while upstream nodes remain negligible. For the 0.05 threshold, FW crosses at 113 s while PN crosses at 117 s. At the lower threshold of 0.01, FW triggers a warning earlier at 110 s, and PN follows at 113 s. 

This result highlights how risk accumulates only when both aircraft are on converging trajectories toward the same conflict node. The model’s selectivity ensures that nodes without simultaneous presence remain at zero risk, avoiding false positives. The FW lead time advantage is once again evident, providing a 3–4 second earlier warning compared to PN. Although the absolute time window is shorter than in the previous case, the relative lead time difference remains operationally significant, offering a larger buffer for mitigation actions.

\subsection{Case III: Tenerife Runway Collision \label{subsec: case-3}}

\begin{figure}
\centering
\includegraphics[width=0.85\textwidth]
{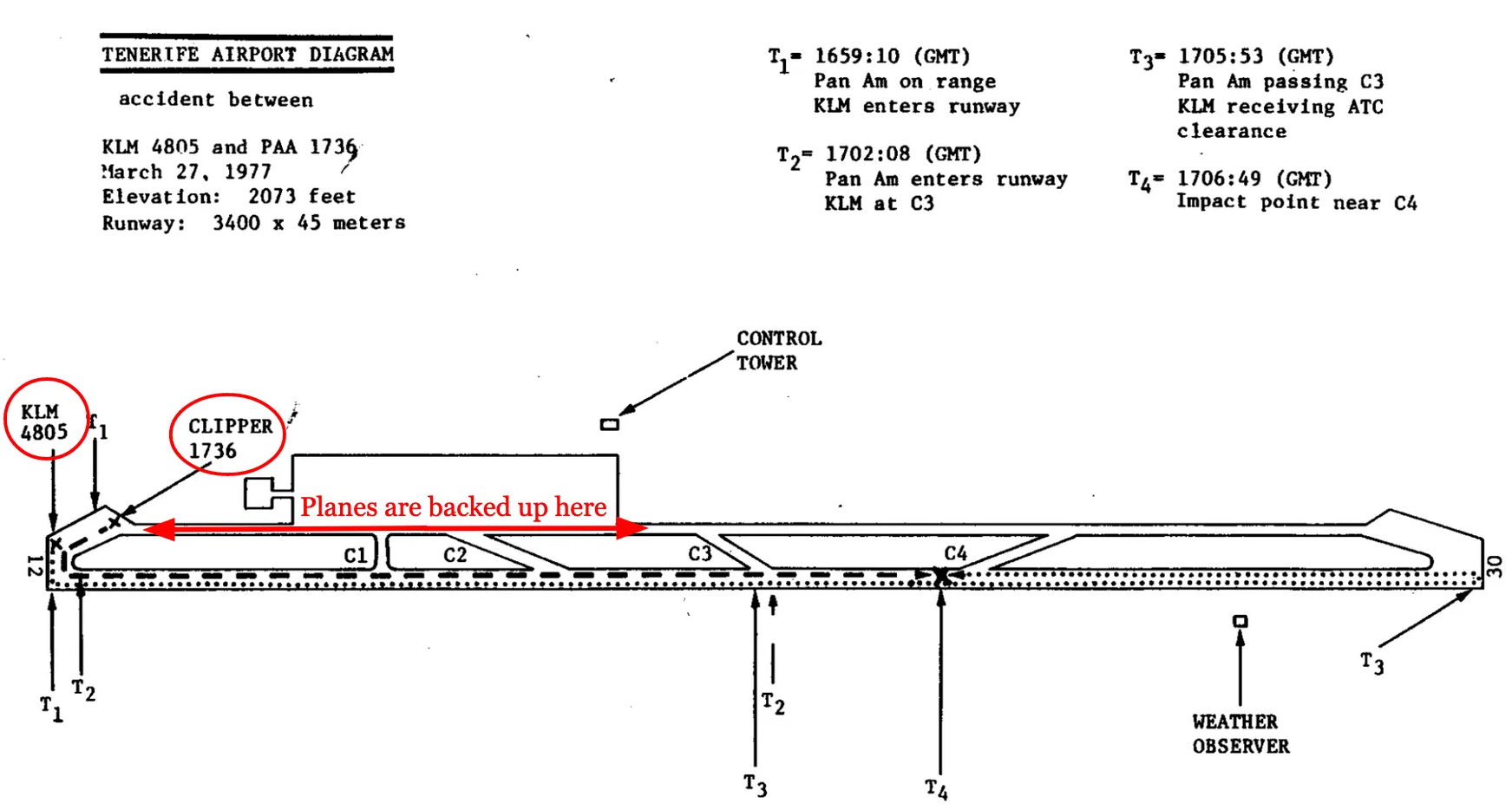}
\caption{The debriefing of the Los Rodeos airport disaster \citep{weick1990vulnerable}, which is one of deadliest accidents in aviation history. This accident happened between a take-off aircraft and a taxiing aircraft on the runway, caused by a series of incidents and failures. This simulation complements the previous two case studies with a head-to-head collision scenario.}
\label{fig: case3-overview}
\end{figure}

The Los Rodeos airport disaster at Tenerife happened on March 27, 1977, and it is one of the deadliest accidents in aviation history, where KLM Flight 4805 and Pan Am Flight 1736 collided on the runway and killed 583 people \citep{weick1990vulnerable, ziomek2018collision, casassa2023collision}. \Cref{fig: case3-overview} provides a debriefing of the accident, while \citep{Krock2006FinalEightMinutes} provides the open-source transcribed communication recordings. 

The accident stemmed from a series of external disruptions, human errors, and communication failures that converged under poor weather conditions, but the chain of miscommunication proved decisive. While KLM awaited clearance, the KLM pilot advanced throttles before explicit takeoff permission was granted \citep{roitsch1978human}. KLM First Officer’s radio call, \textit{we are now at takeoff}, overlapped with Pan Am’s transmission that they were still on the runway, producing a heterodyne effect that blocked critical ATC instructions \citep{casassa2023collision}. KLM only heard \textit{OK} from the tower, misinterpreting it as clearance. Meanwhile, Pan Am’s were unsure about where taxiway exit C3 is as instructed by the tower, and they continued to C4. Although Pan Am called that they were still on the runway, it was never received by KLM due to radio interference. This disaster reshaped civil aviation operations, as English proficiency requirements for ATC were reinforced and standard phraseology was adopted so that \textit{takeoff} is used only when clearance is explicitly granted, with \textit{departure} substituted in all other contexts \citep{casassa2023collision}. 

\begin{table}[H]
\centering
\caption{Key ATC transmissions extracted by the knowledge-enhanced hybrid learning model for the Los Rodeos (Tenerife) disaster with open-sourced communication transcript \citep{Krock2006FinalEightMinutes}}
\label{tab: case-3}
\resizebox{0.9\textwidth}{!}{%
\begin{tabular}{c|c|c|c|c}
\hline
\textbf{TIME} & \textbf{CALLSIGN} & \textbf{AC\_STATE}    & \textbf{DEST\_RUNWAY} & \textbf{DESTINATION} \\ \hline
1658:25.7     & KLM 4805          & backtrack,takeoff     & 30                    & Rwy\_12\_006           \\ \hline
1658:30.4     & KLM 4805          & taxi                  & 30                    & Rwy\_12\_006           \\ \hline
1658:47.4     & KLM 4805          & entering              & 30                    & Rwy\_12\_006           \\ \hline
1658:55.3     & KLM 4805          & taxi                  & 30                    & runway               \\ \hline
1659:28.4     & KLM 4805          & approach              & 30                    & Charlie 1(Rwy\_12\_001)\textbf{?} \\ \hline
1701:57.0     & Clipper 1736      &                       &                       &                \\ \hline
1702:03.6     & \textbf{Clipper 1736}      & taxi         & 30                    & runway\textbf{?}   \\ \hline
1702:16.4     &                   &                       &                       & third (Rwy\_12\_003) \\ \hline
1702:55.6     & KLM 4805          & pass                  & 30                    & Charlie 4(Rwy\_12\_004) \\ \hline
1705:44.6     & KLM 4805          & \textbf{ready for} takeoff & 30               & \\ \hline 
1706:09.6     & \textbf{KLM 4805} & cleared,right turn    & 30                    &   \\ \hline
1706:12.25    & \textbf{KLM 4805} & go                    & 30                    & \\ \hline
1706:20.08    & Clipper 1736      & taxi                  & 30                    & runway \\ \hline
1706:50.00    & KLM 4805          & collision             &                       & \\ \hline
1706:50.00    & Clipper 1736      & collision             &                       & \\ \hline
\end{tabular}
}
\end{table}

\Cref{tab: case-3} shows the NER screening outputs. Due to the frequent use of non-standard ATC phraseology and frequent readbacks, the quality of the meta table is worse than the previous two case studies. The transcript reflects non-standard phraseology and period-specific practices (not uniformly adopted then), such as overlapping transmissions and ambiguous acknowledgments, which blur speaker identity, callsign boundaries, and clearance semantics. Thus, we perform several post-processing steps to further refine the output. In this table, assumed entities are filled and shown in bold. For instance, rows with a question mark indicate the entities are extracted from a question from the pilot to the controller.

\begin{figure}
    \centering
    \begin{subfigure}[t]{0.45\textwidth}
        \centering
        \includegraphics[width=\textwidth]
        {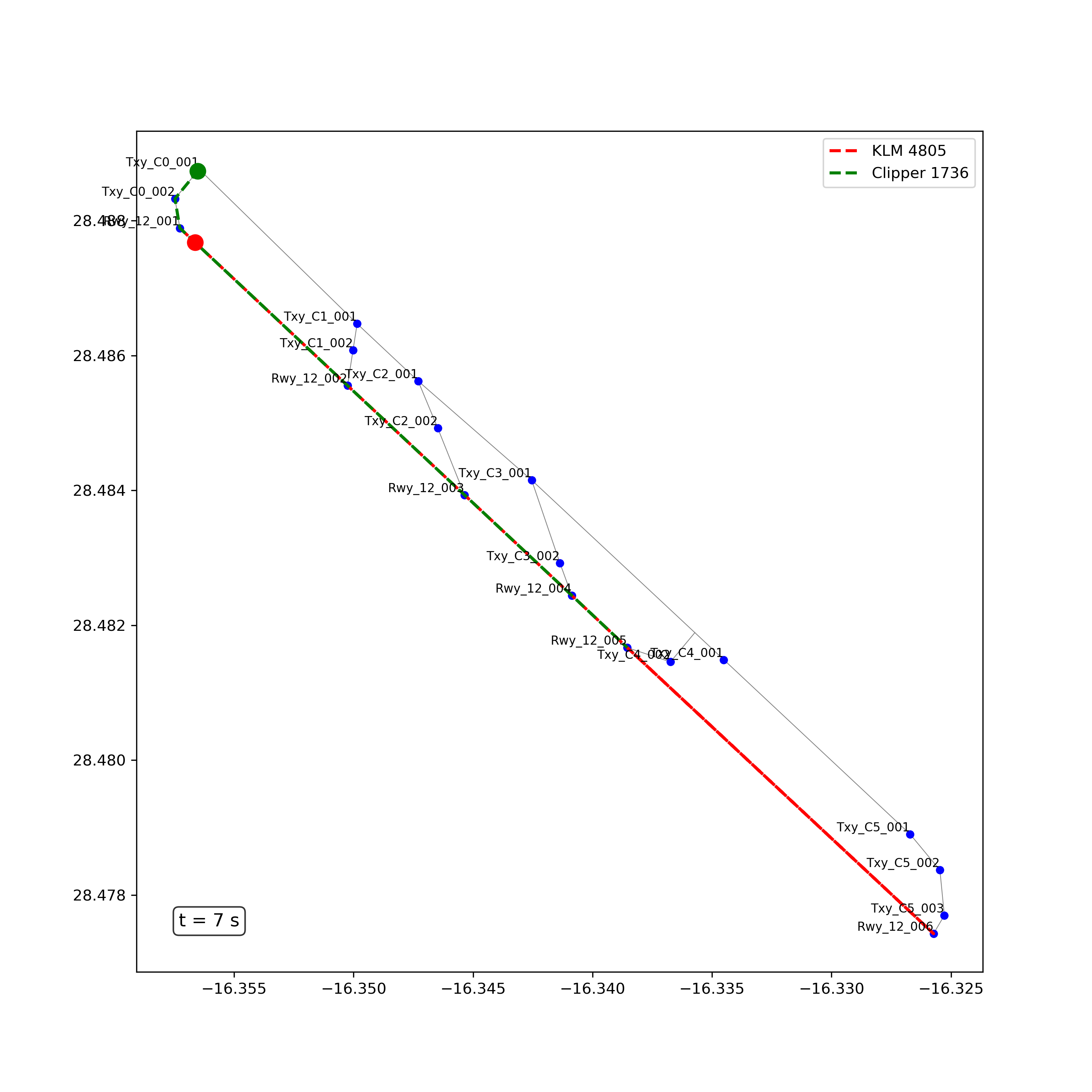}
    \end{subfigure}
    \begin{subfigure}[t]{0.45\textwidth}
        \centering
        \includegraphics[width=\textwidth]
        {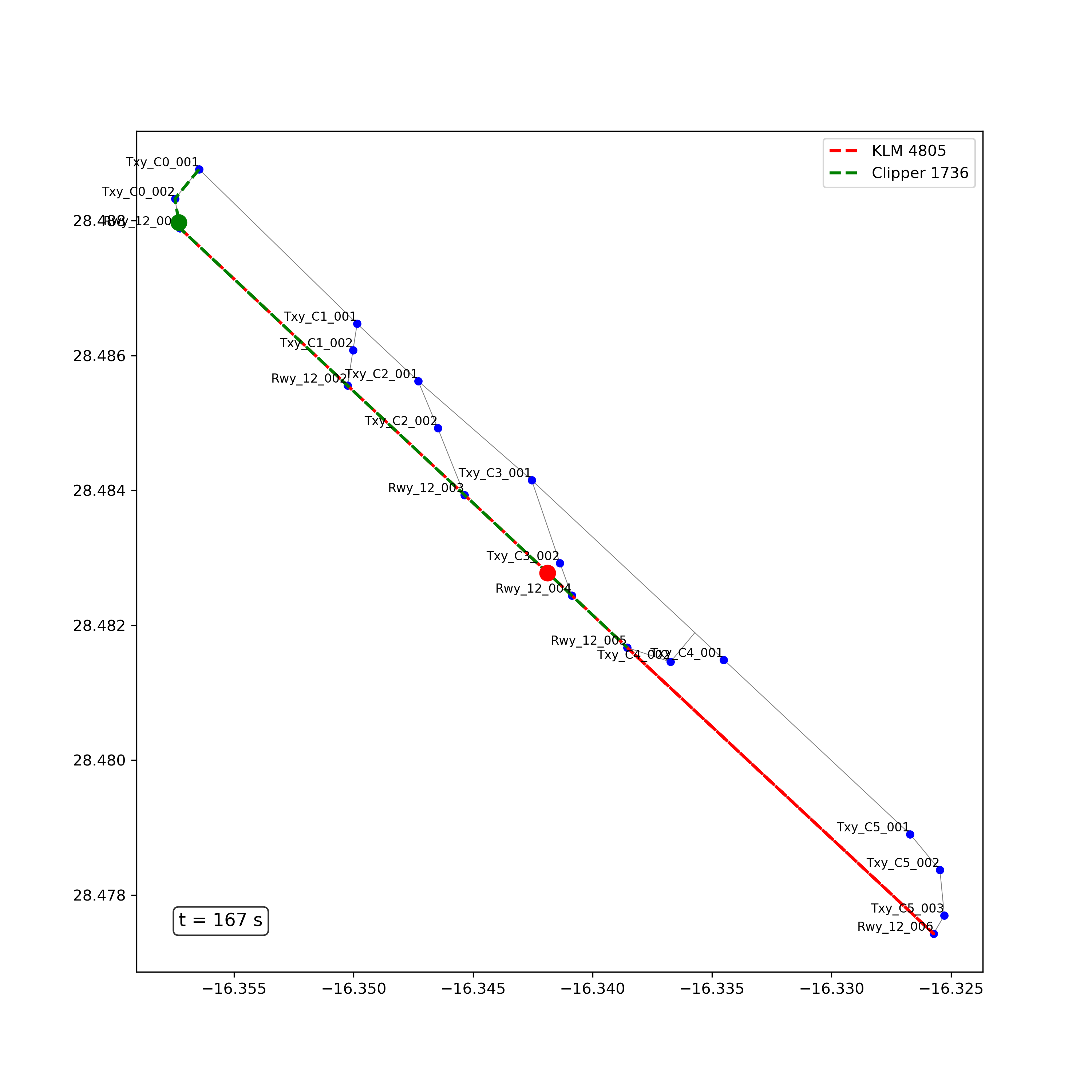}
    \end{subfigure}
    \\~\\
    \begin{subfigure}[t]{0.45\textwidth}
        \centering
        \includegraphics[width=\textwidth]
        {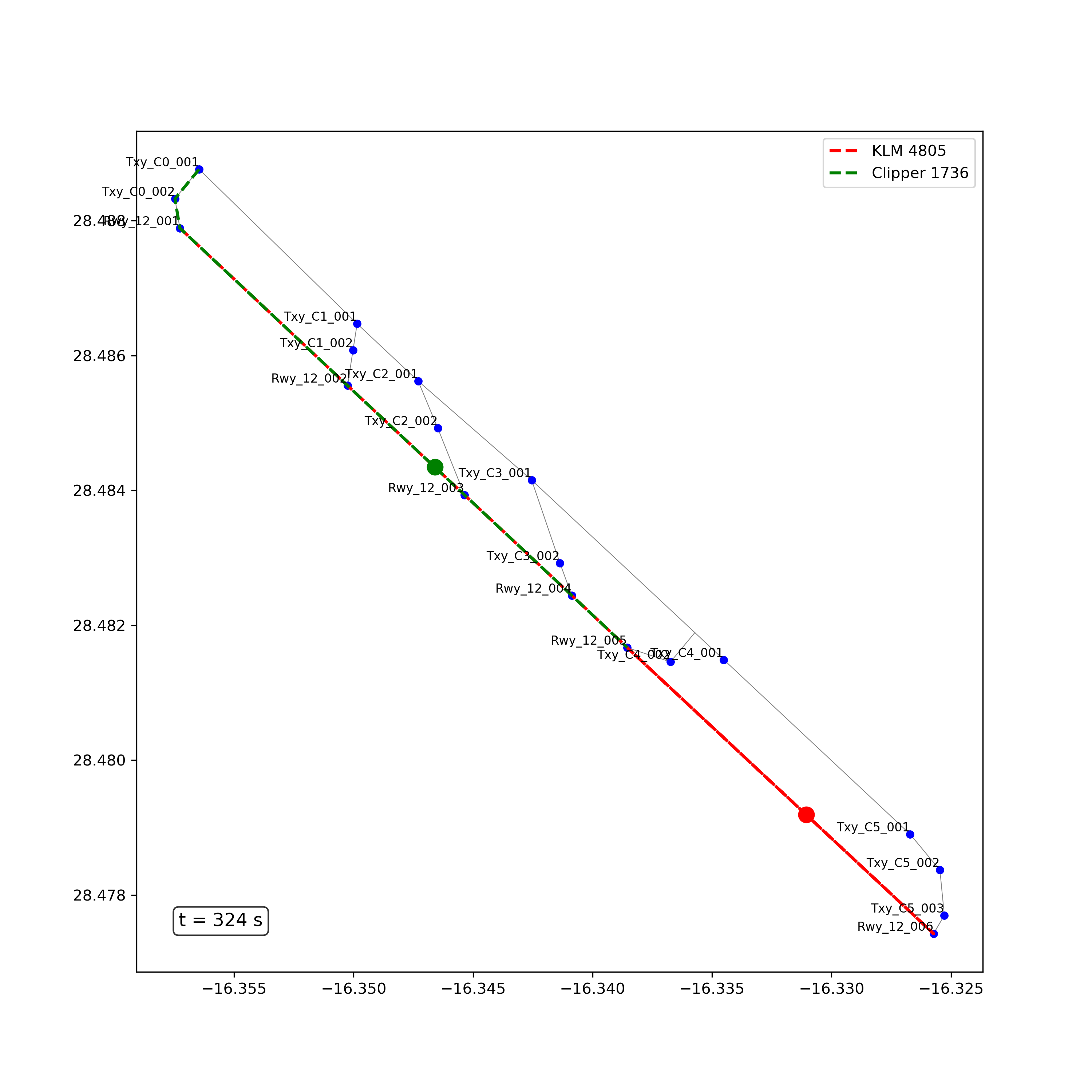}
    \end{subfigure}
    \begin{subfigure}[t]{0.45\textwidth}
        \centering
        \includegraphics[width=\textwidth]
        {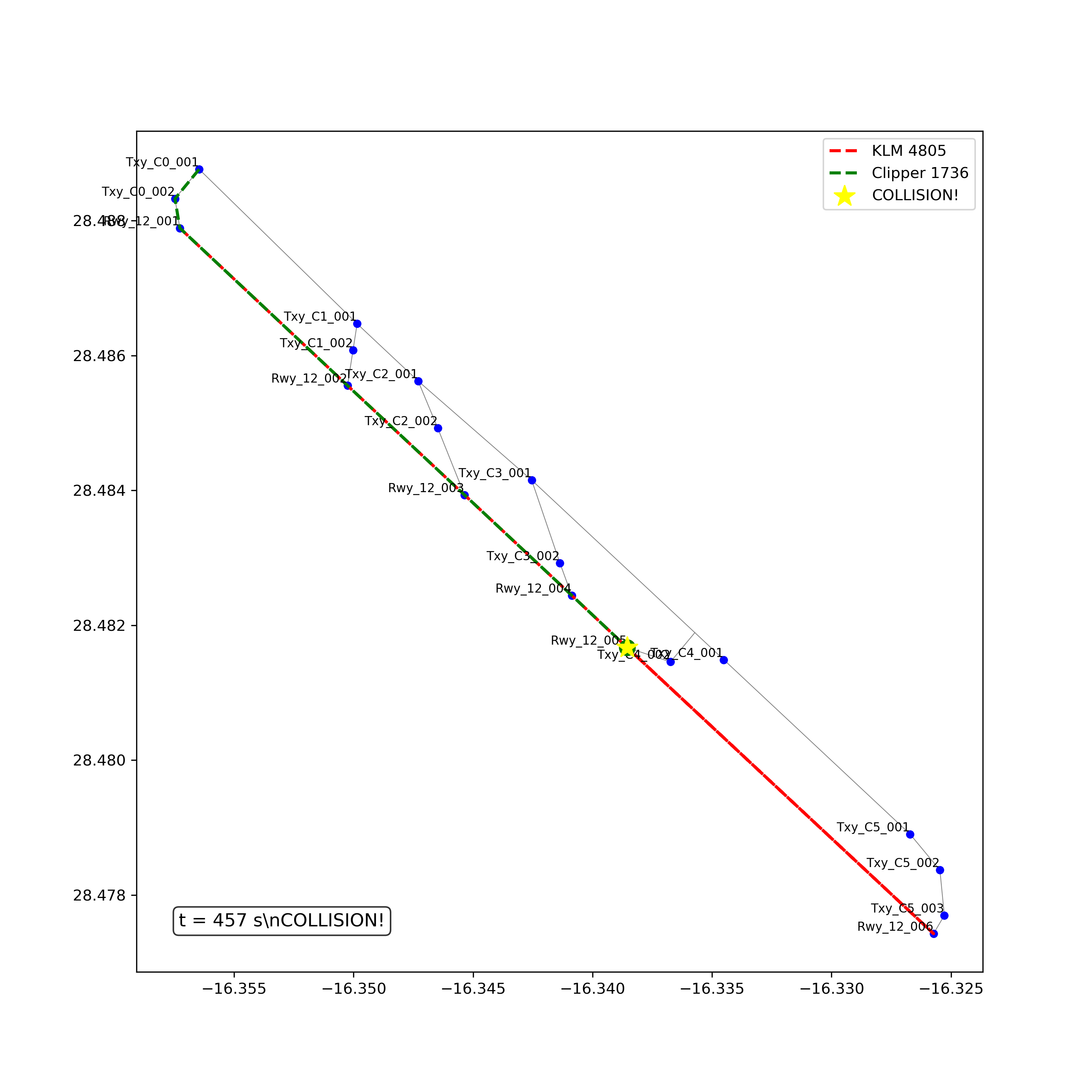}
    \end{subfigure}
\caption{Node-link simulation of the accident happened at the Los Rodeos Airport on March 27, 1977.}
\label{fig: case-3-progress}
\end{figure}

Similar to the previous two case studies, we construct a simulation of the Tenerife accident using the node-link graph representation shown in \Cref{fig: case-3-progress}. For this case, we develop a custom node-link graph of Los Rodeos Airport (now Tenerife North Airport), since it is not available in NASA FACET. The graph is designed to replicate the 1977 airport layout and naming conventions to ensure consistency with historical conditions (i.e., the taxiway name changes from Charlie to Echo now). 

\begin{figure}
\centering
    \begin{subfigure}[t]{0.75\textwidth}
        \centering
        \includegraphics[width=\textwidth]{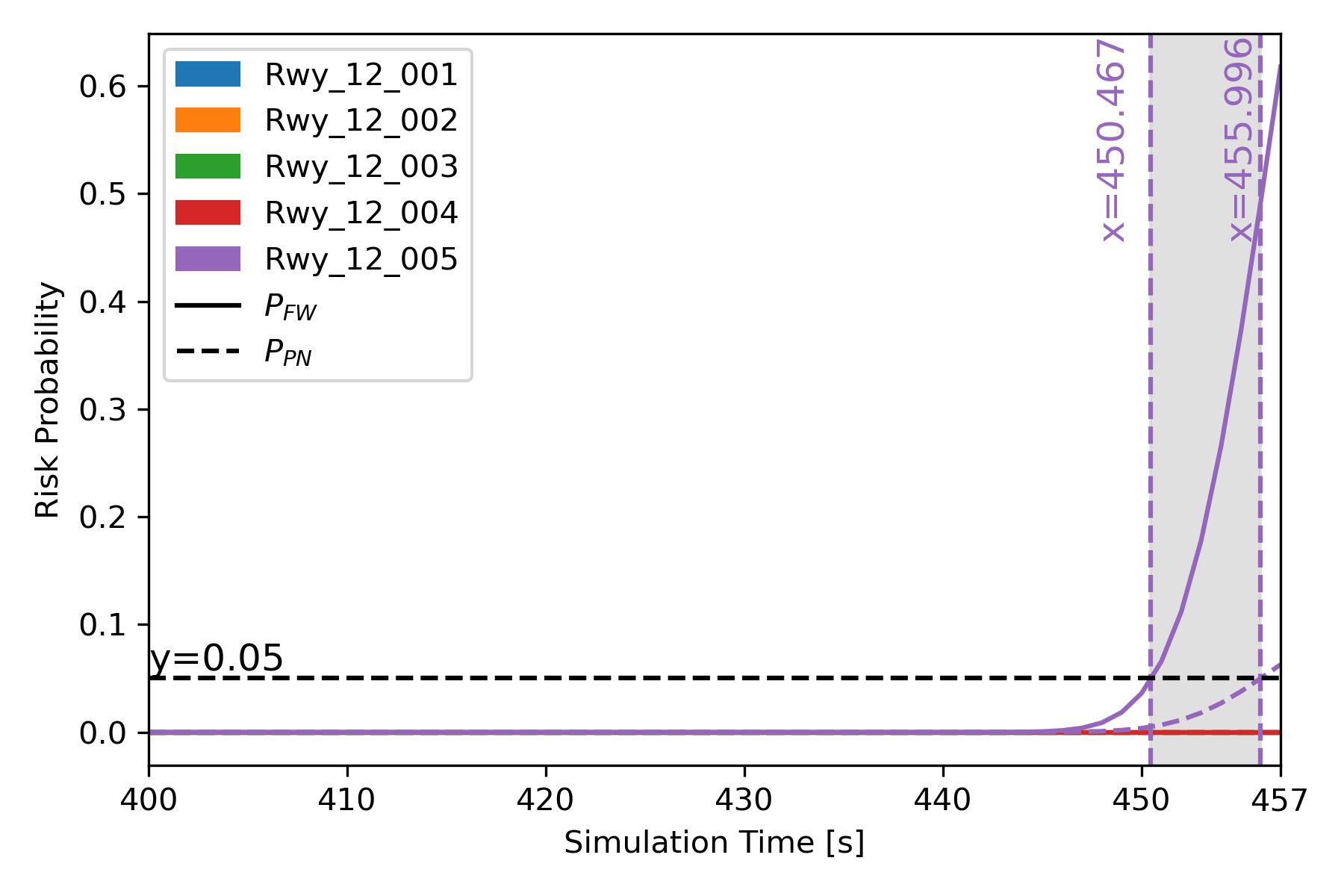}
        \caption{Tenerife Simulation with safety threshold at 0.05.}
        \label{fig: case3-0.05}
    \end{subfigure}
    \begin{subfigure}[t]{0.75\textwidth}
        \centering
        \includegraphics[width=\textwidth]{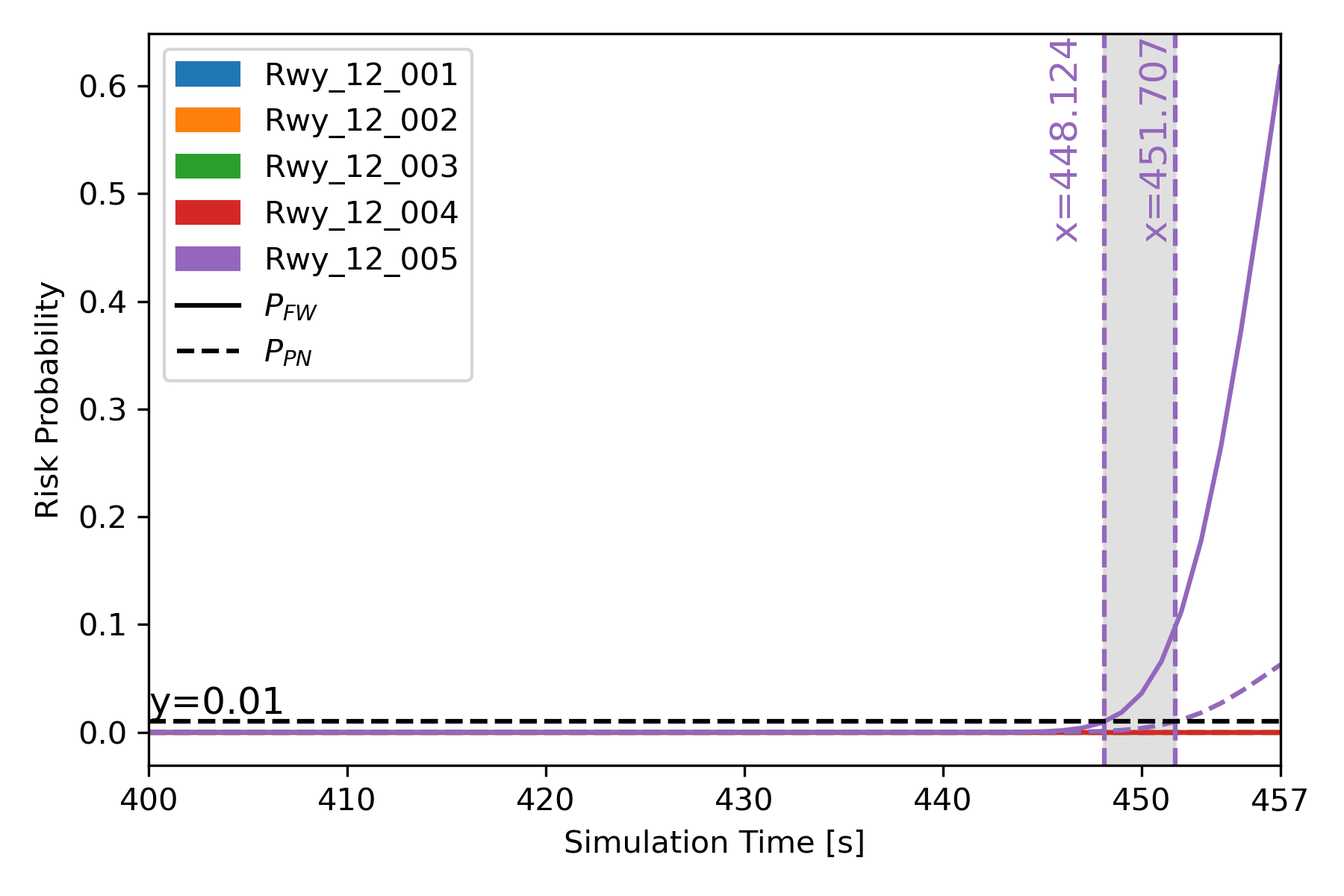}
        \caption{Tenerife Simulation with safety threshold at 0.01.}
        \label{fig: case3-0.01}
    \end{subfigure}
\caption{The real-time risk probability calculated based on the Tenerife airport disaster simulation. FW and PN risk profiles are shown for each overlapping node, with vertical lines marking the threshold crossings.}
\label{fig: case3-risk}
\end{figure}

\Cref{fig: case3-risk} visualizes the risk assessment results with the formulation in \Cref{subsec: riskmodel}. Specifically, the dynamic risk is calculated with the approach introduced in \Cref{subsubsec: real-time-risk}, where the risk is repeatedly calculated at a fixed time interval. The real-time risk is shown at each overlapping node in \Cref{fig: case3-risk}. In this scenario, overlapping nodes span \texttt{Rwy\_12\_001} to \texttt{Rwy\_12\_005}, but the collision occurs at the final overlapping node \texttt{Rwy\_12\_005}. Throughout most of the 457 second simulation, the risk curves remain near zero. A sharp rise appears only in the final $\sim$10 seconds, when the two aircraft converge on the same segment. With a risk threshold of 0.05, FW triggers a warning at 450 s while PN triggers at 456 s. Under the more conservative threshold of 0.01, the warning occurs even earlier (448 s for FW, 452 s for PN). 

These results emphasize two features. First, the model correctly localizes the risk to the true collision node, showing negligible risk at upstream intersections. Second, the FW formulation consistently provides a larger lead time compared to PN, in line with its scaling by aircraft geometry and speed. The steep terminal increase mirrors the operational reality: risk was negligible until both aircraft were committed to the final segment, at which point it rose rapidly to critical levels.

\section{Conclusions \label{sec: conclusion}}
In this paper, we introduce a novel Language AI-powered framework for understanding pilot-ATC communication to enhance surface collision risk assessment in ground movement. Our work bridges the gap between traditional surface safety systems (i.e., ASSC) and advanced natural language processing techniques by integrating NER with a surface collision risk model. The proposed approach demonstrates how language-extracted insights from ATC communications can significantly improve surface movement collision risk estimation, as well as provide a reference for compliance monitoring. 

Our hybrid learning framework lies in the ATC Rule-Enhanced Named Entity Recognition (NER) model, which utilizes domain-specific rules from two ATC manuals regulated by the FAA. Experimental results show that integrating these rules substantially improves NER performance across multiple token-level embedding models, with multilingual RoBERTa and BART models achieving the highest F1 scores but sacrificing the inference time. The study on time-space complexity trade-off of such a hybrid model not only enhances the recognition of critical entities such as flight callsigns, aircraft states, and destination intents but also maintains computational efficiency suitable for near real-time applications.

Our surface collision risk model leverages node-link graph structures of airport layouts and models aircraft taxi speeds with log-normal distributions. Through probabilistic convolution techniques, the model estimates the likelihood of aircraft reaching potential collision nodes simultaneously. Moreover, we propose the real-time risk assessment framework to obtain the collision risk time series at every overlapping node, and demonstrate effectiveness through three real-world case studies. Three case studies demonstrate the effectiveness of our methodology. The risk maps generated in all cases accurately highlighted high-risk nodes and demonstrated the practical utility of the in-time collision risk warning in real-world scenarios.

We acknowledge several limitations and future directions for this research work. First, the NER component can be strengthened by adopting aviation-domain embeddings (e.g., Aviation-BERT \citep{chandra2023aviation}) with a tailored classifier such as a Bi-LSTM-CRF head \citep{chandra2024aviation}. Second, a broader and better-engineered ATC rule dictionary is needed to ensure completeness and support real-world deployment. Third, our current risk model monitors pairwise surface conflicts only; extending it to multi-aircraft interactions is essential for comprehensive safety assessment. Fourth, the model presently assumes static taxi-speed distributions; developing dynamic, context-aware speed models that reflect real-time traffic and weather should improve predictive accuracy. Fifth, computer vision can augment language-based predictions by detecting and tracking aircraft to cross-check compliance with clearances and enhance the risk-assessment pipeline in \Cref{fig: flowchart}. Finally, since embedding security is an increasing concern \citep{seyyar2022attack, li2023sentence}, differentially private fine-tuning \citep{anil2021large} offers a promising path to mitigate inversion attacks and bolster the safety of the overall framework.

This work presents a significant step toward harnessing the capabilities of Language AI in aviation safety. By coupling advanced NLP techniques with established risk modeling methodologies, this work enhances situational awareness and provides a robust tool for mitigating surface collision risks in increasingly complex airport environments. The integration of language understanding into surface movement risk assessment introduces new avenues for enhancing airport safety. By enabling automated detection of miscommunications and deviations from ATC instructions, our approach contributes to proactive real-time incident prevention to enhance aviation safety.

\section*{Acknowledgment}
This work was supported by the National Aeronautics and Space Administration (NASA) University Leadership Initiative (ULI) program under project “Autonomous Aerial Cargo Operations at Scale”, via grant No. 80NSSC21M071 to the University of Texas at Austin. Any opinions, findings, conclusions, or recommendations expressed in this material are those of the authors and do not necessarily reflect the views of the project sponsor.

\bibliography{ref}

\appendix

\section{Spatiotemporal Risk Error Estimate \label{sec: error-proof}}
This section is a discussion to analysis the approximation error in the spatiotemporal risk formulation. In \Cref{eq: fw-formulation}, the linearized equation,

\begin{equation}
    P(\text{c}) = \int_0^\infty f_{\Gamma_1}(t|x_c)\,  \left[ \int_{x_c-r_c}^{x_c+r_c} f_{X_2}(x|t)\, dx \right] dt.
\end{equation}

We do a change of variables to normalize $x$ by $x_c$, and set the integral limits to $\epsilon = \frac{r_c}{x_c}$, the non-dimensional ratio of collision radius to distance traveled.

\begin{equation}
\begin{aligned}
P(\text{c}) &= \int_0^\infty f_{\Gamma_1}(t|x_c)\,  \left[ \int_{-\epsilon}^{\epsilon} f_{X_2}(x_c(1+x)|t)\, x_c\, dx \right] dt\\
&=\int_0^\infty f_{\Gamma_1}(t|x_c)\,  \left[ \int_{-\epsilon}^{\epsilon} \mathbb{E}(v^{-1}_2) f_{\Gamma_2}\big(t|x_c(1+x)\big)\,  x_c\, dx \right] dt\\
&\approx\mathbb{E}(v^{-1}_2)\int_{-\infty}^{\infty} f_{\Gamma_1}(t|x_c)\bigg(2r_c f_{\Gamma_2}(t|x_c)+\frac{r_c}{3}\epsilon^2 \frac{\partial^2}{\partial x^2} f_{\Gamma_2}\big(t|x_c(1+x)\big) \bigg|_0 + o\big(r_c\epsilon^4\big) \bigg)\, dt\\
&\leq \int_{-\infty}^{\infty} f_{\Gamma_1}(t|x_c)\bigg(2r_c f_{\Gamma_2}(t|x_c)+\frac{r_c}{3}\epsilon^2 \sup_t\frac{\partial^2}{\partial x^2}f_{\Gamma_2}\big(t|x_c(1+x)\big) + o\big(r_c\epsilon^4\big) \bigg)\, dt\\
&\leq 2r_c\int_{-\infty}^{\infty} f_{\Gamma_1}(t|x_c)\ f_{\Gamma_2}(t|x_c)dt \\&+\frac{r_c}{3}\epsilon^2 \sup_t\frac{\partial^2}{\partial x^2}f_{\Gamma_2}\big(t|x_c(1+x)\big) + o\big(r_c\epsilon^4\big) \\
\end{aligned}
\label{eq: supintegral}\end{equation}

The $\Gamma_2$ distribution is computed with a convolution, as it is the sum of the time to reach the previous node from the starting location $\Gamma_{2,0}$, and the time to reach the collision location from the previous node $\tau_2$. The supremum of the second spatial derivative of $\Gamma_2$ may be bounded by using the supremum of the second spatial derivative of $\tau_2$.

\begin{equation}
\begin{aligned}
\frac{\partial^2}{\partial x^2} f_{\Gamma_2}(t|x) &= \frac{\partial^2}{\partial x^2}\int_{-\infty}^\infty f_{\Gamma_{2,0}}(\tau)f_{\tau_2}(t-\tau|x)d\tau\\
&= \int_{-\infty}^\infty f_{\Gamma_{2,0}}(\tau)\frac{\partial^2}{\partial x^2}f_{\tau_2}(t-\tau|x)d\tau\\
&\leq \int_{-\infty}^\infty f_{\Gamma_{2,0}}(\tau) \sup_t\frac{\partial^2}{\partial x^2}f_{\tau_2}(t|x)d\tau\\
&\leq  \sup_t\frac{\partial^2}{\partial x^2}f_{\tau_2}(t|x)
\end{aligned}
\end{equation}

We perform a change of variables from the time distribution to the log-normal velocity distribution parameterized by $\mu$ and $\sigma$, and take the second derivative with respect to space.

\begin{equation}
\begin{aligned}
\frac{\partial^2}{\partial x^2}f_{\tau_i}(t|x)&=\frac{\partial^2}{\partial x^2}(f_{V_i}(\frac{x}{t})|\frac{x}{ t^2}|)\\
&=\frac{1}{tx^2\sigma^5\sqrt{2 \pi} }\exp[-\frac{(\mu -\ln x +\ln t)^2}{2\sigma^2} ]\bigg( \ln^2t  \\ &+\ln t \big(2\mu -\sigma^2 -2 \ln x \big) \\
&+ \big( \mu^2 -\sigma^2 - \mu\sigma^2 + \ln^2 x  -2\mu \ln x +\sigma^2 \ln x \big) \bigg)
\end{aligned}
\end{equation}

It is apparent that this function may be reorganized as the product of a second-order polynomial with the exponential of another second-order polynomial in terms of $\tau = \ln t$. The $x^2$ term in the denominator is kept aside to provide a non-dimensional ratio $\frac{r_c}{x_c}$.

\begin{equation}
\begin{aligned}
a &= 1\\
b &= 2\mu -\sigma^2 -2 \ln x\\
c &= \mu^2 -\sigma^2 - \mu\sigma^2 + \ln^2 x  -2\mu \ln x +\sigma^2 \ln x\\
\alpha &= \frac{1}{2\sigma^2}\\
\beta &= \frac{\ln x-\mu-\sigma^2}{\sigma^2}\\
\gamma &= -\frac{(\mu-\ln x)^2}{2\sigma^2}\\
K &= \frac{1}{\sigma^5\sqrt{2 \pi} }\\
\end{aligned}
\end{equation}

\begin{equation}
\begin{aligned}
\frac{\partial^2}{\partial x^2}f_{\Delta T_i}(t|x)&=\frac{K}{x^2}\exp\big[-\alpha \tau^2+\beta\tau + \gamma \big]\big(a \tau^2 +b \tau +c \big)
\end{aligned}
\end{equation}

We complete the square and perform a change of variables to leave only a quadratic term in the exponential.
\begin{equation}
\begin{aligned}
m &= \frac{\beta}{2\alpha}\\
y &= \tau-m\\
K' &= K\exp \big[\frac{\beta^2}{4\alpha}+\gamma\big]\\
A_2 &= a\\
A_1 &= b+2am\\
A_0 &= c+bm+am^2
\end{aligned}
\end{equation}

\begin{equation}
\begin{aligned}
-\alpha \tau^2+\beta\tau + \gamma &= -\alpha y^2 + \frac{\beta^2}{4\alpha}+\gamma
\end{aligned}
\end{equation}
\begin{equation}
\begin{aligned}
a \tau^2 +b \tau +c &= A_2 y^2 + A_1 y +A_0
\end{aligned}
\end{equation}

\begin{equation}
\begin{aligned}
\frac{\partial^2}{\partial x^2}f_{\tau_i}(t|x) = \frac{K'}{x^2}(A_2 y^2 + A_1 y +A_0)e^{-\alpha y^2}
\end{aligned}
\end{equation}

We use known suprema for the individual terms of the expression to bound the supremum for positive $t$.

\begin{equation}
\begin{aligned}
\sup_{y\in \mathbb R} e^{-\alpha y^2} = 1, \quad \sup_{y\in \mathbb R} |y|e^{-\alpha y^2} = \frac{e^{-\frac{1}{2}}}{\sqrt{2\alpha}},\quad \sup_{y\in \mathbb R} y^2e^{-\alpha y^2} = \frac{e^{-1}}{\alpha},
\end{aligned}
\end{equation}

\begin{equation}
\begin{aligned}
\sup_{t\in \mathbb{R}^+}\frac{\partial^2}{\partial x^2}f_{ \tau_i}(t|x)=\frac{C(x,\mu,\sigma)}{x^2}\leq \frac{K'}{x^2} \bigg(\frac{A_2e^{-1}}{\alpha}+\frac{|A_1|e^{-\frac{1}{2}}}{\sqrt{2\alpha}}+A_0^+\bigg)
\end{aligned}
\label{eq: supbound}\end{equation}

\begin{figure}
\centering
\includegraphics[width=0.85\textwidth]
{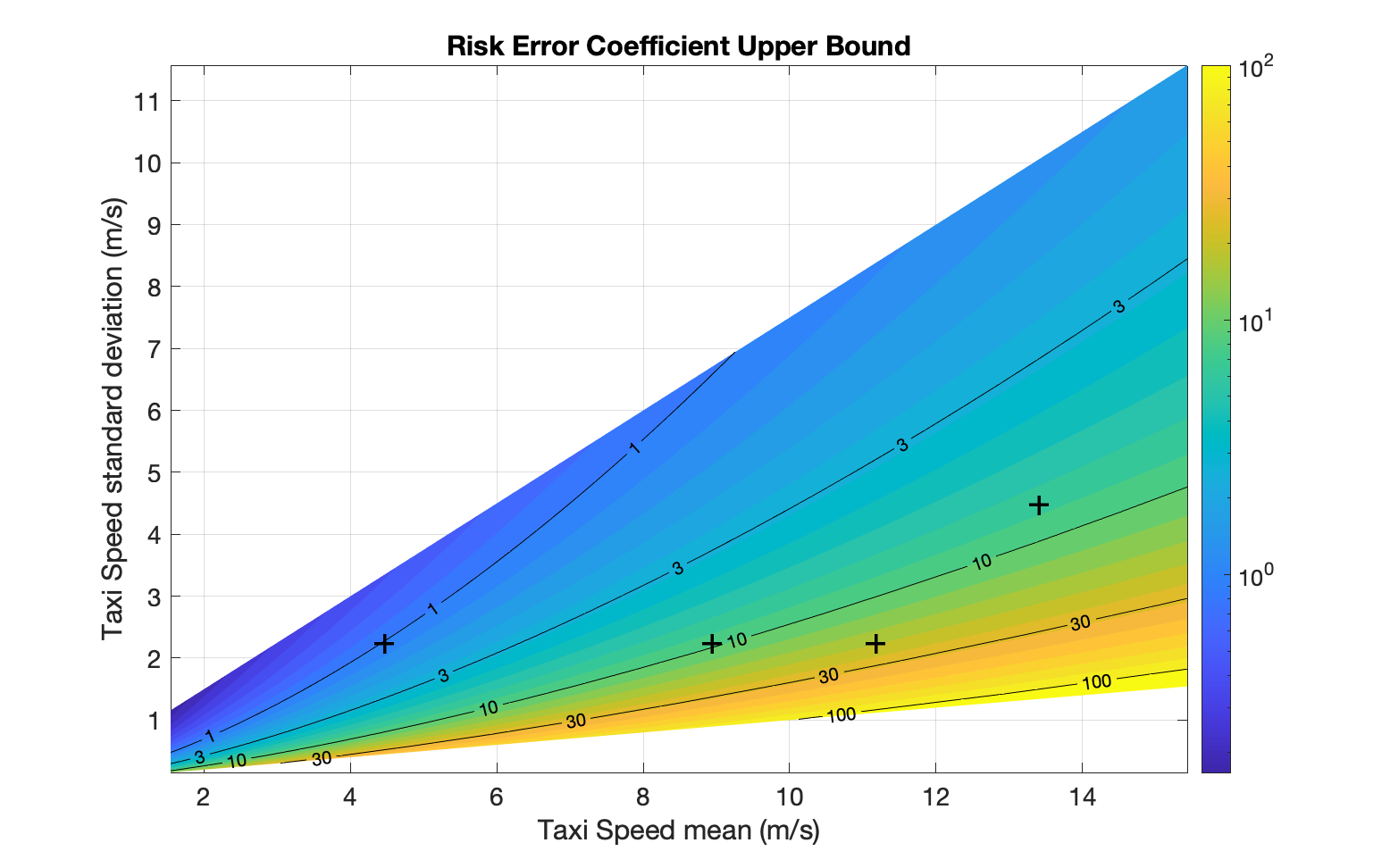}
\caption{The upper bound on risk error coefficient $C(x,\mu,\sigma)$ from \eqref{eq: supbound} is computed at a reference distance of 50 meters. Markers indicate the sets of distribution parameters used in case studies}
\label{fig: riskerrorbound}
\end{figure}

The dependence of risk error coefficient $C$ on taxi speed distribution parameters is shown in \Cref{fig: riskerrorbound}. Short distances with relatively low speed uncertainty lead to higher error bounds. The upper bound is evaluated at $x_c$ and plugged back into \eqref{eq: supintegral}. The $x_c$ denominator term is canceled because the second derivative is with respect to the position normalized by $x_c$. 

\begin{equation}
\begin{aligned}
P(\text{c}) 
&\leq 2r_c\int_{-\infty}^{\infty} f_{\Gamma_1}(t|x_c)\ f_{\Gamma_2}(t|x_c)dt \\&+\frac{r_c}{3}\epsilon^2K' \bigg(\frac{A_2e^{-1}}{\alpha}+\frac{|A_1|e^{-\frac{1}{2}}}{\sqrt{2\alpha}}+A_0^+\bigg) + o\big(r_c\epsilon^4\big) \\
\end{aligned}
\label{eq: supintegral}\end{equation}

In conclusion, the approximation error is dependent on $\epsilon$, the ratio of collision radius to taxi distance, with an order of 2. 
\end{document}